\documentclass[12pt, preprint]{aastex}

\shorttitle{Cluster Structure in Simulations}

\shortauthors{Jeltema et al.}

\begin{document}

\title{ Cluster Structure in Cosmological Simulations I: Correlation to Observables, Mass Estimates, and Evolution }

\author{Tesla E. Jeltema\altaffilmark{1,2}, Eric J. Hallman\altaffilmark{3,4}, Jack O. Burns\altaffilmark{4}, and Patrick M. Motl\altaffilmark{5}}

\altaffiltext{1}{The Observatories of the Carnegie Institution of Washington, 813 Santa Barbara St., Pasadena, CA 91101; tesla@ucolick.org}
\altaffiltext{2}{Morrison Fellow, UCO/Lick Observatories, 1156 High St., Santa Cruz, CA 95064}
\altaffiltext{3}{NSF Astronomy and Astrophysics Postdoctoral Fellow}
\altaffiltext{4}{Center for Astrophysics and Space Astronomy, University of Colorado, Boulder, CO 80309}
\altaffiltext{5}{Department of Physics and Astronomy, Louisiana State University, Baton Rouge, LA 70803}

\begin{abstract}

We use \textit{Enzo}, a hybrid Eulerian AMR/N-body code including non-gravitational heating and cooling, 
to explore the morphology of the
X-ray gas in clusters of galaxies and its evolution in current generation cosmological simulations.  We employ 
and compare two observationally motivated structure measures: power ratios and centroid 
shift.  Overall, the structure of our simulated clusters compares remarkably well to 
low-redshift observations, although some differences remain that may point to incomplete 
gas physics.  We find no dependence on cluster structure in the mass-observable scaling 
relations, $T_X-M$ and $Y_X-M$, when using the true cluster masses.  However, estimates 
of the total mass based on the assumption of hydrostatic equilibrium, as assumed in 
observational studies, are systematically low.  We show that the hydrostatic mass bias
strongly correlates with cluster structure and, more weakly, with cluster mass.  
When the hydrostatic masses are used, the mass-observable scaling relations and gas mass 
fractions depend significantly on cluster morphology, and the true relations are not 
recovered even if the most relaxed clusters are used.  We show that cluster structure, 
via the power ratios, can be used to effectively correct the hydrostatic mass estimates and 
mass-scaling relations, suggesting that we can calibrate for this systematic effect in 
cosmological studies.  Similar to observational studies, we find that cluster structure, 
particularly centroid shift, evolves with redshift.  This evolution is mild but will lead 
to additional errors at high redshift.  Projection along the line of sight leads to 
significant uncertainty in the structure of individual clusters: less than 50\% of 
clusters which appear relaxed in projection based on our structure measures are truly 
relaxed.

\end{abstract}


\section{ INTRODUCTION }

The growth of large-scale structure in the universe is a powerful probe of cosmology.  In particular, the largest collapsed structures, clusters of galaxies, are often used as a cosmological probe (see Voit 2005 for a recent review).  Clusters are approximately self-similar objects, and in the standard paradigm (a cold dark matter universe with structure forming from Gaussian initial density perturbations), their bulk properties can largely be predicted simply from gravity and the composition of the universe (e.g. the fractions of baryons, dark matter, and dark energy).  However, as with other cosmological probes (e.g. SN Ia) the accuracy of cosmological studies with clusters is currently limited by systematics, and the precise constraints now being sought on parameters like dark energy require a more detailed understanding of cluster formation and evolution (e.g. Voit 2005).  In particular, cosmological constraints from cluster studies are limited by the accuracy to which we can determine cluster mass.

One major source of uncertainty in cluster mass estimates is the cluster's dynamical state.  Most methods of estimating cluster mass (e.g. from the velocity dispersion of the cluster galaxies, X-ray temperature, or gas mass) assume that the cluster and its gas are in equilibrium.  However, massive clusters formed relatively recently and are often dynamically young; clusters form and grow through relatively frequent mergers with other clusters and groups of galaxies (e.g. Cohn \& White 2005 and references therein).  The common nature of cluster mergers is supported by the high fraction of clusters observed to have significant substructure or disturbed, asymmetric morphologies.  The fraction of substructured/disturbed clusters observed depends on the method and wavelength considered, but for X-ray observations at low redshift this fraction is 40-70\% (Mohr et al. 1995; Jones \& Forman 1999; Schuecker et al. 2001; Kolokotronis et al. 2001).  Recent X-ray observations with \textit{Chandra} of clusters out to $z\sim1$ also show that cluster structure evolves with redshift: clusters are more disturbed at higher redshifts \cite{J05, M07, H07}.  This evolution is expected in hierarchical structure formation, but could lead to additional biases in the selection of clusters and their observable properties at high redshift.  The evolution of cluster structure in simulations is less clear, but higher merger rates are found at high redshift (e.g. Cohn \& White 2005; Burns et al. 2007) and morphological evolution is detected in a couple of recent simulations (Rahman et al. 2006; Kay et al. 2007).

Cluster structure is also a powerful probe of the physics underlying cluster formation.  The formation epoch of clusters and thus the frequency of mergers depends on cosmology (e.g. Richstone, Loeb, \& Turner 1992).  In low density universes, clusters form earlier and will be on average more relaxed in the present epoch.  In a $\Lambda$CDM universe, the growth of clusters slows when the universe transitions to dark energy dominated.  Cluster structure also depends on gas physics.  Cooling, for example, increases the central concentration of halos and makes them more spherical (Kazantzidis et al. 2004; Rahman et al. 2006).

In this paper, we study cluster structure in hydrodynamical simulations based on \textit{Enzo}, a hybrid Eulerian adaptive mesh refinement (AMR)/N-body code \cite{NB99,O05}, allowing us to directly probe the relationship between cluster structure and observable cluster properties.  The structure of simulated clusters compared to observations tests the reality of the models of cosmology and gas physics employed in current simulations.  Most previous studies have looked at the effects of mergers on clusters using simulations of idealized mergers (e.g. Schindler \& Mueller 1993; Roettiger, Burns, \& Loken 1996; Ricker \& Sarazin 2001; Ritchie \& Thomas 2002; Poole et al. 2007), individual mergers pulled from cosmological simulations (e.g. Mathiesen \& Evrard 2001; Rowley, Thomas, \& Kay 2004; Motl et al. 2005), or simulated clusters selected based on a visually disturbed appearance (e.g. Kravtsov, Vikhlinin, \& Nagai 2006; Nagai, Vikhlinin, \& Kravtsov 2007).  These studies show that mergers lead to significant variation (usually boosts but it depends on the phase of the merger) in cluster luminosity, temperature, and central Sunyaev-Zeldovich (SZ) flux.  In many cases, mergers are found to boost clusters along the scaling relations rather than increasing the scatter in these relations, but the exact behavior depends on the type and phase of the merger.  Promisingly, some scaling relations, like $Y_X-M$ and $Y_{SZ}-M$, appear to be fairly insensitive to mergers \cite{M05, K06, P07}.

In this work, we use observationally motivated measures of cluster structure rather than the time evolution of individual mergers to allow a quantitative comparison to the observed degree of substructure or asymmetry.  The use of a quantitative measure of structure also allows us to explore whether these measures can be used either as a tool to select clusters or as a third parameter to improve cluster mass estimates.  The large volume of these simulations (256 $h^{-1}$ Mpc on a side) allows us to study a large number of massive clusters in a cosmological setting, giving numbers directly comparable to cluster surveys.  We choose to focus on simulated X-ray observations of clusters.  X-ray observations currently provide both the best depth and resolution for studies of cluster structure (compared to the low-resolution of weak lensing and SZ observations and the sparse distribution of cluster galaxies), and X-ray surveys are the most commonly used to constrain cosmology.  We utilize two common measures of structure based on the X-ray surface brightness distribution: the power ratios (e.g. Buote \& Tsai 1995) and centroid shift (e.g. Mohr et al. 1993), which are described in \S 2 along with the details of the simulations.  

Our results are presented in section 3.  We first explore the effect of projection along the line of sight on observable cluster structure (\S 3.1).  We then look for correlations between cluster morphology and overall cluster properties like mass, luminosity, temperature and $Y_X$ as well as the effects of cluster structure on the scaling relations used to estimate cluster mass (\S 3.2).  In particular, we find a strong correlation between cluster structure and inaccuracy in hydrostatic mass estimates.  Finally, we compare the structure of our simulated clusters to low-redshift observations (\S 3.3), and look for evolution in cluster structure with redshift (\S 3.4).

\section{ METHODOLOGY }

\subsection{ Numerical Simulations }

In this study, we use numerical galaxy clusters simulated using the cosmological hydro/N-body
adaptive mesh refinement (AMR) code {\it Enzo} \cite{O05}
(http://cosmos.ucsd.edu/enzo). {\it Enzo} evolves both the dark matter and
baryonic fluid in the clusters utilizing the piecewise parabolic
method (PPM) for the hydrodynamics. The initial dark matter power spectrum from Eisenstein \& Hu (1999) was
initialized on the grid at $z=30$. An initial
low-resolution simulation was used to identify clusters in a volume of
256 $h^{-1}$ Mpc on a side using 128$^{3}$ dark matter particles and
grid zones.  High-resolution simulations were then
performed that evolved the entire volume but adaptively refined 50
smaller regions around the largest clusters identified on the low
resolution grid. Each of these 50 regions first is refined with two
nested static grids, with double the spatial resolution of their
parent grid (thus 4 times more resolved than the root grid). Within
the nested grids,  the dark matter particles have a
mass resolution of $9 \times 10^9 h^{-1} M_{\odot}$. Then, within the
nested static grids, we evolve the simulation with 5 additional levels
of adaptive refinement, again with a factor of two increase in spatial
resolution at each level. These simulated clusters have a peak resolution of
15.6$h^{-1}$kpc on the finest simulation grid. Refinement of high
density regions is done as described in Motl at al. (2004). The criteria for refinement are both
baryonic and dark matter overdensity thresholds. We assume a
concordance $\Lambda$CDM cosmological model with the following
parameters: $\Omega_b = 0.026, \Omega_m = 0.3, \Omega_\Lambda = 0.7,
h = 0.7,$ and $\sigma_8 = 0.928$.  

At each of 18 redshift outputs, clusters with masses greater than $10^{14} M_{\odot}$ are identified in the refined regions, but unless otherwise noted in this work we chose to use a slightly higher mass cut of $2 \times 10^{14} M_{\odot}$.  We consider only redshifts less than or equal to 1.5, and our final sample includes 61 clusters at $z=0.04$, 25 clusters at $z=1.0$, and 5 clusters at $z=1.5$.

The simulations used for this analysis were described in
Motl et al. (2005) and Hallman et al. (2006), and include models for radiative cooling and star
formation with thermal feedback from supernovae using the
Cen \& Ostriker (1992) algorithm. Radiative cooling is calculated from a tabulated cooling curve derived
from a Raymond-Smith plasma emission model \cite{Br95} assuming a constant
metallicity of 0.3 relative to solar.  The cooling curve is truncated
below a temperature of $10^4$ K.  Every timestep, we calculate the
energy radiated from each cell and remove that amount of energy from
the cluster gas \cite{M04, B07}. A full description of the
simulated cluster catalog will be presented in Hallman et al.~(2007). 

We have generated synthetic X-ray images for each cluster in order to apply
our structure measures to the projected X-ray morphology as has
been performed in observational analyses. Images are created by
calculating a Raymond-Smith X-ray emissivity in the 0.3-7 keV band at
each grid cell in a 6$h^{-1}$ Mpc box centered on each numerical
cluster. The values of the electron density and gas temperature at
each zone are used in this calculation. Then the emissivities are
integrated along a succession of rays covering the image plane at the
maximum physical resolution of the cluster. We then convert the
physical scale to the angular scale that the cluster would subtend at
the redshift appropriate to its evolutionary state in the
simulation. We also enforce a maximum angular resolution of 0.5
arcseconds in the images, the maximum resolution of \textit{Chandra}
images. Additionally, in order to more directly compare the absolute
X-ray flux values of simulated clusters to observed clusters, we have
rescaled the X-ray surface brightness to account for the difference
between the value of $\Omega_b$ used in our simulations and that
preferred by the WMAP III results. Images are created to include at least the virial radius for each cluster.  Clusters are considered to be a single object when they overlap within their projected virial radii.

In the following analysis, X-ray luminosities are determined by integrating the flux in the
X-ray images over the appropriate aperture and are given for the 0.3-7 keV band.  Temperatures refer to 
average spectral temperature and exclude a cool core if present.  Cool cores are excluded using the temperature profile; 
we remove regions where the temperature decreases by $\ge 20$\% compared to the surrounding region.  Gas masses, 
including those used to calculate $Y_X$, are estimated through deprojection of the X-ray images under the assumption of spherical symmetry and using a universal temperature profile (Hallman et al.~2006).  Since we are interested in the effects of substructure, we do not remove clumps from our determination of average cluster properties, as is sometimes done by observers.  The removal of substructure in observations will necessarily depend on the depth of the data and will not be possible in large shallow surveys, and, in addition, structure along the line of sight and shock regions are not removed.  We seek to calibrate for these effects rather than to remove them.  With the substructures included the average temperatures of some clusters may be different, most likely lower, than they would be with the substructures removed.

\subsection{ Power Ratios }

The power ratios have been described in detail elsewhere (Buote \& Tsai 1995, 1996; Jeltema et al. 2005); here we give a brief summary of this method.  Essentially, this method entails calculating the multipole moments of the X-ray surface brightness in a circular aperture centered on the cluster's centroid.  The moments, $a_m$ and $b_m$ given below, are sensitive to asymmetries in the surface brightness distribution and are, therefore, sensitive to substructure.
The physical motivation for this method is that it is related to the multipole expansion of the two-dimensional gravitational potential.  The multipole expansion of the two-dimensional gravitational potential is
\begin{equation}
\Psi(R,\phi) = -2Ga_0\ln\left({1 \over R}\right) -2G
\sum^{\infty}_{m=1} {1\over m R^m}\left(a_m\cos m\phi + b_m\sin
m\phi\right). \label{eqn.multipole}
\end{equation}
and the moments $a_m$ and $b_m$ are
\begin{eqnarray}
a_m(R) & = & \int_{R^{\prime}\le R} \Sigma(\vec x^{\prime})
\left(R^{\prime}\right)^m \cos m\phi^{\prime} d^2x^{\prime}, \nonumber \\
b_m(R) & = & \int_{R^{\prime}\le R} \Sigma(\vec x^{\prime})
\left(R^{\prime}\right)^m \sin m\phi^{\prime} d^2x^{\prime}, \nonumber
\end{eqnarray}
where $\vec x^{\prime} = (R^{\prime},\phi^{\prime})$ and $\Sigma$ is the surface mass density.  In the case of X-ray studies, X-ray surface brightness replaces surface mass density in the calculation of the power ratios.  
 X-ray surface brightness is proportional to the gas density squared and generally shows the same qualitative structure as the projected mass density, allowing a similar quantitative classification of clusters.  

The powers are formed by integrating the magnitude of $\Psi_m$, the \textit{m}th term in the multipole expansion of the potential given in equation (1), over a circle of radius $R$,
\begin{equation}
P_m(R)={1 \over 2\pi}\int^{2\pi}_0\Psi_m(R, \phi)\Psi_m(R, \phi)d\phi.
\end{equation}
Ignoring factors of $2G$, this gives
\begin{equation}
P_0=\left[a_0\ln\left(R\right)\right]^2 \nonumber
\end{equation}
\begin{equation}
P_m={1\over 2m^2 R^{2m}}\left( a^2_m + b^2_m\right). \nonumber
\end{equation}
Rather than using the powers themselves, we divide by $P_0$ to normalize out flux forming the power ratios, $P_m/P_0$.  

For each cluster in the simulations, we calculate $P_2/P_0$, $P_3/P_0$, and $P_4/P_0$ over a circular aperture centered on the centroid of cluster emission (with the origin at the centroid, $P_1$ vanishes by definition).  $P_2/P_0$ measures ellipticity combined with radial fall-off in the X-ray surface brightness ($P_4/P_0$ is similar but probes smaller scales).  $P_3/P_0$, as an odd moment, is sensitive instead to deviations from mirror symmetry and insensitive to ellipticity.  A high $P_3/P_0$ is a clear indication of an asymmetric cluster structure, and this ratio shows the strongest and most unambiguous evolution with redshift in observations \cite{J05}.  In this work, we consider three choices of aperture radius: two radii of fixed physical size, 0.5 Mpc and 1 Mpc, for ease of comparison to observations and $r_{500}$, the radius at which the mean cluster density is 500 times the critical density.

Table 1 lists the median power ratios of the simulated clusters for each radius as well as the range in power ratios spanned by clusters with the central 80\% of power ratios.  These ranges reflect the overall range of power ratios while eliminating the most extreme cases.  For the aperture radius of 1 Mpc, we only include clusters whose virial radii are greater than 1 Mpc, and it should be noted that this leads to a sample with slightly higher average masses than is used for the other two radii.  The cluster virial radii all exceed 0.5 Mpc.  The average size of $r_{500}$ for clusters in our sample is 0.8 Mpc, but this radius ranges from 0.4 Mpc up to 1.6 Mpc depending on cluster mass and redshift.  The overall range of power ratios displayed by the simulated clusters compares well to the range of power ratios observed in clusters \cite{J05, B96}(see section 3.3 for further discussion).  Investigation of Table 1 reveals that the power ratios of typical clusters decrease with increasing radius as enclosed substructure becomes less significant on the scale of the aperture size.  The same trend is not present for the clusters with the highest power ratios due to the fact that increasing radius can lead to the inclusion of additional structures within the aperture.

The power ratios of the simulated clusters are plotted in Figure 1 for both the 0.5 Mpc (left) and $r_{500}$ (right) aperture radii, and the distributions look similar for $R=1$ Mpc.  Here we show the relations between sets of two of the power ratios.  Highly significant correlations exist between the power ratios, particularly in the $P_2/P_0$ - $P_4/P_0$ plane.  Similar correlations are present for observed clusters \cite{J05, B96}.  Such correlations imply that the level of structure on different scales increases and decreases together.  

\clearpage
\begin{figure}
\begin{center}
\epsscale{1.0}
\plottwo{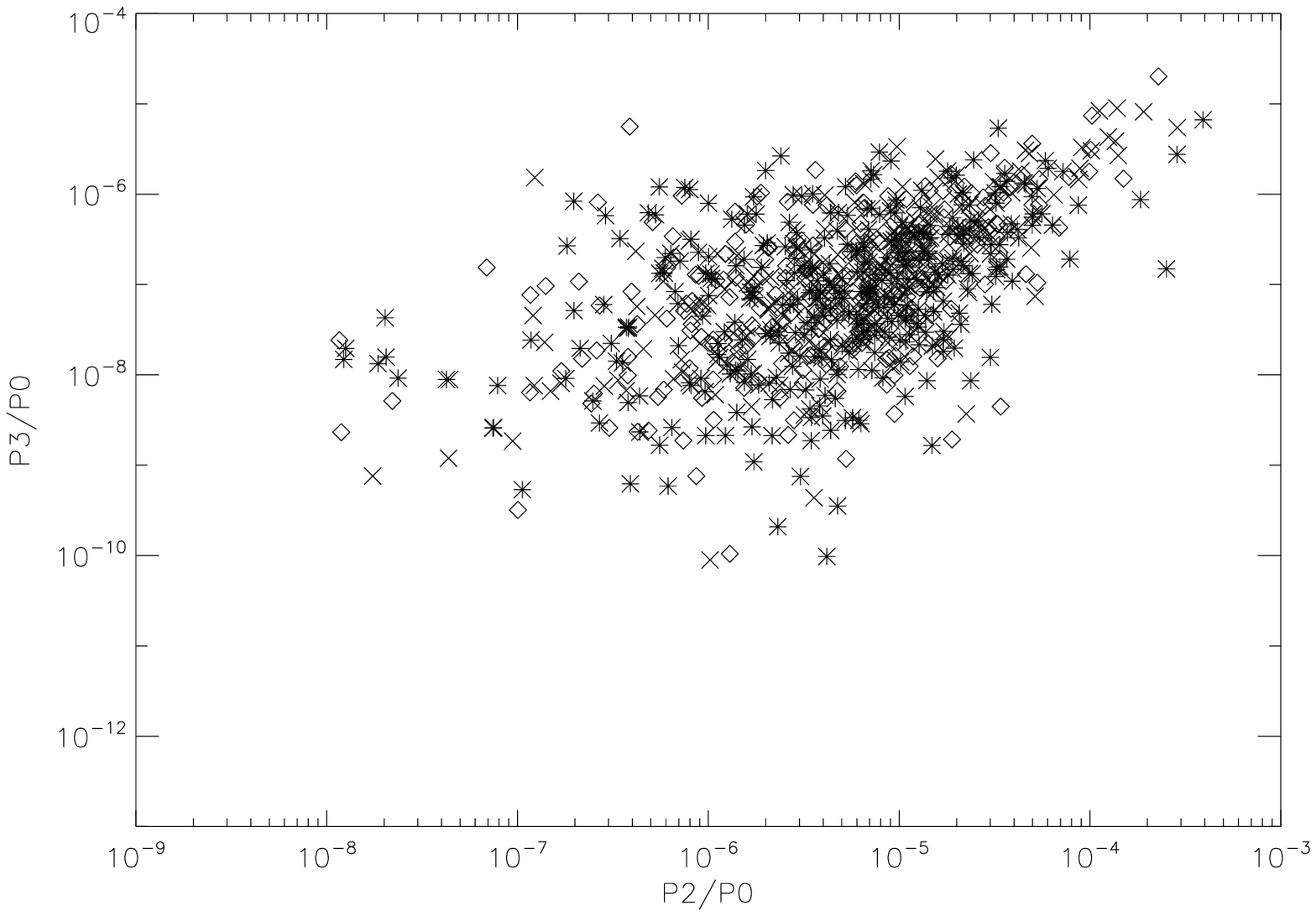}{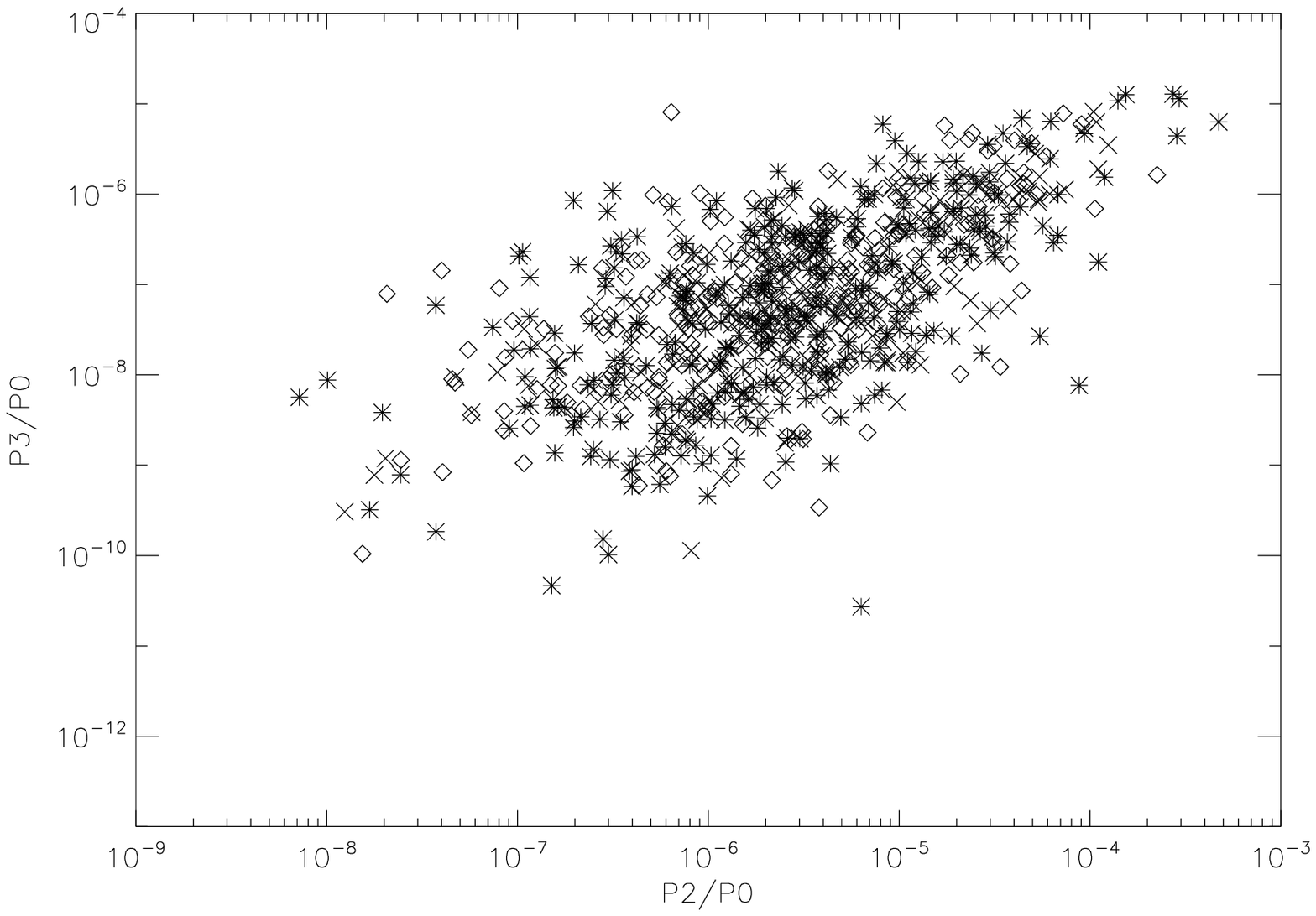}
\plottwo{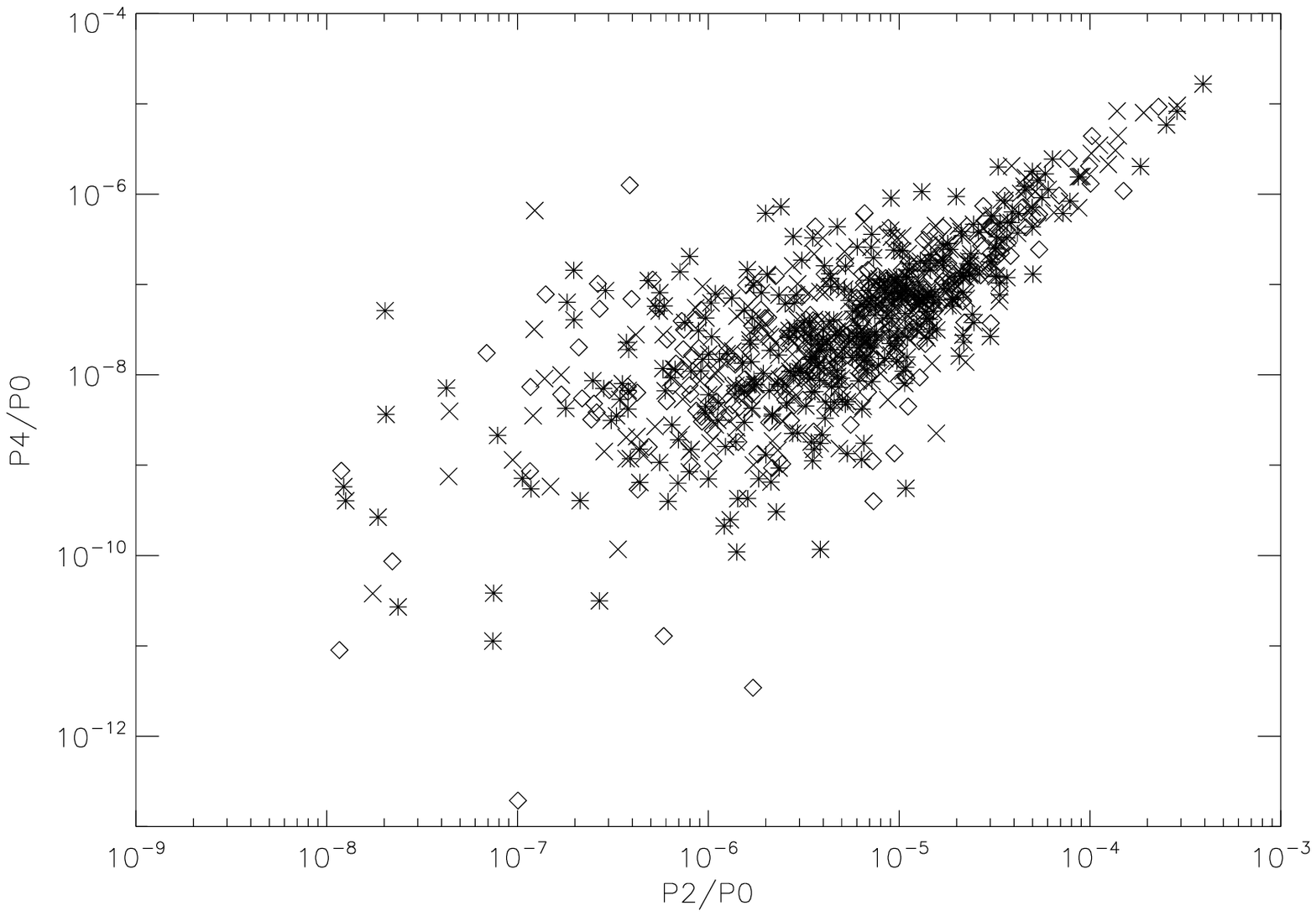}{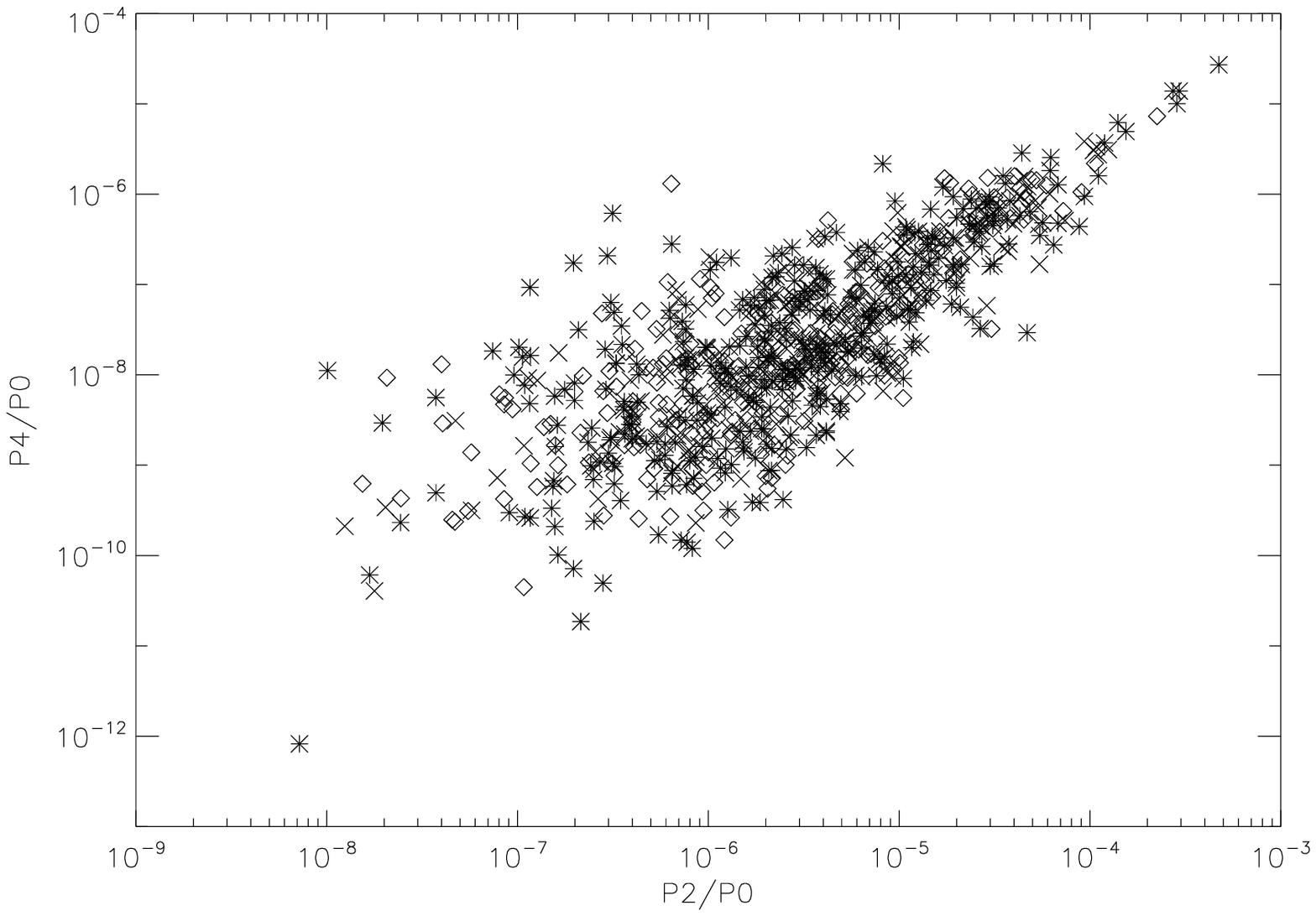}
\plottwo{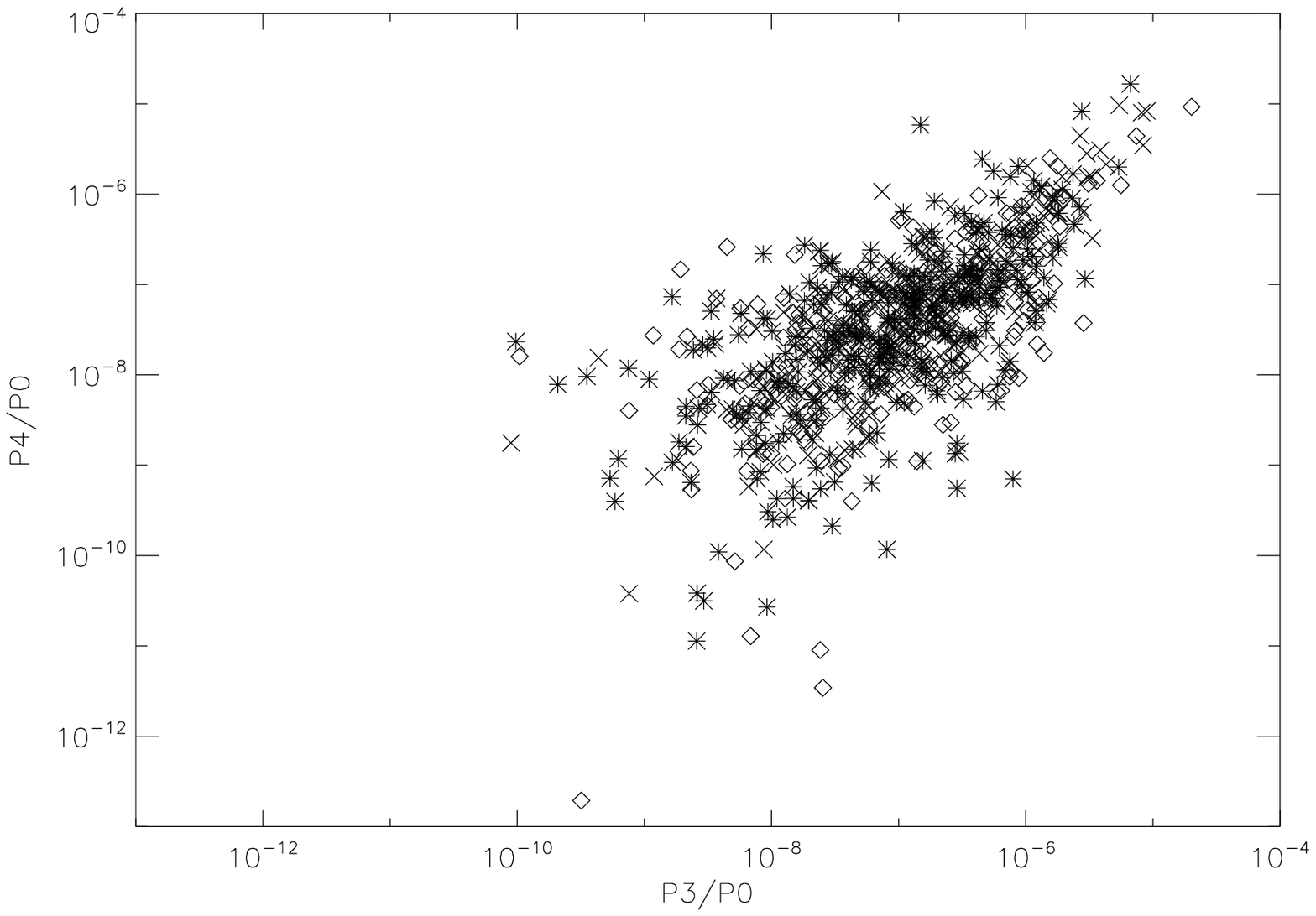}{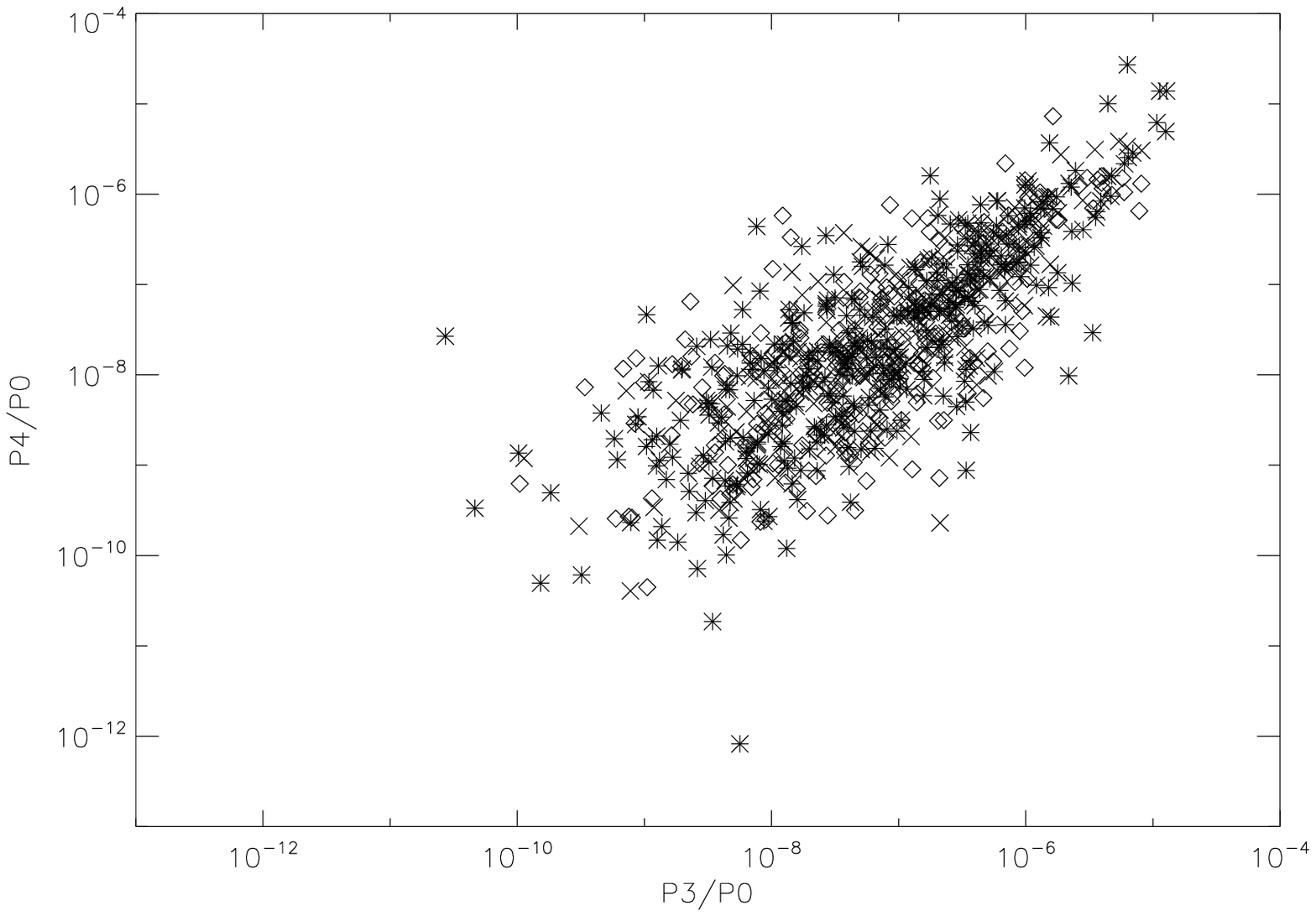}
\caption{ Correlations among the power ratios for aperture radii of 500 kpc (\textit{left}) and $r_{500}$ (\textit{right}).  Clusters are divided in to three roughly equal time bins: $z \le 0.25$ (asterisks), $0.25 < z \le 0.65$ (diamonds), and $0.65 < z \le 1.5$ (crosses). }
\end{center}
\end{figure}
\clearpage

\subsection{ Centroid Shifts }

Another common measure of observed cluster morphology, based on the first moment of the X-ray surface brightness, is centroid shifting (e.g. Mohr et al. 1993; O'Hara et al. 2006).  In this method the displacement between a cluster's core or peak from the cluster center on large scales is used to indicate deviation from dynamical equilibrium.  Many different methods have been employed to measure centroid shifts.  In all methods, the centroid is measured in apertures of varying size which could be based on radial steps of set physical or angular size (e.g. Mohr et al. 1993), contours of set surface brightness (e.g. O'Hara et al. 2006), or radial steps of a set fraction of $r_{500}$ (e.g. Poole et al. 2006; Maughan et al. 2007).  The centroid shift is then calculated from the variation of the centroid with radius.  For example, Mohr et al. (1993) calculated the emission-weighted variation of the centroid in circular apertures of set angular size, and Poole et al. (2006) use the RMS offset of the centroid from the cluster peak in circular apertures with radii between $0.05 r_{500}$ and $r_{500}$ in steps of $0.05 r_{500}$.

We chose to use a method similar to those used by Poole et al. (2006) and Maughan et al. (2007), but here we make a couple of small adjustments.  First, we choose to center our apertures on the centroid calculated within $0.05 r_{500}$ rather than the cluster peak.  The cluster peak is difficult to define robustly both for simulated clusters which often have small, brightly peaked substructures and for observations where the peak is smeared by noise.  In practice, we find that the centroid within $0.05 r_{500}$ typically picks out the peak of the main cluster.  In order to pick the right center for complex cluster morphologies, the centroid calculation is begun at large radii with the center allowed to vary and iterated to smaller radii to find the centroid within $0.05 r_{500}$.  We then find the centroid in circular apertures all centered on this point with radii increasing in steps of $0.05 r_{500}$ out to $r_{500}$.  The overall centroid shift, $\langle w \rangle$, is then the standard deviation of the distance between the aperture center and the centroid within these apertures normalized by $r_{500}$.  

Poole et al. (2006) and Maughan et al. (2007) also remove the central 30 kpc surrounding the X-ray peak from the centroid calculation in order to remove the effects of a bright cool core and to increase the sensitivity to distortions in the cluster distribution.  We also investigated the effect of removing the core, but instead we remove the region within our smallest aperture, i.e. the region within a radius of $0.05 r_{500}$.  For our simulated sample, which encompasses a large range in redshift and mass, 30 kpc represents a varying fraction of each cluster.  It should be noted that both 30 kpc and $0.05 r_{500}$ represent a significantly smaller region than the extent of typical cool cores ($\sim 0.15 r_{500}$; Vikhlinin et al. 2005).  However, we do not wish to completely remove the cluster core as core structure and central concentration relate to cluster dynamical state, but we do remove the very strong peak which could prevent the detection of deviations from symmetry.  For the majority of the clusters, the centroid shifts are very similar regardless of whether the peaks are removed.

Table 1 lists the median and ranges of the centroid shifts both with and without the core region.  Figure 2 shows the relationship between $P_3/P_0$ and $\langle w \rangle_{nocore}$.  The comparison of these two structure measures shows a lot of scatter, but they are significantly correlated.  For comparison, in Table 1 we also list centroid shifts calculated within the physical aperture radii of 0.5 Mpc and 1.0 Mpc, which may be more appropriate for the comparison to observations.  Here the calculation of $\langle w \rangle$ is quite similar, but the aperture radii are varied in steps of 0.05 Mpc, and the centroid shifts are normalized to 0.5 Mpc and 1.0 Mpc, respectively.

\clearpage
\begin{figure}[h]
\epsscale{0.5}
\plotone{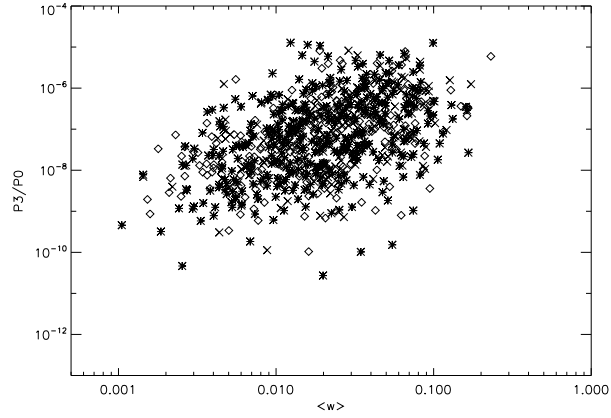}
\caption{ Correlation between $P_3/P_0$ and centroid shift (with the cluster peak removed), $\langle w \rangle_{nocore}$, for an aperture radius of $r_{500}$.  Clusters are divided in to three roughly equal time bins: $z \le 0.25$ (asterisks), $0.25 < z \le 0.65$ (diamonds), and $0.65 < z \le 1.5$ (crosses). }
\end{figure}
\clearpage

\section{ RESULTS }

\subsection{ Effect of Projection }

Clearly, one important consideration in determining cluster structure from observations is projection along the line of sight.  Projection can have the effect of either hiding structure or projecting unrelated structures to lie close to each other in the plane of the sky.  The latter case is less concerning in that it can be distinguished observationally with sufficient spectroscopy, but this may not be possible for large surveys.  In this paper, we typically treat clusters which overlap within their projected virial radii as one cluster.  With the simulations, we create X-ray cluster images for three orthogonal projections and investigate the scatter in structure that results for individual clusters.  For example, Figure 3 compares the power ratios for two projections at both $R=500$ kpc (left) and $R=r_{500}$ (right).  The typical deviation in the power ratios between projections is less than a factor of two, which is not large for the power ratios (notice that the plots are log-log), but the scatter due to projection is significant.  The overall variation with projection depends on which ratio and which radius are chosen, but for $\sim$90\% of clusters the power ratios change by a small fraction (less than 10\%) of the total observed ranges in the power ratios presented in Table 1.  A similar scatter with projection is seen for the centroid shifts, shown in Figure 4 for an aperture radius of $r_{500}$.  Again the variation of the centroid shifts with projection for 90\% of clusters is less than 10\% of the total ranges in Table 1.

\clearpage
\begin{figure}
\begin{center}
\epsscale{1.0}
\plottwo{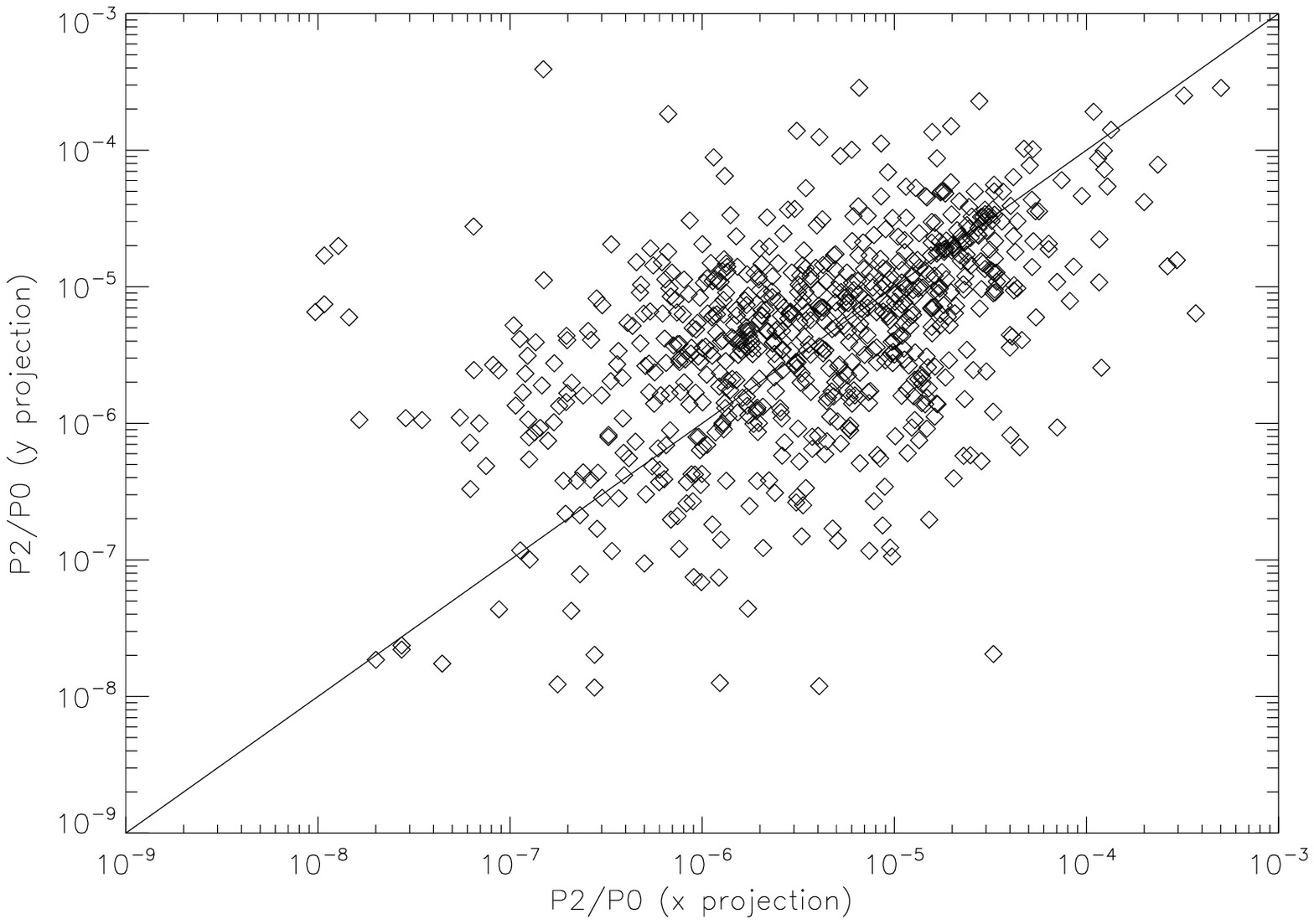}{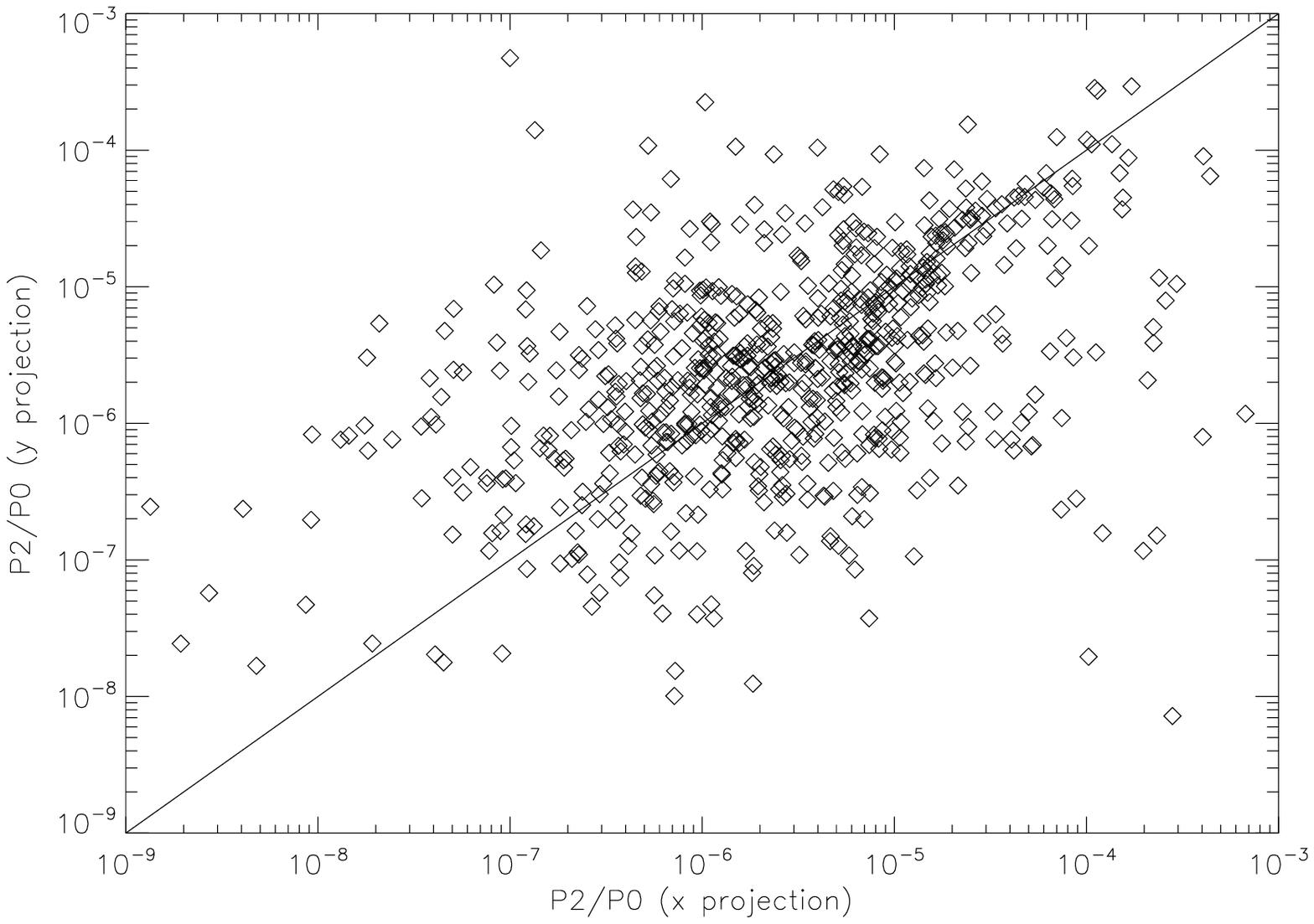}
\plottwo{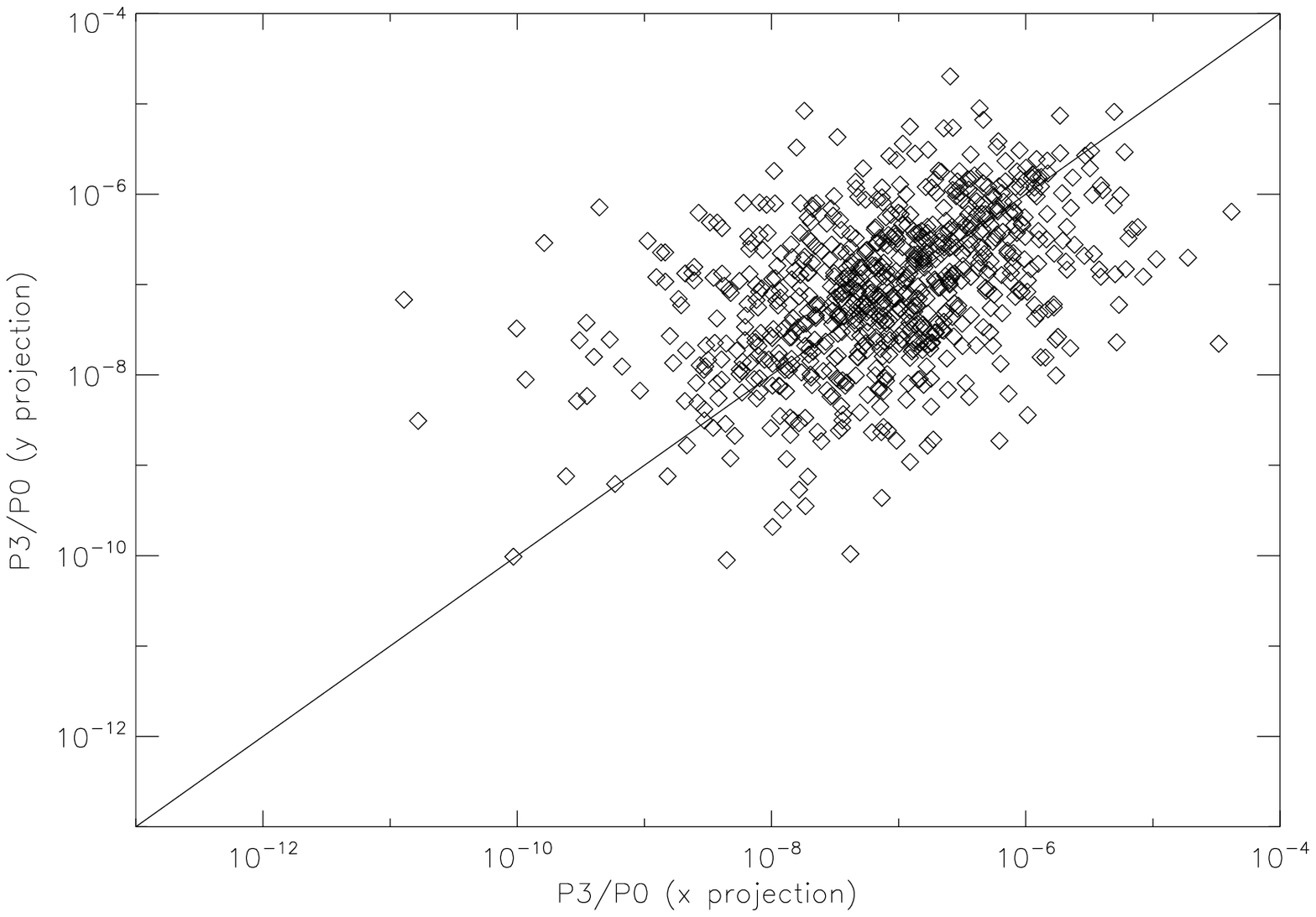}{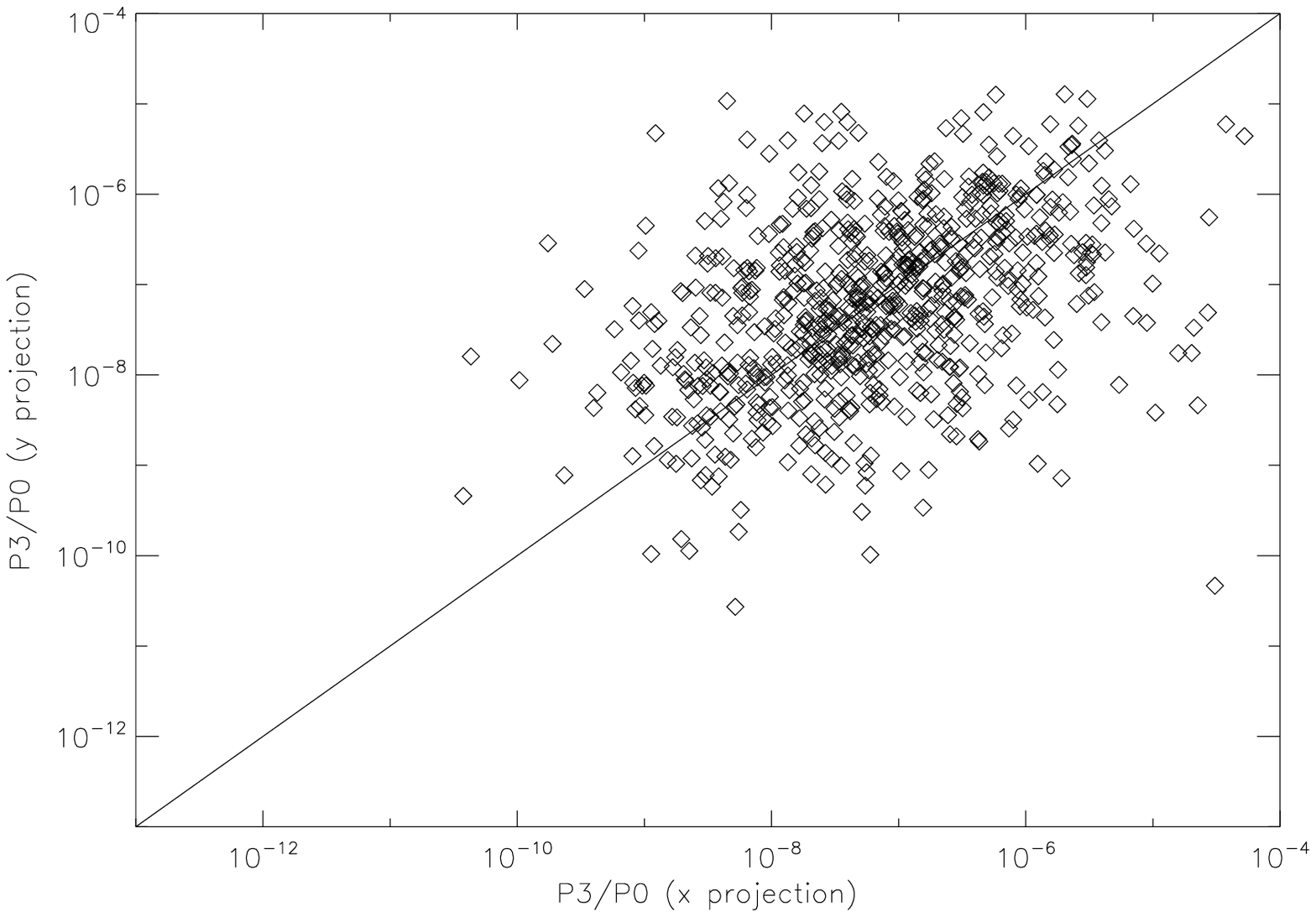}
\plottwo{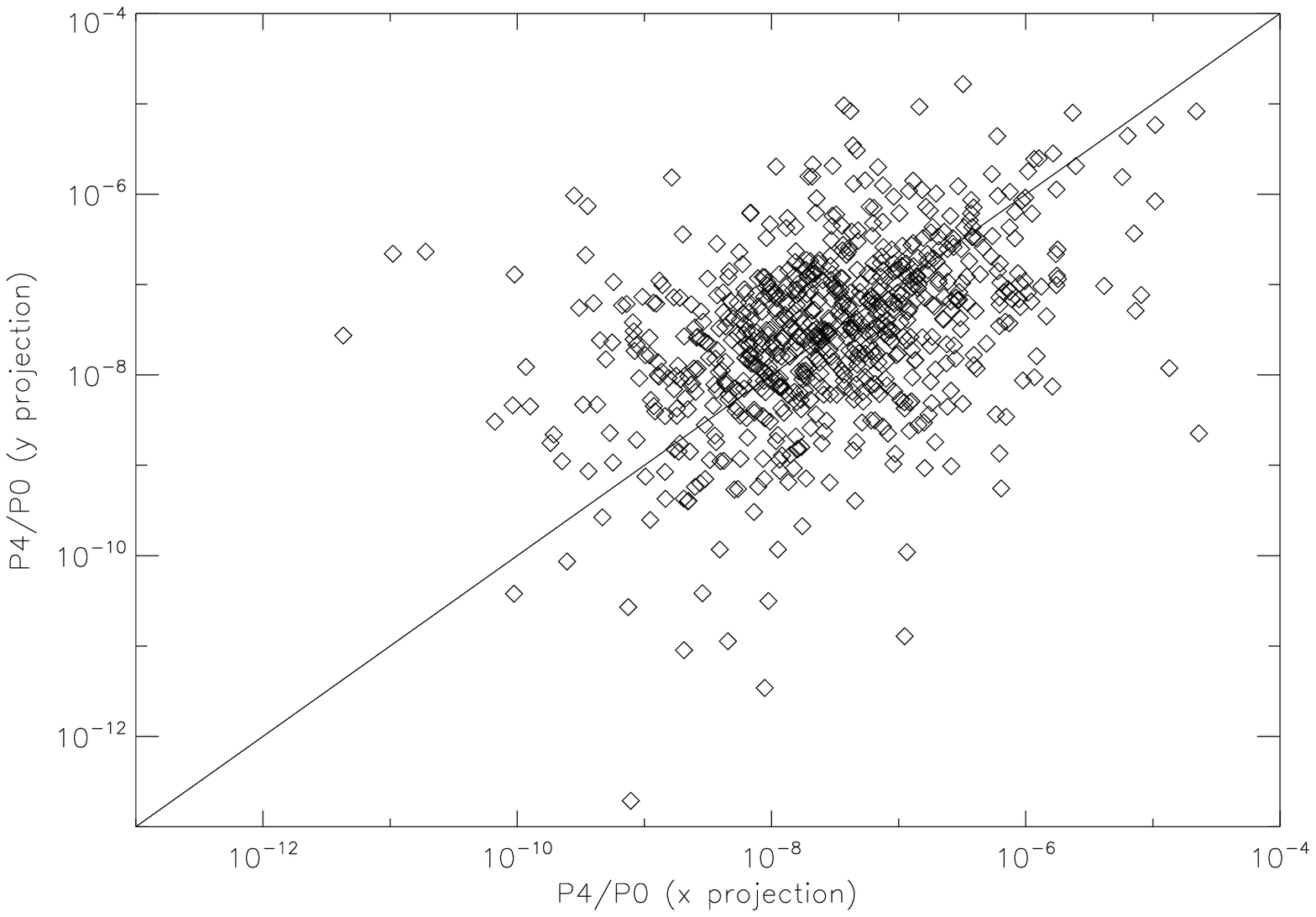}{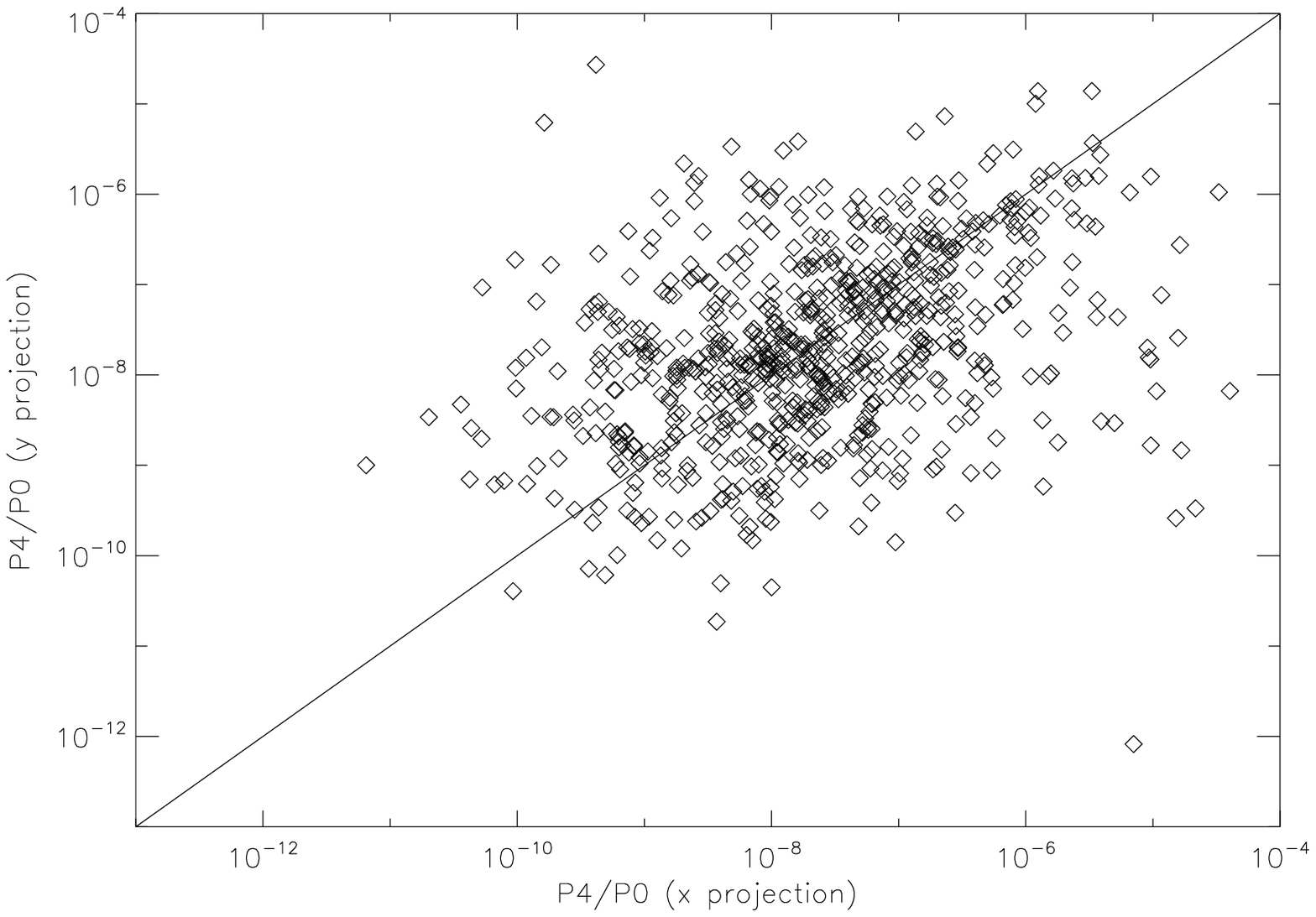}
\caption{ Scatter in the power ratios due to projection for two orthogonal projections with aperture radii of (\textit{left}) $R=500$ kpc and (\textit{right}) $R=r_{500}$.  The solid line shows a one-to-one correlation. }
\end{center}
\end{figure}

\begin{figure}[h]
\begin{center}
\epsscale{1.0}
\plottwo{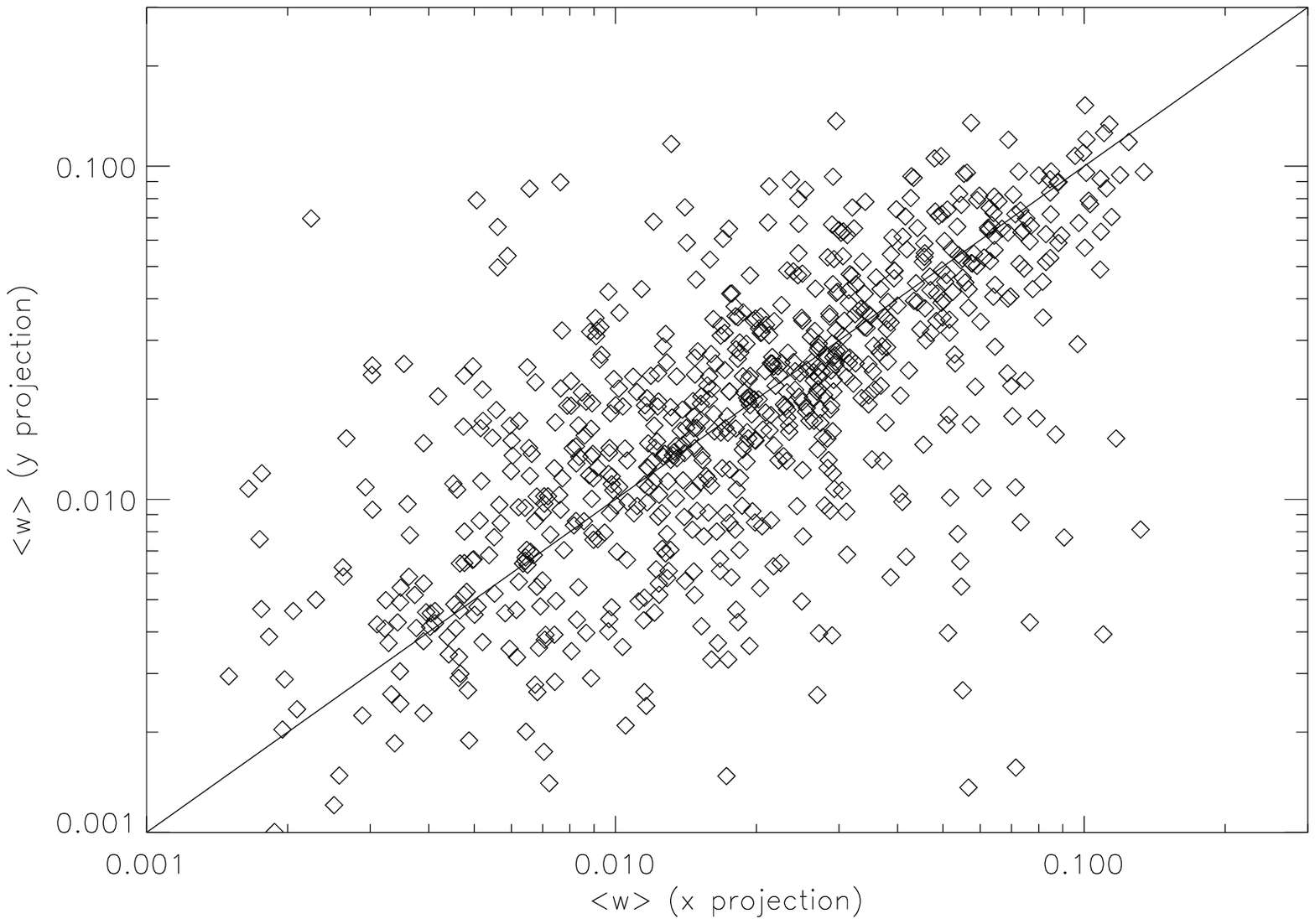}{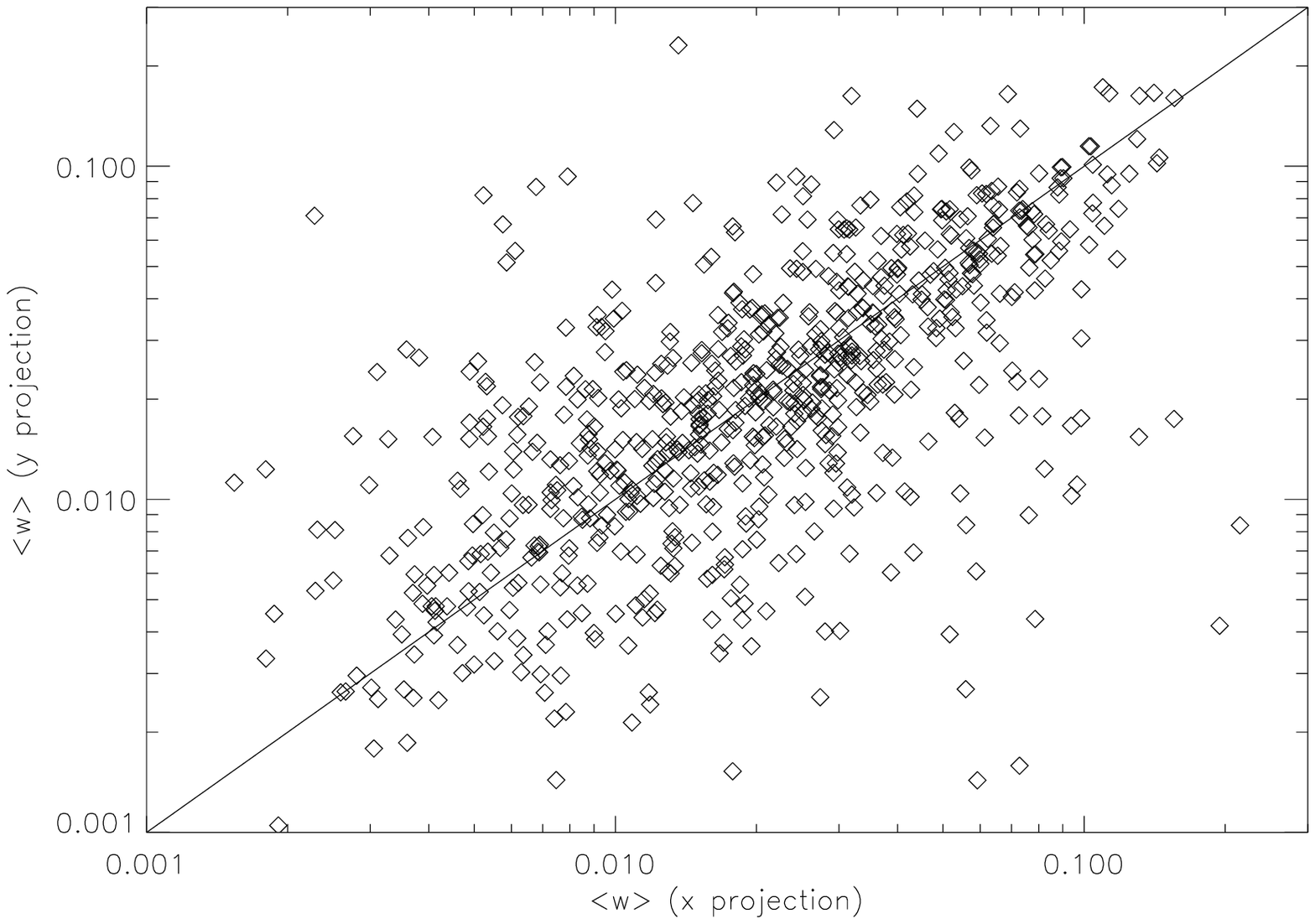}
\caption{ Scatter in centroid shifts due to projection for two orthogonal projections.  Centroid shifts are calculated within $R=r_{500}$ both with (\textit{left}) and without (\textit{right}) the cluster peak.  The solid line shows a one-to-one correlation. }
\end{center}
\end{figure}
\clearpage

Interesting questions are how frequently does a disturbed cluster, for example one undergoing a major merger, appear relaxed in projection.  And what fraction of clusters we observe to be relaxed is actually disturbed?  Clusters exhibit a very large range of morphologies and structure on many scales, so here we will consider only clusters at the two extremes selected based on their power ratios.  For each projection, ``disturbed'' clusters are selected based on having $P_2/P_0$, $P_3/P_0$, or $P_4/P_0$ above $1.5 \times 10^{-5}$, $4.5 \times 10^{-7}$, and $1.5 \times 10^{-7}$, respectively.  These limits correspond to the power ratios of observed clusters undergoing major mergers and select clusters with roughly the highest 20\% of a given power ratio in each projection ($R=500$ kpc or $R=r_{500}$).  ``Relaxed'' clusters have $P_2/P_0$, $P_3/P_0$, or $P_4/P_0$ below $1.0 \times 10^{-6}$, $1.5 \times 10^{-8}$, and $6.5 \times 10^{-9}$, corresponding to observations of relaxed, single clusters and clusters with the lowest $\sim$20\% of each ratio.  Here we need to be careful to account for clusters which may appear disturbed in projection while actually not overlaping in 3-dimensions.  Using the 3D information from the simulations, we find cases where a second cluster (including the region covered by its virial radius) is projected to lie within an aperture radius of a given cluster.  These cases are then removed from the list of disturbed clusters for a given projection.  For $R=500$ kpc, $R=r_{500}$, and $R=1$ Mpc we find that 2\%, 3\%, and 4\% of clusters have chance projections with another cluster, respectively, although not all of these projections lead to disturbed power ratios.

We first divide clusters, including all redshifts, in to relaxed and disturbed subsamples using the information available from all three projections.  Clusters are considered to be disturbed if they have power ratios above our limits in one or more of the projections (after removing chance projections).  For an aperture radius of $r_{500}$, approximately 31\% of all clusters are disturbed.  Relaxed clusters must have power ratios below our limits in all three projections, and only about 11\% of clusters are relaxed.  Of the disturbed clusters, only about 55\% will look disturbed in any one projection.  Of the clusters that appear relaxed when viewed in a single projection, about 38\% are relaxed from all angles while 8\% are disturbed (i.e. above our defined power ratio limits) in one of the other two projections.  The true fraction of relaxed clusters may be even lower, because at the moment of core passage during a merger the two cores overlap and the cluster can appear relaxed from all angles (Jeltema et al. 2008; Rowley et al. 2004; Poole et al. 2007).  Table 2 lists the above pecentages for the three aperture radii considered.  Percentages are averaged over the three power ratios and where appropriate the three projections, but the variations among these are not large.  In general, these numbers are similar for different radii with the possible trend that relaxed clusters are better distinguished at larger radii.

We ran a similar experiment using the centroid shifts.  Table 2 lists the percentages for the centroid shifts with the peaks removed, but the numbers are similar if the peaks are included.  We again define disturbed clusters to be those with the highest 20\% of centroid shifts ($\langle w \rangle_{no core} > 0.047$) and relaxed clusters to be those with the lowest 20\% of centroid shifts ($\langle w \rangle_{no core} < 0.008$).  It can be seen from Table 2 that the centroid shifts perform a bit better than the power ratios at distinguishing disturbed and relaxed clusters in a single projection.  For example, only about 4\% of clusters which appear relaxed in a single projection turn out to be disturbed in another projection, a factor of two better than the power ratios.

Using the 3D information from the simulations it is also possible to determine the fraction of clusters which overlap with another cluster within a given radius.  Here both halos must have masses above $10^{14} M_{\odot}$ and be sufficiently separated to be identified in the simulation as individual halos.  We find that 33\%, 37\%, and 40\% of clusters overlap another cluster within radii of 0.5 Mpc, $r_{500}$, and 1 Mpc, respectively.  Here we define overlap to be when the 3D distance of the center of a second cluster is within a virial radius plus the aperture radius.  This test does not probe exactly the same types of cluster morphologies as the selection of disturbed clusters based on high power ratios or centroid shifts, because minor mergers may not result in highly disturbed morphologies and in the later stages of mergers the clusters will not be selected as separate halos.  It is, however, encouraging that both give a similar fraction of merging clusters.

\subsection{ Correlations with Other Cluster Properties }

\subsubsection{Overall Properties}

We now turn our attention to the correlation between cluster morphology and other cluster properties.  Do clusters of different masses show more or less structure?  If cluster mergers affect observable cluster properties like luminosity and temperature, do we observe a correlation between these properties and the power ratios?  Specifically, we look for correlations between cluster structure and X-ray luminosity, total mass, X-ray temperature, and $Y_X$ within a radius of $r_{500}$.  $Y_X$, the product of the gas mass and the temperature, is proportional to the total thermal energy of the cluster and is the X-ray equivalent of the integrated Sunyaev-Zeldovich (SZ) flux.  This measure and the total SZ flux have been shown in simulations to be very low scatter proxies for cluster mass and to be fairly insensitive to mergers \cite{M05, K06, P07}.  

Not surprisingly there is quite a bit of scatter in the relations between cluster structure and these properties; a cluster of a given mass can have a large range of structures.
Only cluster luminosity shows a significant correlation with the power ratios.  This correlation is shown in Figure 5 for $P_3/P_0$ versus luminosity with dashed lines at the locations of the limits on $P_3/P_0$ defined above for relaxed and disturbed clusters.  Table 3 gives the probability of a correlation between the three power ratios and each cluster property considered.  Probabilities are derived from a Spearman Rank-Order Correlation test (e.g., Press et al. 1992, \S 14.6), and a probability of one indicates no correlation.  The correlation with luminosity is in the sense that clusters with higher luminosities tend to have lower power ratios.  Since we do not find a correlation between mass and the power ratios this trend can not simply be explained as a trend for more massive clusters to be more relaxed.  The trend of structure with luminosity is in the opposite direction of what one might expect given that simulations show that mergers can significantly boost cluster luminosity (e.g. Ricker \& Sarazin 2001; Rowley et al. 2004; Poole et al. 2007).  However, relaxed, cooling clusters show very bright cores and high luminosities.  We also find that during a merger, the cluster luminosity peaks at the moment of core passage; during this phase the two cores overlap and the cluster appears relaxed (see also Jeltema et al. 2008; Rowley et al. 2004; Poole et al. 2007).  We expect the core passage phase to be short lived, so this case should be relatively rare, and it may be possible to distinguish these cases through temperature variations.  $P_2/P_0$ also correlates with $Y_X$, but the slope is greater than zero at less than $3\sigma$ significance.
\clearpage
\begin{figure}
\begin{center}
\epsscale{0.5}
\plotone{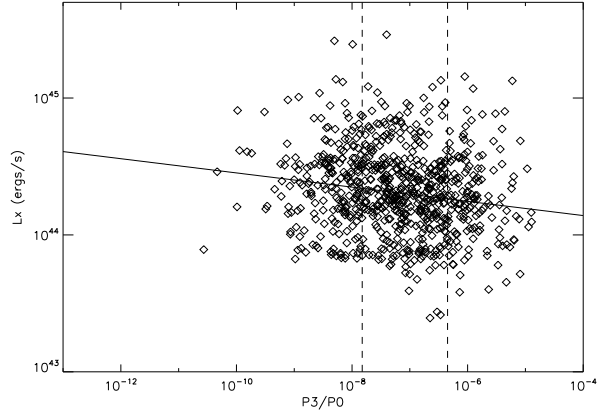}
\caption{ Correlation between $P_3/P_0$ and cluster luminosity within $r_{500}$.  Solid line shows the best fit relation.  Vertical dashed lines indicate the limits defined in section 3.1 for relaxed and disturbed clusters.  No significant correlation is seen between the power ratios and total mass, temperature, and $Y_X$. }
\end{center}
\end{figure}

\begin{figure}
\begin{center}
\epsscale{0.5}
\plotone{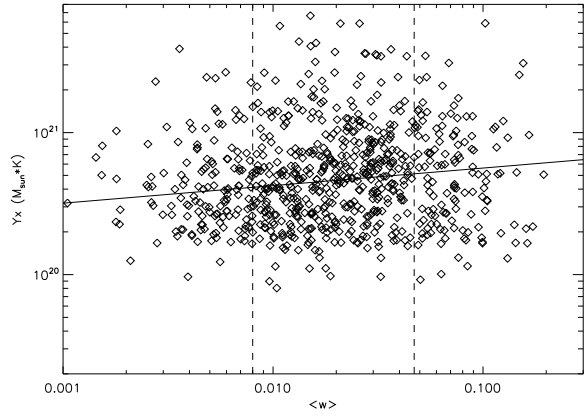}
\caption{ Correlation between $\langle w \rangle_{nocore}$ and total mass within $r_{500}$.  Solid line shows the best fit relation.  Vertical dashed lines indicate the limits defined in section 3.1 for relaxed and disturbed clusters.  Mass shows the strongest correlation with $\langle w \rangle_{nocore}$, but the trend is mild. }
\end{center}
\end{figure}
\clearpage

We also look for possible correlations with centroid shift.  Table 3 reveals that based on the rank-order probabilities there are significant correlations between centroid shift and all of the cluster properties except for $L_X$, particularly mass and $Y_X$.  However, the best fit relations are all fairly flat indicating that these trends are weak.  For example, the most significant trend is with $Y_X$, shown in Figure 6; clusters with large $\langle w \rangle_{no core}$ tend to have slightly higher $Y_X$, but again the slope is greater than zero at less than $3\sigma$ significance.  It is interesting that we do not see a trend of centroid shift (either $\langle w \rangle$ or $\langle w \rangle_{no core}$) with $L_X$, the one property that correlates with the power ratios, possibly due to a difference in the types of structure probed by these methods.  For example, the significant radial dependence in the power ratios may make them more sensitive to cool cores.  However, in the next section we will show that for both the power ratios and the centroid shifts, relaxed clusters have luminosities which are high for their masses.  We note that the trends presented here also depend on the aperture radius chosen.  For $R=1$ Mpc, we do find significant but mild trends in mass and $T_X$ with the power ratios.  

Finally, we note that for any choice of structure measure or radius the change (if present) in mass, luminosity, temperature, or $Y_X$ with cluster structure is small; the average cluster properties change by less than a factor of 1.5 over the full range in cluster structure.  Over the range of mass considered here ($2 \times 10^{14} M_{\odot} < M_{vir} < 2 \times 10^{15} M_{\odot}$), clusters of different masses have a similar range of cluster structures.  Below we turn our attention to the effect of cluster structure on correlations between different cluster properties, in particular the mass-scaling relations.

\subsubsection{Scaling Relations}

Given the well known correlations between cluster properties (e.g. luminosity and temperature, temperature and mass, etc.), we examine where clusters of different structures fall on these scaling relations.  For example, Figure 7a shows the relationship between X-ray luminosity and temperature within $R=r_{500}$.  Here relaxed clusters are shown in red and disturbed clusters are shown in blue; relaxed (disturbed) clusters have $P_3/P_0$ below (above) our limits in the one projection plotted.  The solid line gives the best fit relation for all clusters with the slope fixed at the expected self-similar relation and with the proper evolutionary correction applied (i.e. $log(E_z^{-1}L_X) = 2.0 log(T_X) + B$ where $E_z = (\Omega_m(1+z)^3 + 1 - \Omega_{\Lambda})^{1/2}$).  The dashed and dotted lines give the fits to just the relaxed and disturbed clusters, respectively.  It is clear here that relaxed clusters tend to have higher luminosities for a given temperature, and, in fact, our fits show a significant offset between the relaxed clusters and the overall cluster population.  For the relaxed clusters $B=29.239 \pm 0.014$, while for all of the clusters $B=29.165 \pm 0.007$.  We determine the errors on our fits through 10,000 bootstrap trials where clusters are randomly resampled and the fits repeated.  The disturbed clusters, on the other hand, appear to have a similar distribution in $L_X-T_X$ space to the overall population, and the fit to these clusters gives $B=29.155 \pm 0.017$, very similar to the intercept for the whole cluster sample.  These trends become even more apparent if the selection of disturbed and relaxed clusters is done using the information from all three projections, as in \S 3.1.  We now define relaxed clusters to be clusters that are relaxed in all three projections, and disturbed clusters to be clusters which are disturbed in any one projection after removing chance projections.  Figure 7b shows the $L_X-T_X$ relation with these selections.  The disturbed clusters now comprise $\sim$40\% of the sample and are consistent with the total population, while the relaxed clusters show an even larger offset with $B=29.342 \pm 0.027$.  

Figure 8 shows the relations between $L_X$, $T_X$, $Y_X$ and total cluster mass within $R=r_{500}$ for the same selection of relaxed and disturbed clusters.  The $L_X-M$ relation shows the same trend of relaxed clusters having luminosities that are too high for their masses, while the disturbed clusters show no significant offset from the overall population.  The $T_X-M$ and $Y_X-M$ relations, however, show no significant dependence on cluster structure.  For an aperture radius of 500 kpc, there is a significant trend for the relaxed clusters to have temperatures which are low for their masses; however, this trend disappears at larger radii like $r_{500}$ and 1 Mpc where the contribution of the cluster core to the temperature becomes less important.  In contrast, Kravtsov et al.~(2006) and Nagai et al.~(2007b) find an offset of unrelaxed clusters toward cooler temperatures in the $T_X-M$ relation.  There are several differences between our study and theirs, including sample size, average cluster mass, star formation and feedback formulation, and selection of disturbed clusters, and it is unclear which of these leads to the difference in the derived dependence of the $T_X-M$ relation.  We do, however, confirm their finding that the $Y_X-M$ relation is independent of cluster structure.  In Figure 8, there are two outliers at low mass representing two redshift outputs of a cluster undergoing a major merger where the gas is very offset from the dark matter which leads to a problem in centering.  This cluster is a known outlier and does not represent a general problem.
\clearpage
\begin{figure}
\epsscale{0.5}
\plotone{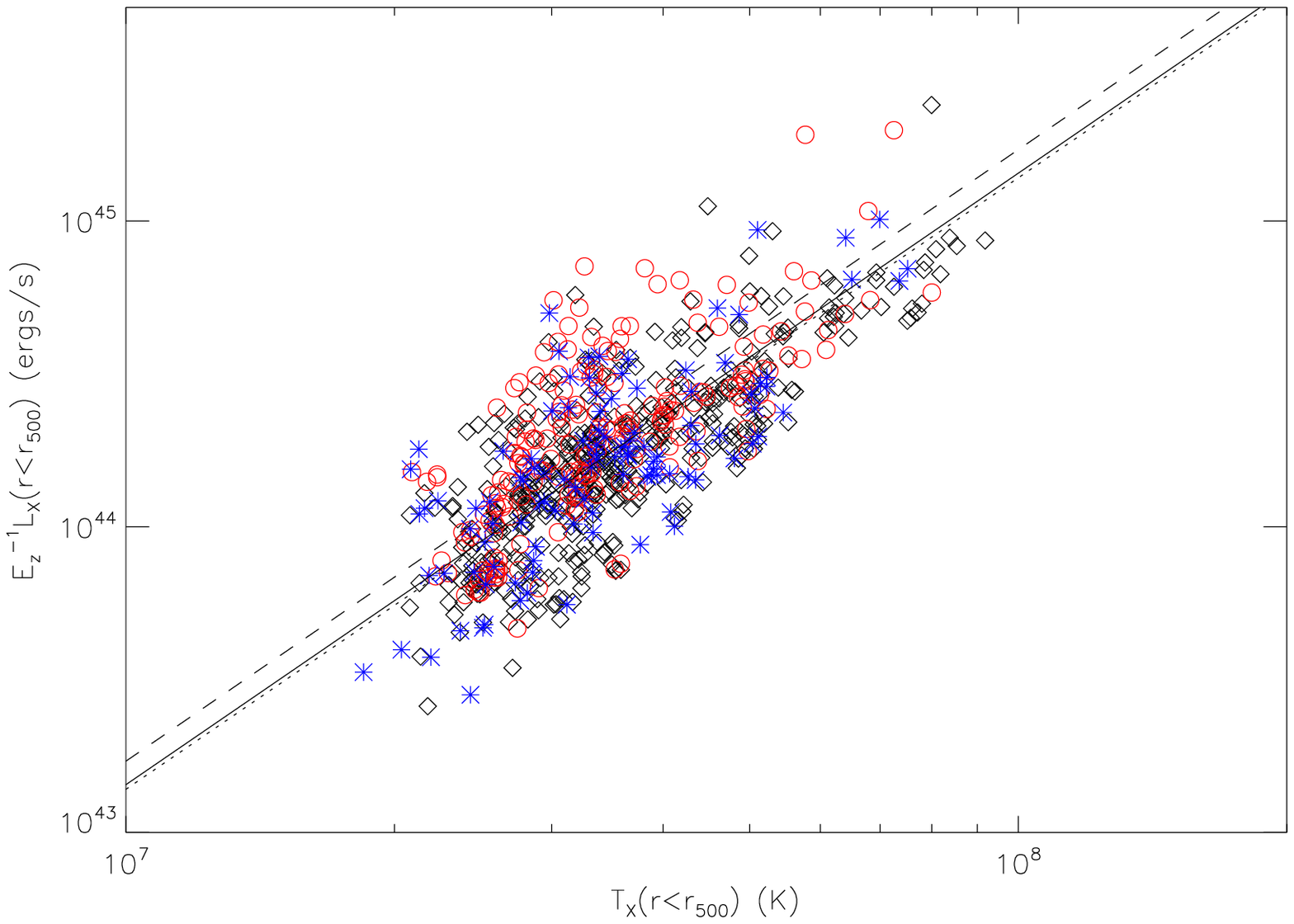}
\plotone{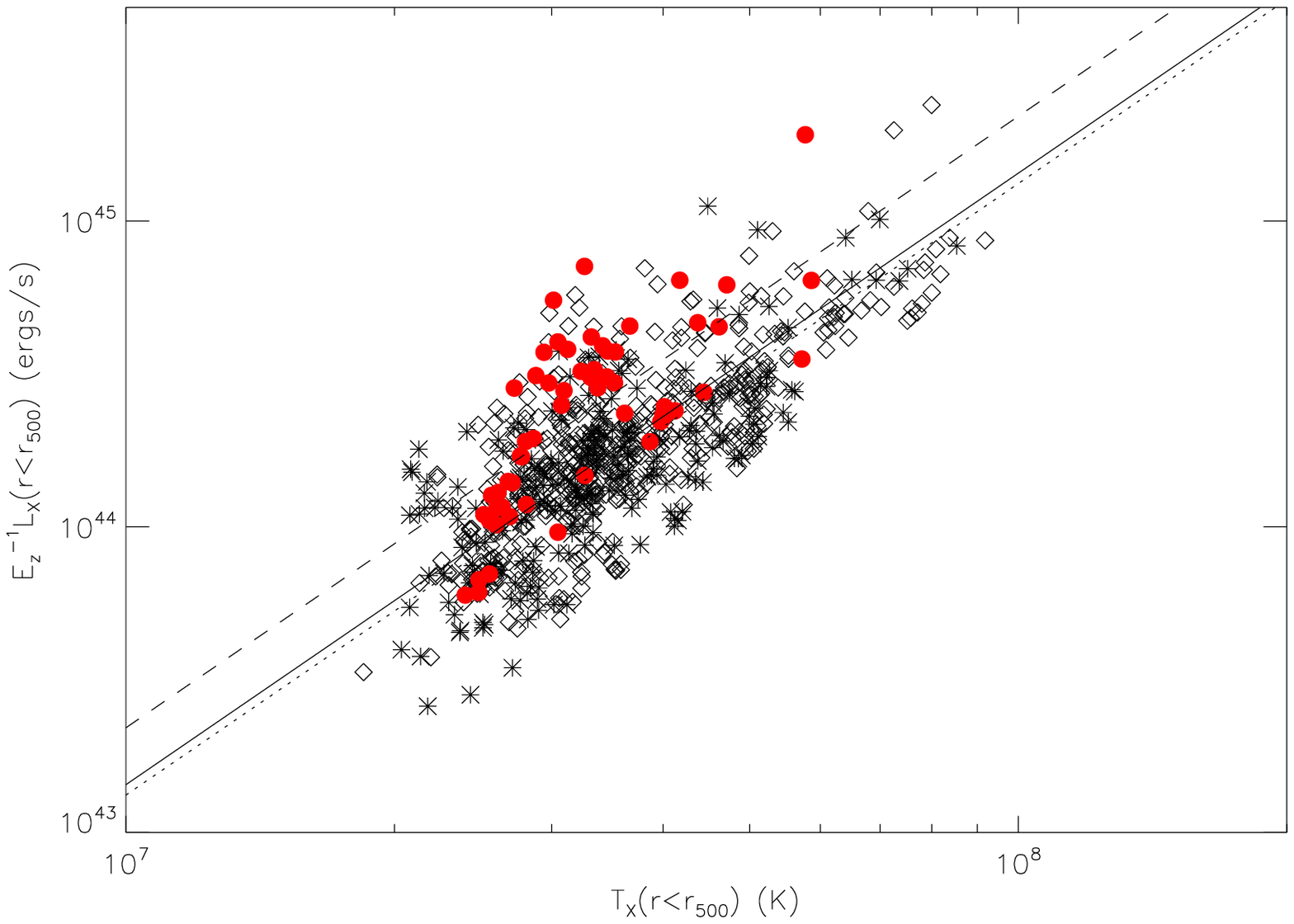}
\plotone{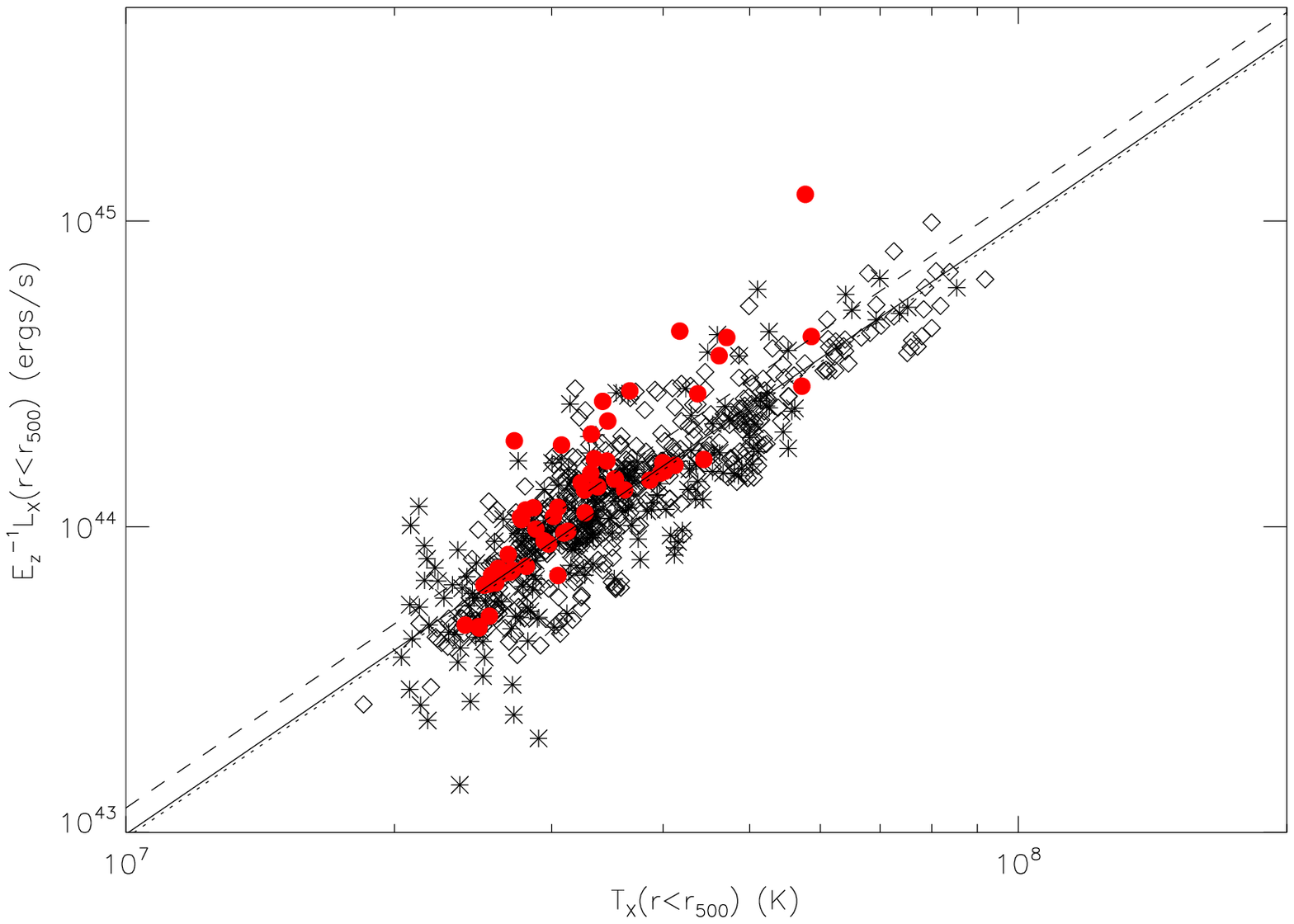}
\caption{ $L_X-T_X$ relation for a radius of $r_{500}$.  \textit{Top:} Red circles show relaxed clusters, blue asterisks show disturbed clusters, and the rest of the sample is plotted with diamonds.  Here the selection of these subsamples was done based on the values of $P_3/P_0$ in one projection. \textit{Middle:} The same for relaxed (solid red circles), disturbed (asterisks), and typical (diamonds) clusters with the selection of clusters based on their $P_3/P_0$ in all three projections.  \textit{Bottom:} The same except that the luminosity of the cluster core within $R \le 0.15 r_{500}$ has been removed.  Also shown are fits of the form $log(E_z^{-1}L_X) = 2.0 log(T_X) + B$ to all clusters (solid line), relaxed clusters (dashed line), and disturbed clusters (dotted line). }
\end{figure}

\begin{figure}
\begin{center}
\epsscale{0.5}
\plotone{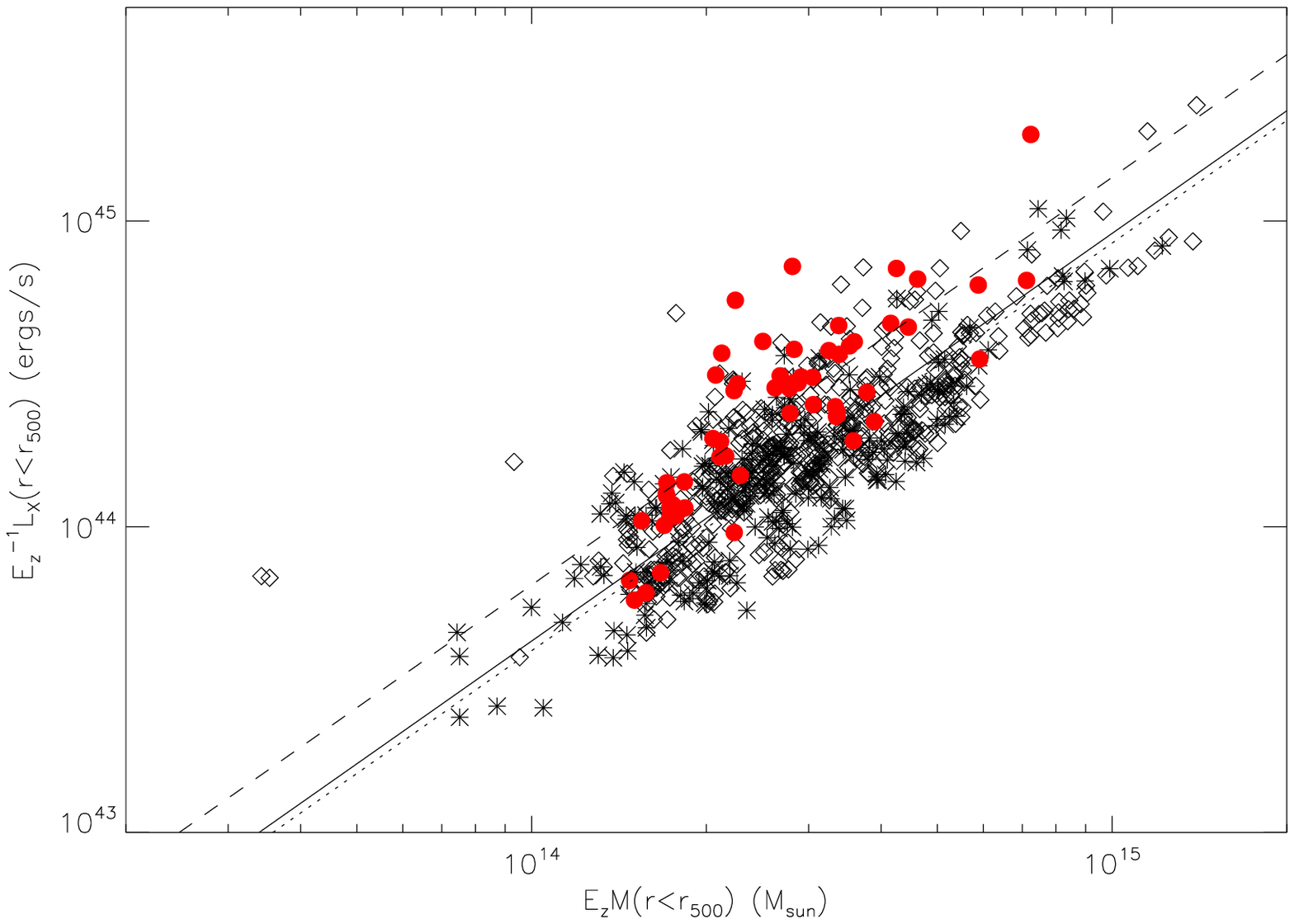}
\plotone{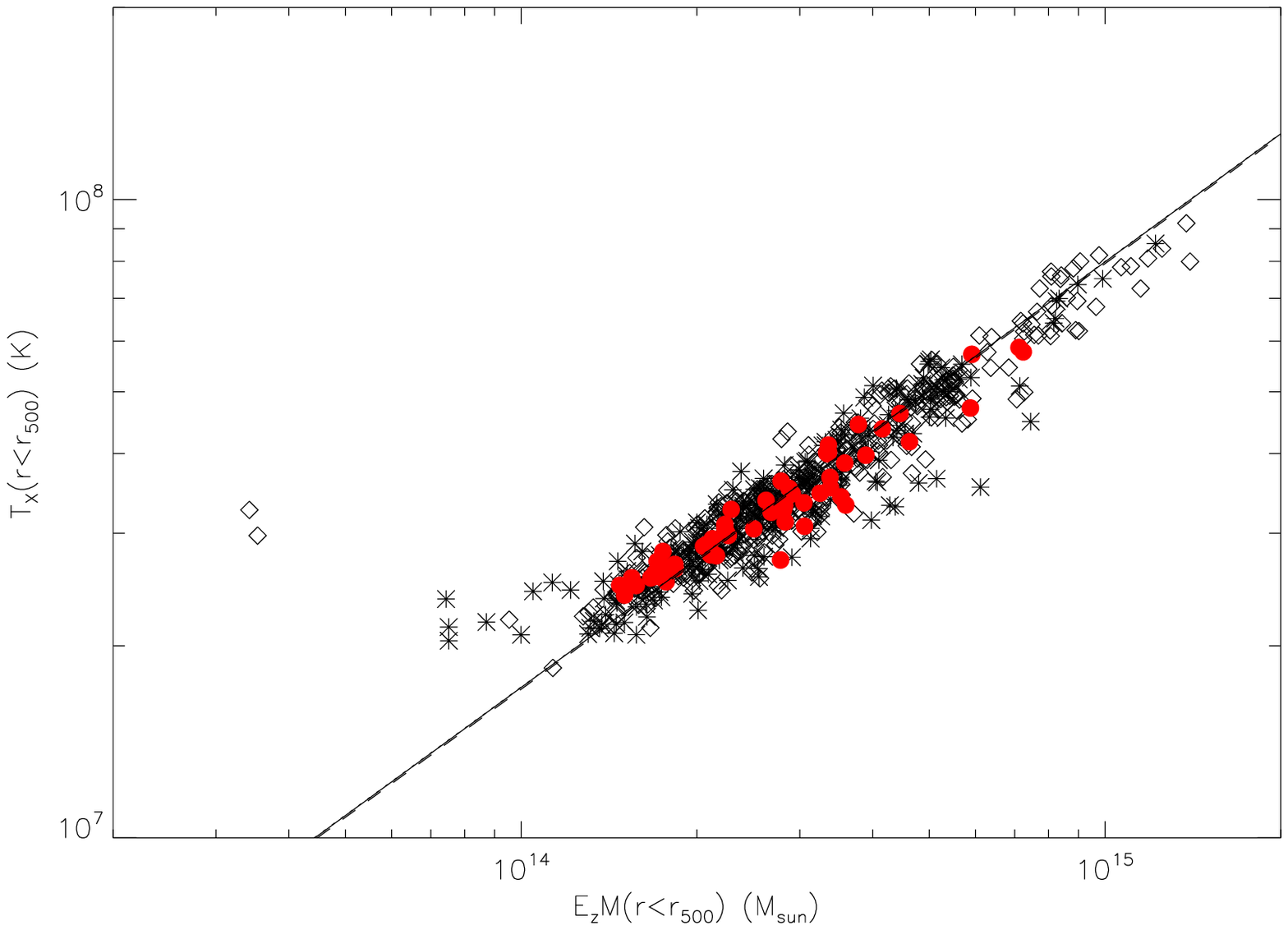}
\plotone{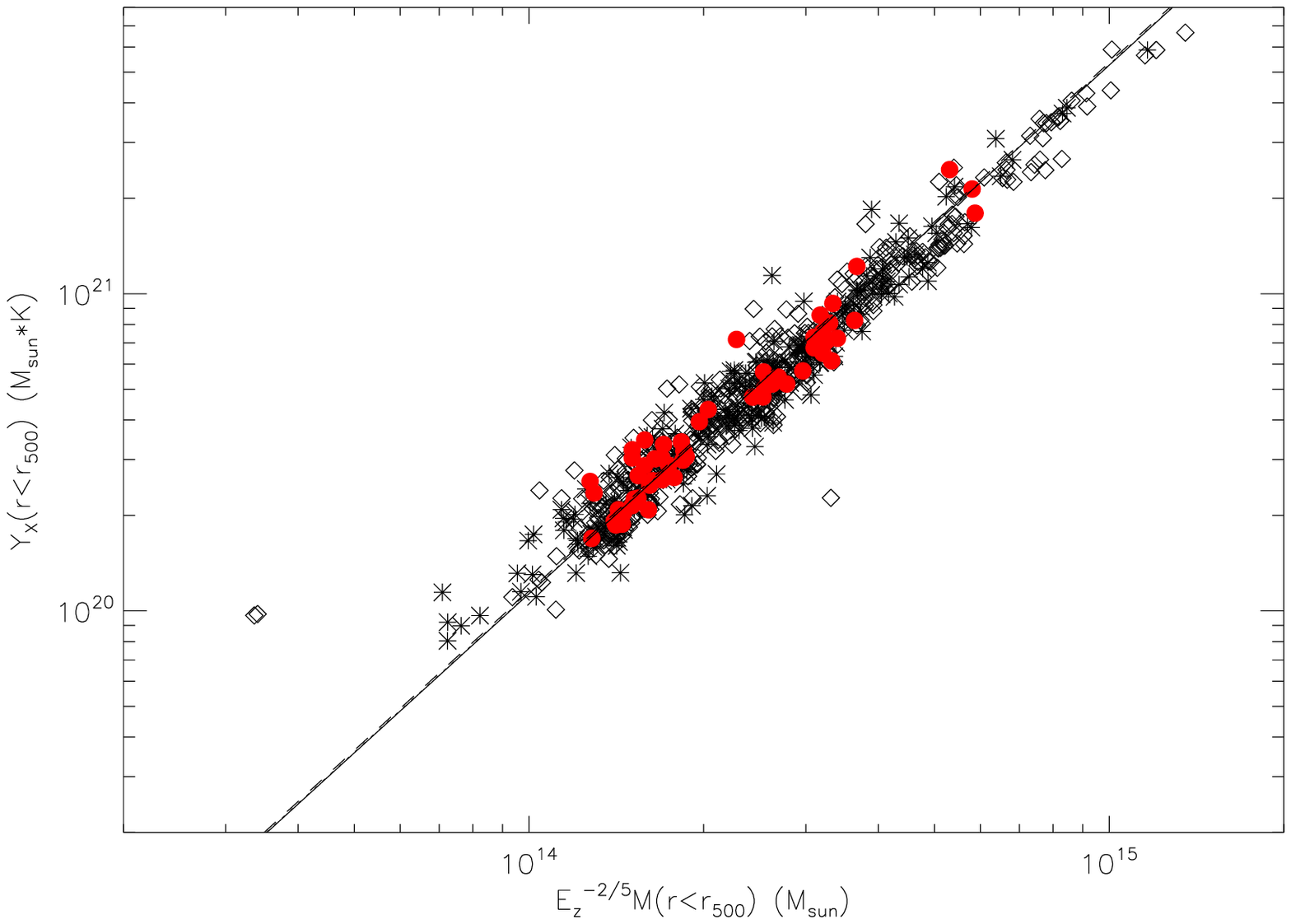}
\caption{ $L_X-M$ (\textit{top}), $T_X-M$ (\textit{middle}), and $Y_X-M$ (\textit{bottom}) relations within a radius of $r_{500}$.  Points show relaxed (solid red circles), disturbed (asterisks), and typical (diamonds) clusters selected based on $P_3/P_0$ in all three projections.  Also shown are fits of the form $log(y) = A log(x) + B$ with the slopes fixed at 4/3 for $L_X-M$, 2/3 for $T_X-M$, and 5/3 for $Y_X-M$.  Lines show the fits to all clusters (solid line), relaxed clusters (dashed line), and disturbed clusters (dotted line). }
\end{center}
\end{figure}
\clearpage

Most of these trends can be explained by the presence of cool cores in the relaxed clusters, although core passage events would also create similar offsets.  As is frequently done in observational studies, we tried removing the cluster cores in the calculation of their luminosities.  This procedure mitigates the effects of both excess luminosity from cool cores and luminosity spikes during the core passage phase of mergers.  We adopt the cluster peak location from the centroid shift determination (i.e. the centroid within $0.05 r_{500}$) and subtract off the luminosity within $R \le 0.15 r_{500}$ from the total luminosity, similar to observational studies \cite{V05,M07b}. The results for the $L_X-T_X$ relation are shown in Figure 7c.  The relaxed clusters are now shifted down in luminosity to have better agreement with the overall cluster population, but they still have luminosities which are significantly high for their temperatures.  The same is true of the $L_X-M$ relation.  The scatter in the $L_X-T_X$ and $L_X-M$ relations is significantly reduced by removing the cluster cores but remains larger than the scatter in the $T_X-M$ and $Y_X-M$ relations.  On the observational side, Hashimoto et al. (2006) find a similar offset between ``distorted'' and ``non-distorted'' clusters in the $L_X-T_X$ relation when the cores are included, while Maughan (2007) finds agreement in the $L_X-Y_X$ and $L_X-M$ relations for relaxed and unrelaxed clusters if the cores are removed.  Our results confirm that the offset between relaxed and unrelaxed clusters as well as the overall scatter can be reduced significantly by removing the core, but the remaining offset indicates that, at least in simulations, increased luminosity in ``relaxed'' clusters is also present at larger radii (see also Burns et al.~2007).

The trends with cluster structure described above are also present, although with typically lower significance, if we select disturbed and relaxed clusters based on centroid shift.

\subsubsection{ Hydrostatic Masses }

Above we presented the correlations of structure with the true, 3D cluster masses, but these are not directly observable.  Most often cluster masses are determined from the brightness and temperature of the X-ray emitting gas under the assumption that the cluster is in hydrostatic equilibrium.  However, our simulations and previous studies show that the hydrostatic masses are systematically biased toward underestimates of the true mass \cite{Ka04,Rs06,N07}.  In our simulations, the hydrostatic masses are lower than the true masses by on average 16\%, similar to what is found by other authors.  Several authors have suggested that this deviation of the hydrostatic mass from the true mass is due to additional pressure support from turbulent gas motions \cite{Rs04,Rs06,Ka04,D05,N07}.  We might then expect a correlation between this kinetic component and cluster dynamical state.  In Figure 9, we show the correlations between the deviation of the hydrostatic masses from the true masses versus $P_3/P_0$ and $\langle w \rangle_{no core}$.  Hydrostatic masses are calculated using the 3D spherically averaged density and temperature profiles.  Our results, therefore, represent the best possible result one could get even with perfect data; the typical errors we find in the hydorstatic masses are very similar to those found by other investigators using simulated observations \cite{N07}.
\clearpage
\begin{figure}
\epsscale{0.8}
\plotone{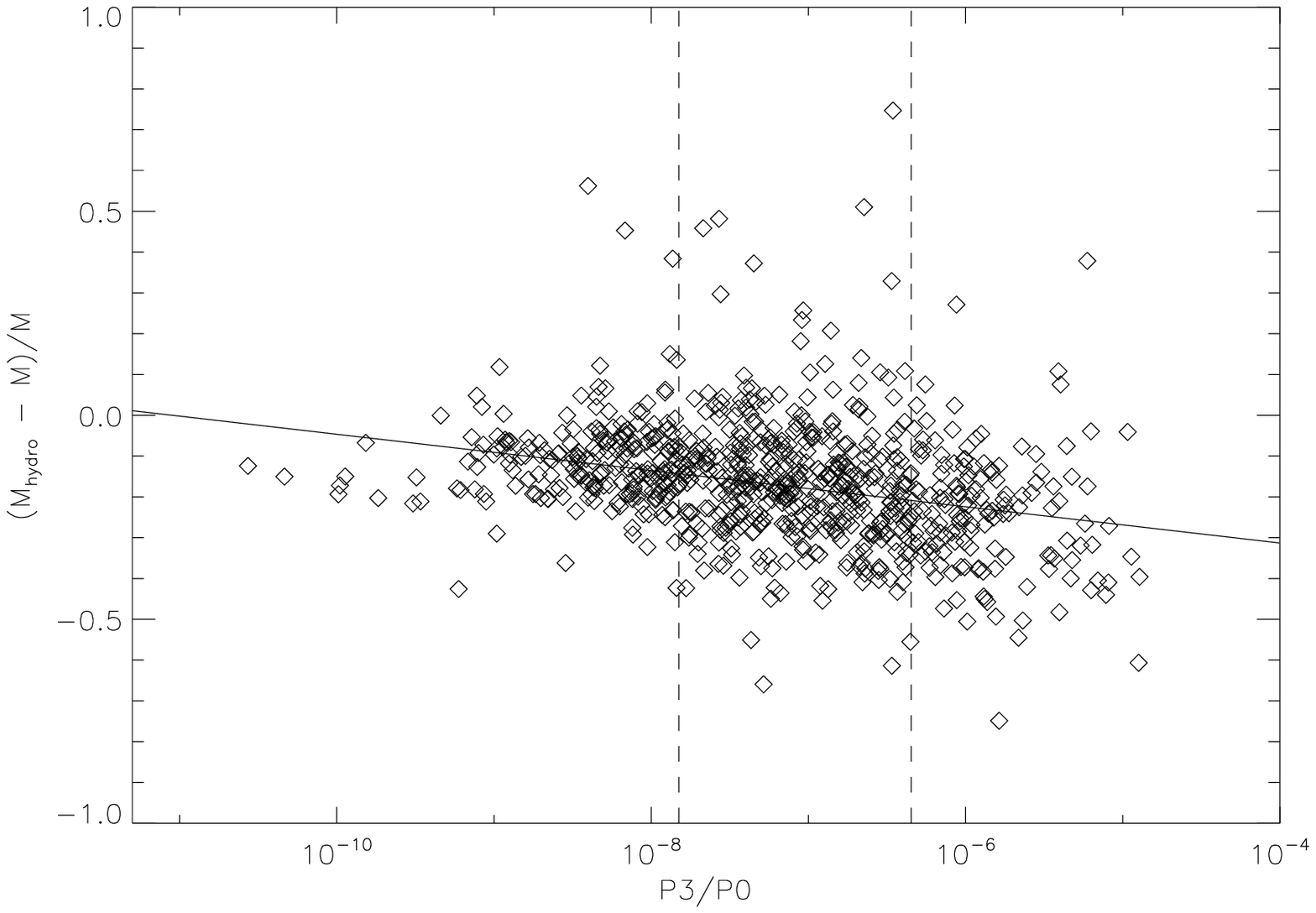}
\plotone{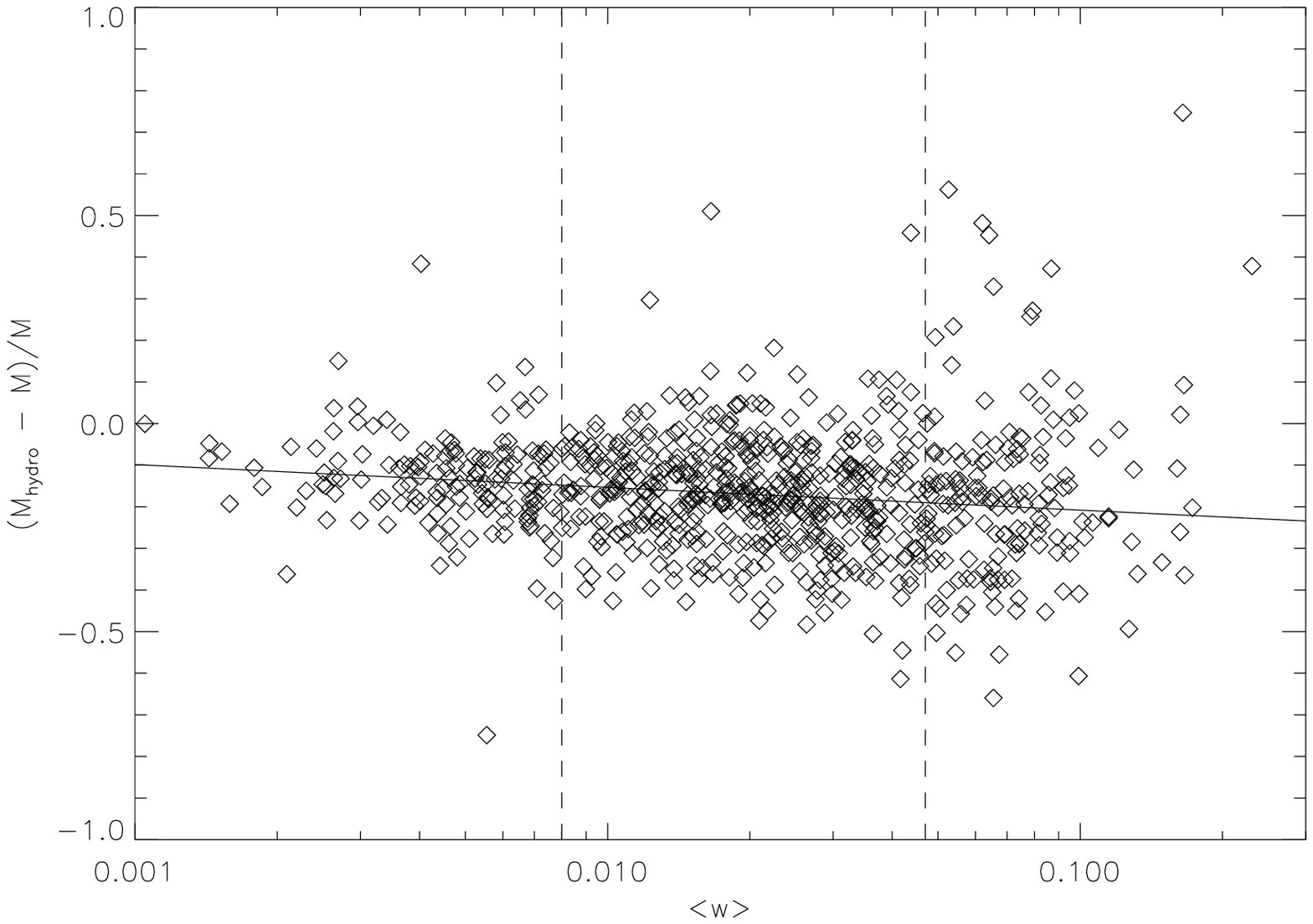}
\caption{ Correlation of $P_3/P_0$ (\textit{top}) and $\langle w \rangle_{no core}$ (\textit{bottom}) with the deviation of the hydrostatic masses from the true masses within $r_{500}$.  Also shown are the best-fit relations. }
\end{figure}
\clearpage

We find a very significant correlation of cluster structure with the inaccuracy of the hydrostatic masses, particularly for $P_3/P_0$.  The hydrostatic masses of disturbed clusters are typically worse underestimates of the true masses with probabilities of no correlation of $8 \times 10^{-23}$ and $8 \times 10^{-6}$ for $P_3/P_0$ and $\langle w \rangle_{no core}$, respectively.  The other power ratios show similar trends to $P_3/P_0$.  Also shown in Figure 9 are the best fit relations of the form $\Delta M/M = A log(X) + B$ based on a least absolute deviation fit (with the significant scatter, fits based on the ordinary least squares estimator appeared to be too biased toward outliers).  For $P_3/P_0$, $A=-0.044$ and $B=-0.49$, and for $\langle w \rangle_{no core}$, $A=-0.055$ and $B=-0.26$.  Similarly, Nagai et al. (2007a) find that the hydrostatic masses of visually unrelaxed systems are more biased than visually relaxed systems.  Here we quantify the correlation with cluster structure, and below we show that this correlation can be used effectively to correct the hydrostatic masses.  We also find a significant (probability of $6 \times 10^{-3}$), but milder correlation of mass underestimate with cluster mass, shown in Figure 10.  The masses of lower mass clusters are typically more underestimated with $A=0.050$ and $B=-0.88$.  These correlations between deviations from hydrostatic equilibrium and both cluster structure and mass can be used as diagnostics for the sources of non-thermal pressure support in clusters.

\clearpage
\begin{figure}
\epsscale{0.8}
\plotone{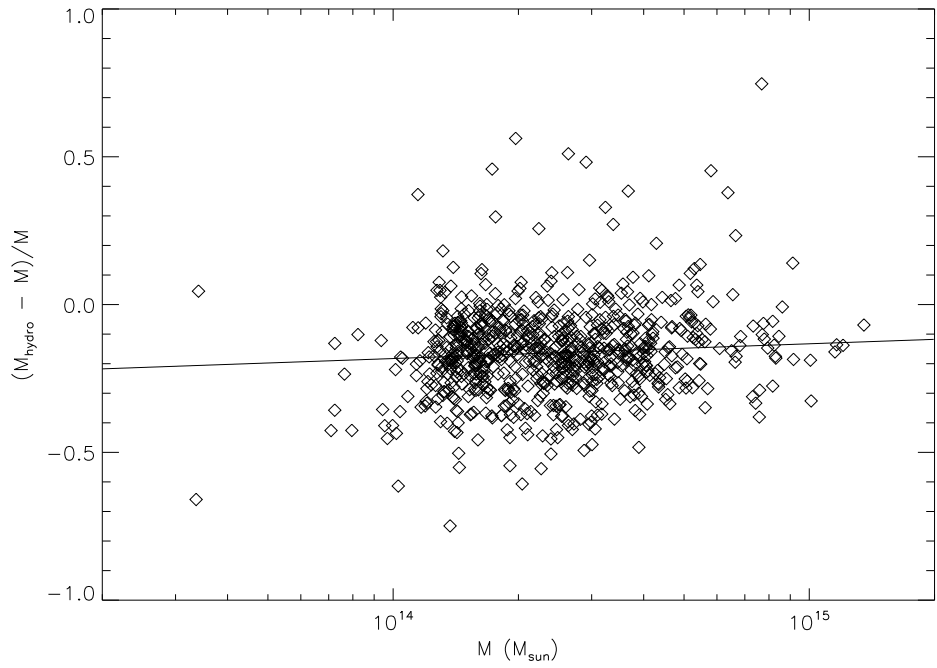}
\caption{ Correlation of the true cluster masses with the deviation of the hydrostatic masses from the true masses within $r_{500}$.  Also shown is the best-fit relation.}
\end{figure}
\clearpage

The results for cluster scaling relations presented above are encouraging, because they show tight relations between observables like $T_X$ and $Y_X$ and cluster mass and that these relations are unbiased with respect to cluster structure, confirming previous results \cite{M05, K06}.  However, the above relations (and previous studies) use the true cluster masses.  As the hydrostatic masses are biased with respect to the true masses, observationally normalized $T_X-M$ and $Y_X-M$ relations will also be biased, even with the best data.  Here we investigate the mass scaling relations with the hydrostatic masses and their dependence on cluster structure.

The left side of Figure 11 shows the $L_X-M_{hydro}$, $T_X-M_{hydro}$, and $Y_X-M_{hydro}$ scaling relations with separate fits to all clusters, relaxed clusters, and disturbed clusters, similar to what was shown for the true masses in Figure 8.  Here the selection of relaxed and disturbed clusters was made using only one projection, corresponding to what is actually observable.  These fits are listed in Table 4 as well as the fits to the total cluster sample using the true masses which we will take as the ``true'' scaling relations.  Again, the slopes have been fixed at their predicted self-similar values (4/3 for $L_X-M$, 2/3 for $T_X-M$, and 5/3 for $Y_X-M$). 

It is clear from Figure 11 that using the hydrostatic masses leads to both a significant increase in the scatter of these relations as well as systematic offsets due to cluster structure.  We find significant offsets in the $T_X-M_{hydro}$ and $Y_X-M_{hydro}$ relations between relaxed and disturbed clusters (the known offset of the relaxed clusters in the $L_X-M$ relation is less obvious here).  In addition, the mass scaling relations are significantly different than those obtained using the true masses even if the most relaxed clusters, which have the least biased hydrostatic masses, are used.  For example, the normalization of the $Y_X-M_{hydro}$ for the relaxed subsample is 27\% higher than the relation for all clusters using the true masses ($11 \sigma$ significance).  This offset easily explains the offset found by Kravtsov et al. (2006) between simulated and observed clusters (see also Nagai et al. 2007b).
This result implies that even if deep observations of relaxed clusters are used to normalize the mass scaling relations the normalizations will be systematically biased.

\clearpage
\begin{figure}
\begin{center}
\epsscale{1.0}
\plottwo{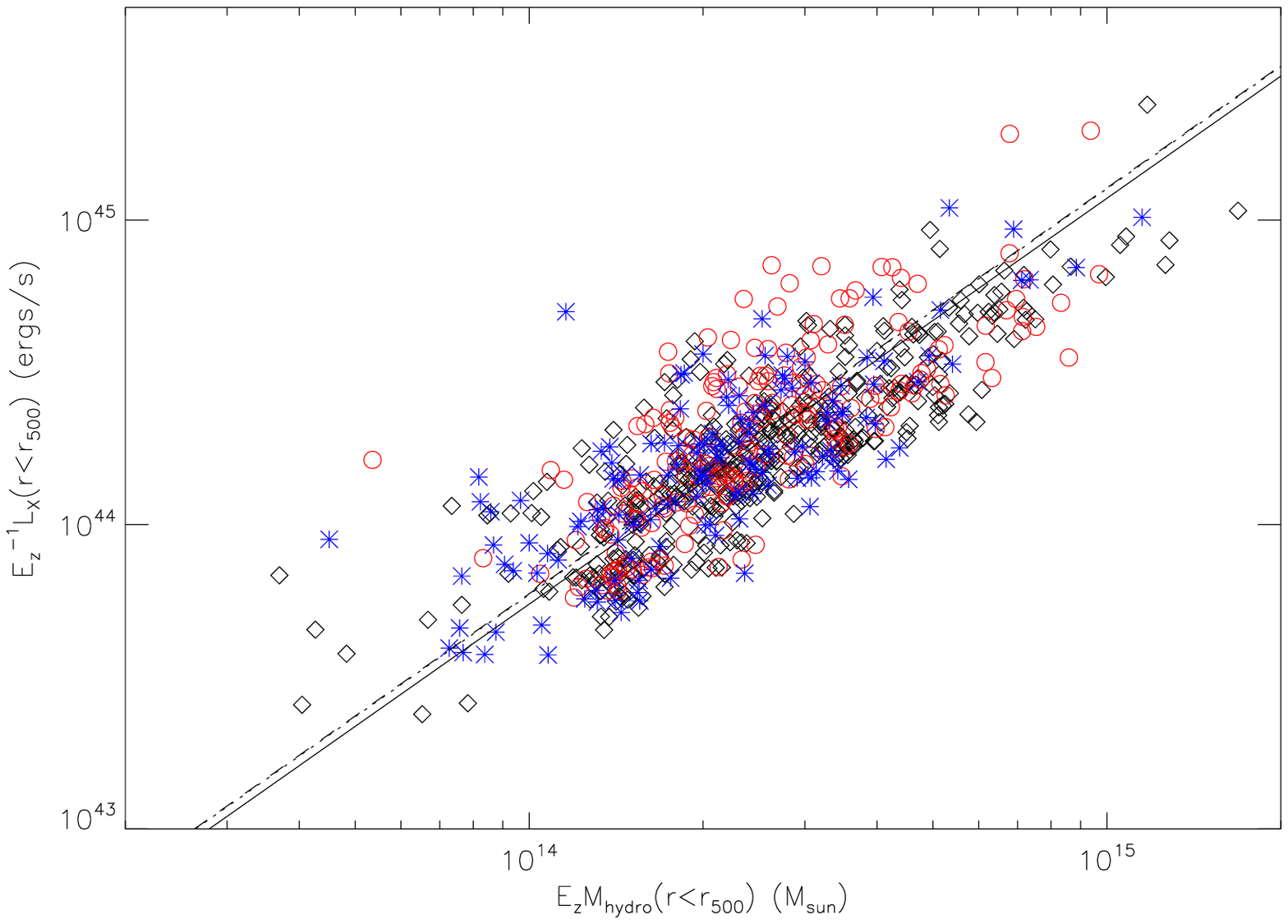}{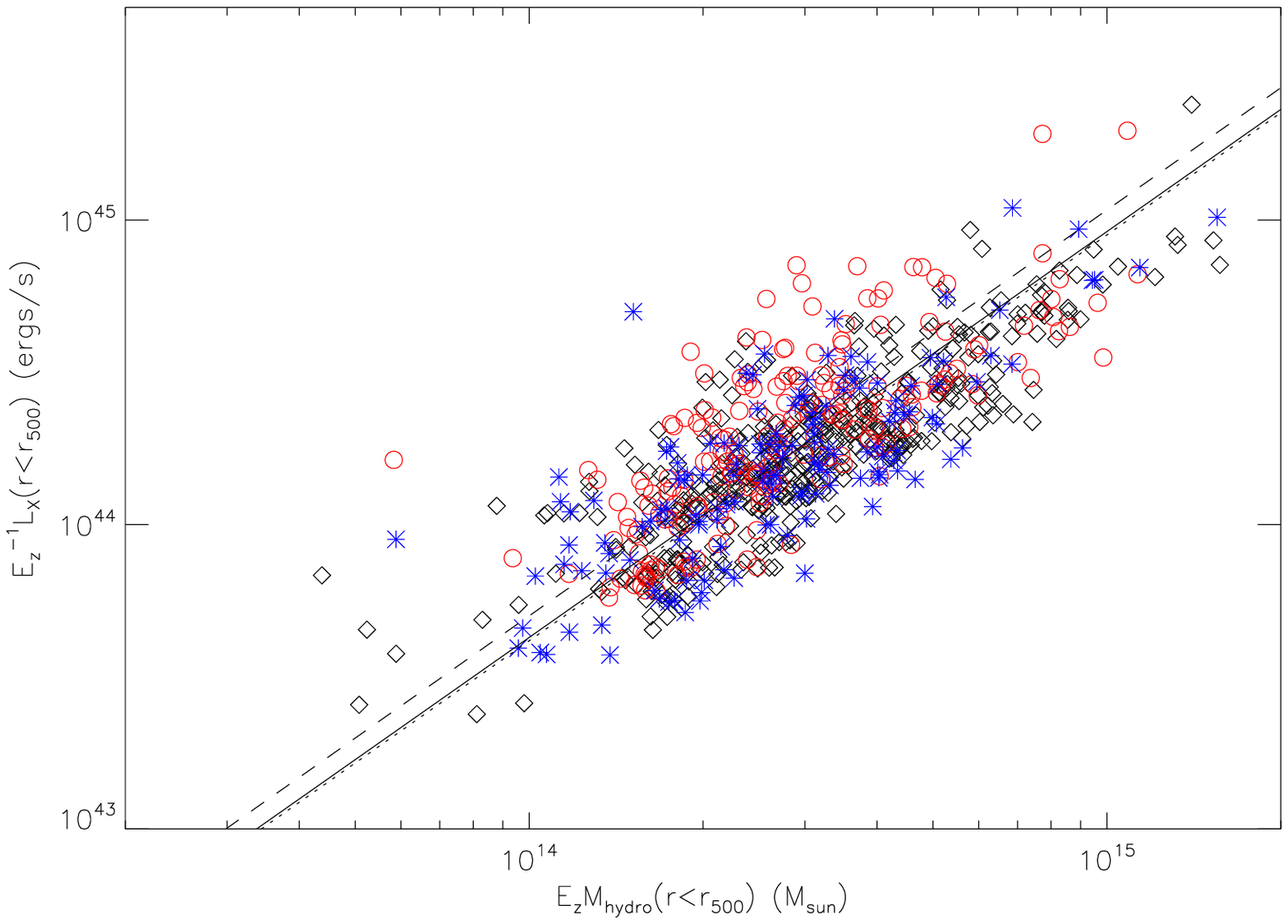}
\plottwo{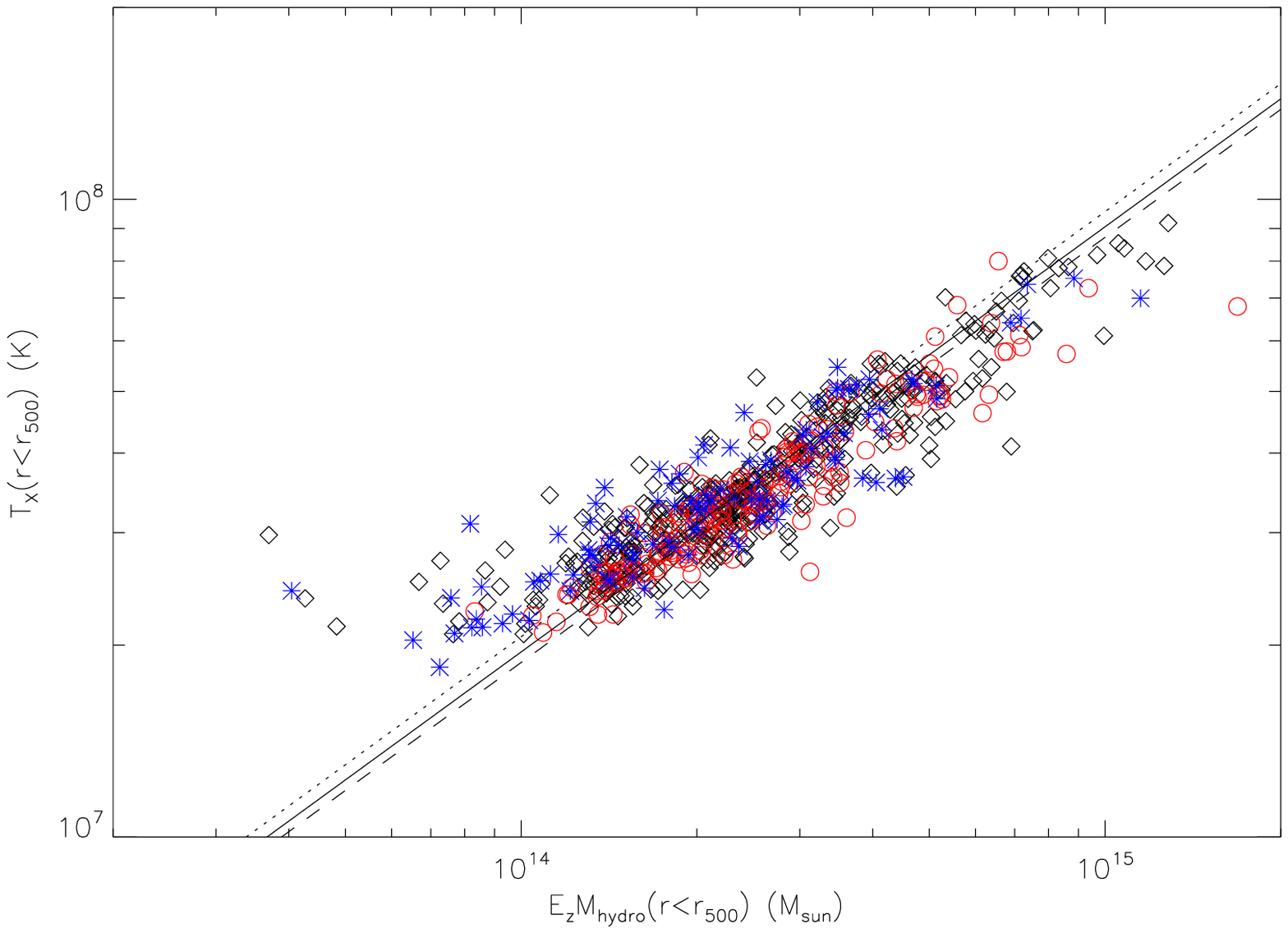}{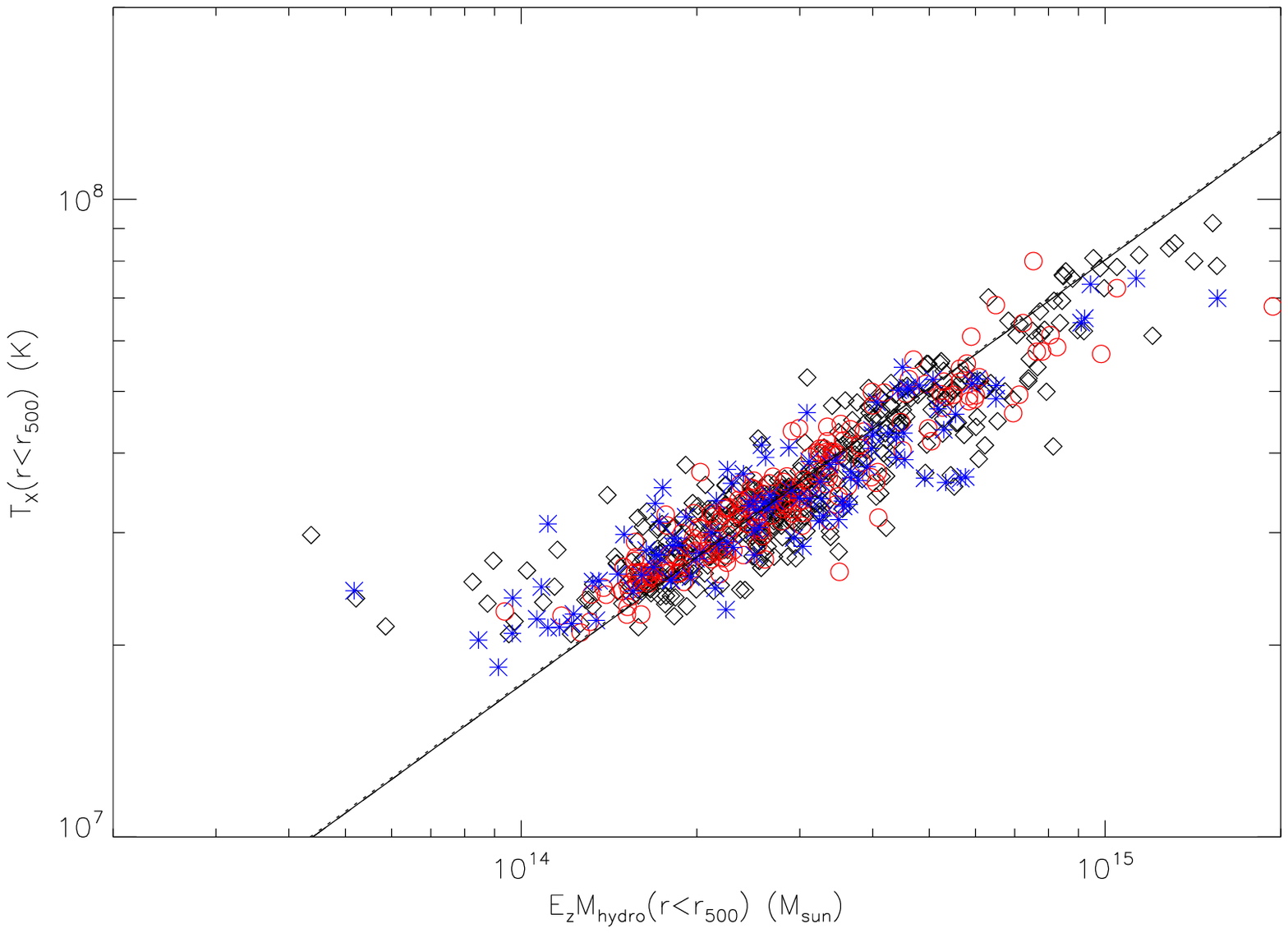}
\plottwo{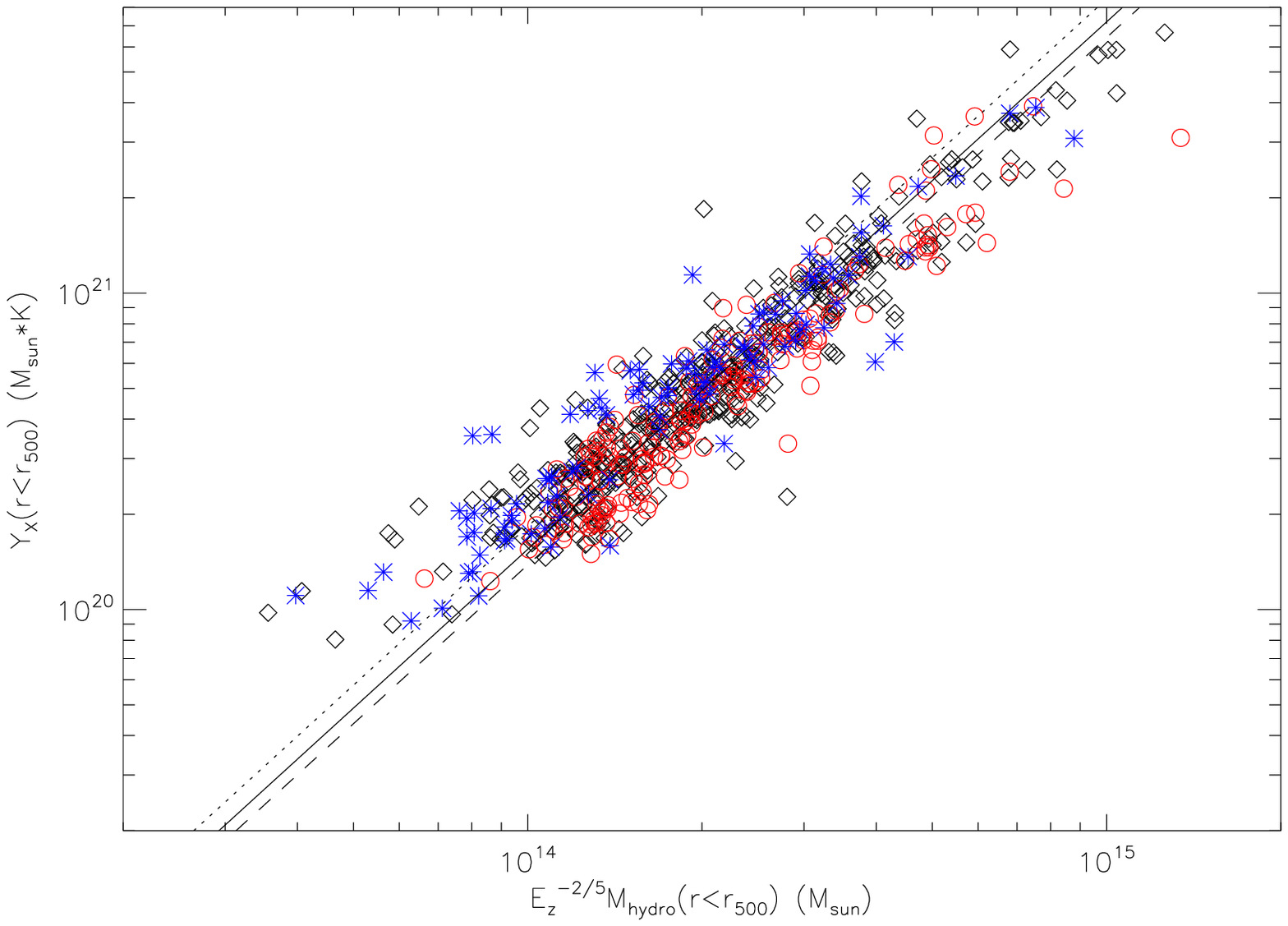}{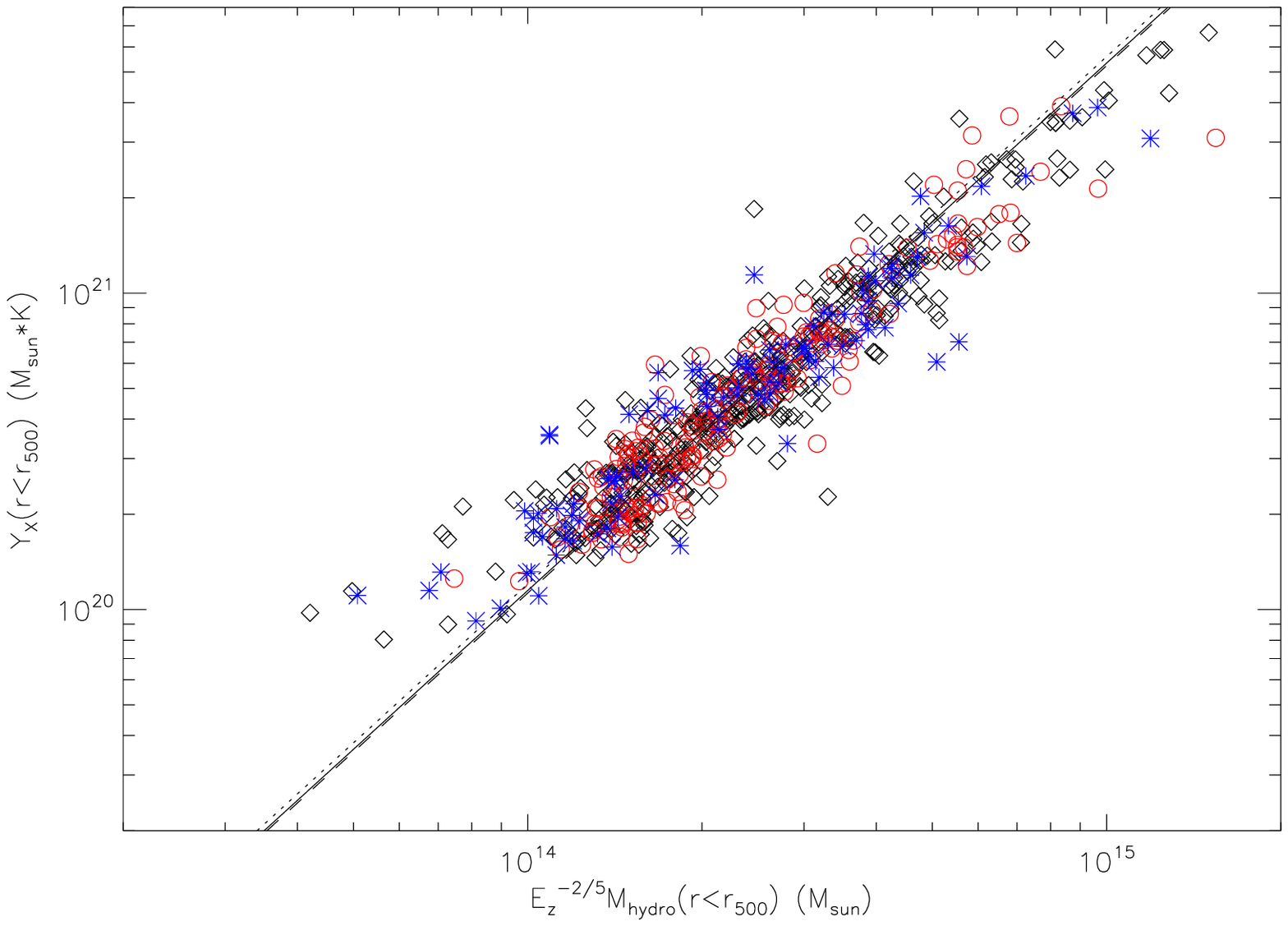}
\caption{ \textit{Left:} $L_X-M_{hydro}$ (\textit{top}), $T_X-M_{hydro}$ (\textit{middle}), and $Y_X-M_{hydro}$ (\textit{bottom}) relations where the masses are calculated assuming hydrostatic equilibrium and using the 3D spherically averaged density and temperature profiles.  All quantities are calculated within a radius of $r_{500}$.  Red circles show relaxed clusters, blue asterisks show disturbed clusters, and the rest of the sample is plotted with diamonds.  The selection of these subsamples was done based on the values of $P_3/P_0$ in one projection.  Lines show the fits with the slopes fixed as in Figure 7 to all clusters (solid line), relaxed clusters (dashed line), and disturbed clusters (dotted line).  \textit{Right:} Same except that the hydrostatic masses have been corrected based on $P_3/P_0$ for each cluster. }
\end{center}
\end{figure}
\clearpage

However, the strong correlation between cluster structure and the deviation of the hydrostatic masses from the true masses shown in Figure 9 means that we can use cluster structure to at least partially correct the hydrostatic mass estimates.  The right side of Figure 11 shows the scaling relations that result if we use the fit in Figure 9a to adjust the hydrostatic cluster masses based on their $P_3/P_0$.  The corresponding fits are listed in Table 4.  The $T_X-M$ and $Y_X-M$ relations are now unbiased with respect to cluster structure showing no significant offsets for the relaxed and disturbed subsamples.  The significant offset of relaxed clusters in the $L_X-M$ relation present for the true masses is now also reproduced.  Even better, the normalizations of all of the scaling relations are now in very good agreement with the relations based on the true masses, although the scatter about these relations remains higher.

The above results are for an assumed slope; however, we also find that the best fit slopes of the mass scaling relations change if we use the hydrostatic masses.  Figure 12 and Table 5 compare fits with variable slope to the mass scaling relations (for all clusters) using the true, hydrostatic, and corrected masses.  These fits use the bisector modification to the BCES method in Akritas \& Bershady (1996), a regression method commonly employed by observers because it allows for intrinsic scatter and non-uniform measurement errors in both variables.  Here it can be seen that the scaling relations based on the hydrostatic masses do differ significantly in both slope and normalization from the true relations.  Correcting the hydrostatic masses based on cluster structure changes the best-fit slopes only slightly, but does shift the normalization to make the relations in overall better agreement with most of the clusters.  The lack of change in slope after mass correction is unsurprising given the lack of correlation between mass and cluster morphology shown in section 3.2.1.  Correcting for the additional correlation between mass and the bias in the hydrostatic mass (Figure 10) would bring the slopes in to better agreement.

Table 5 also lists the scatter about the fit relations in terms of the scatter in $log(X)$ (mass) for fixed observable.  Employing the hydrostatic masses significantly increases the scatter, and the scatter is only slightly reduced by correcting the masses.  We have shown that we can eliminate systematic biases in the mass-scaling relations; the scatter remains higher than the true relations due to the scatter in the relationship between our structure measures and deviations from hydrostatic equilibrium.  As we showed in section 3.1, one of the major sources of scatter in the relationship between observed structure and dynamical state is projection, and the nature of this scatter can be better understood through studies of this kind.  Similar to Kravtsov et al.~(2006), we find the lowest scatter in mass for the $Y_X-M$ relation.

\clearpage
\begin{figure}
\begin{center}
\epsscale{1.1}
\plottwo{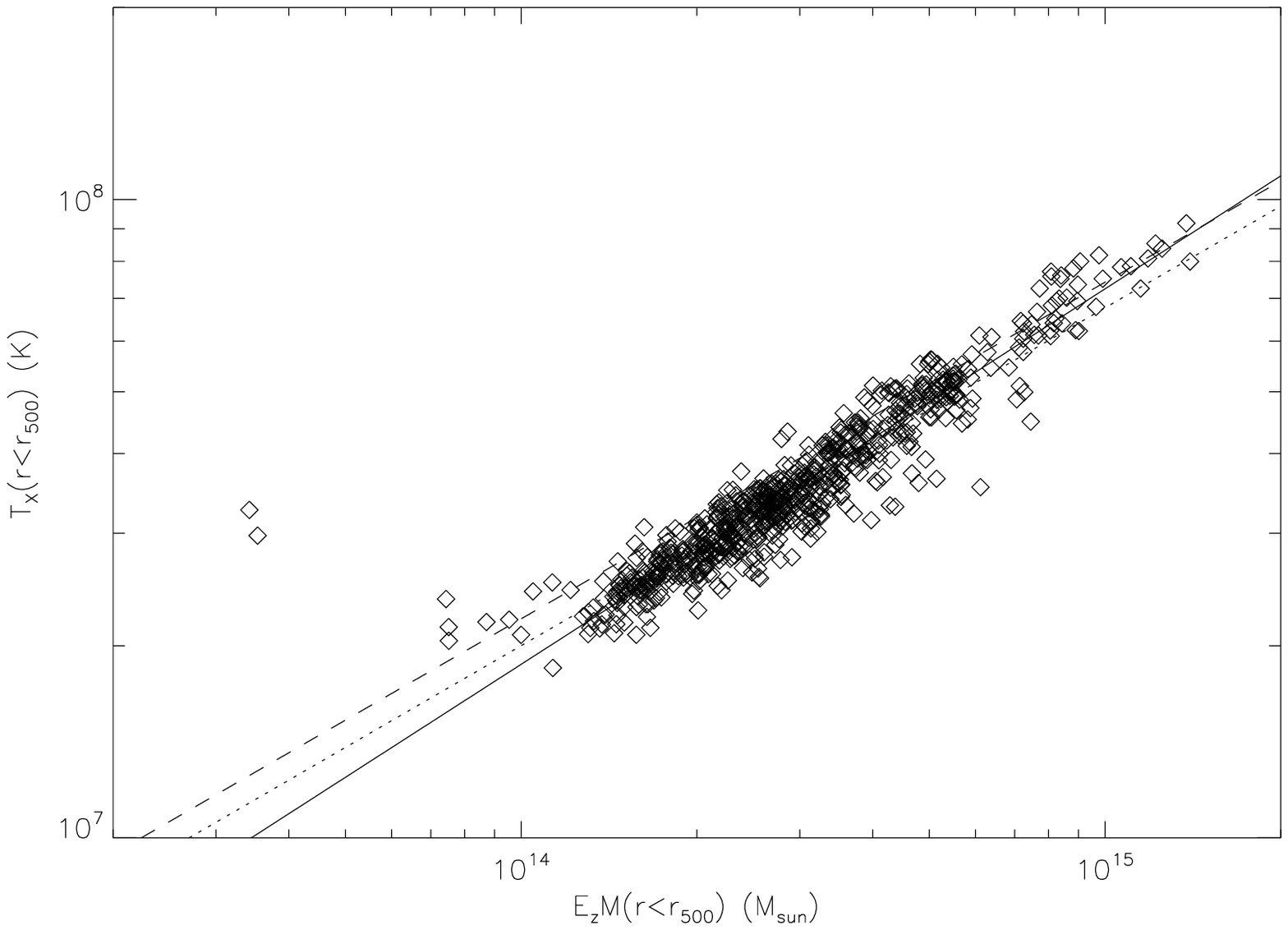}{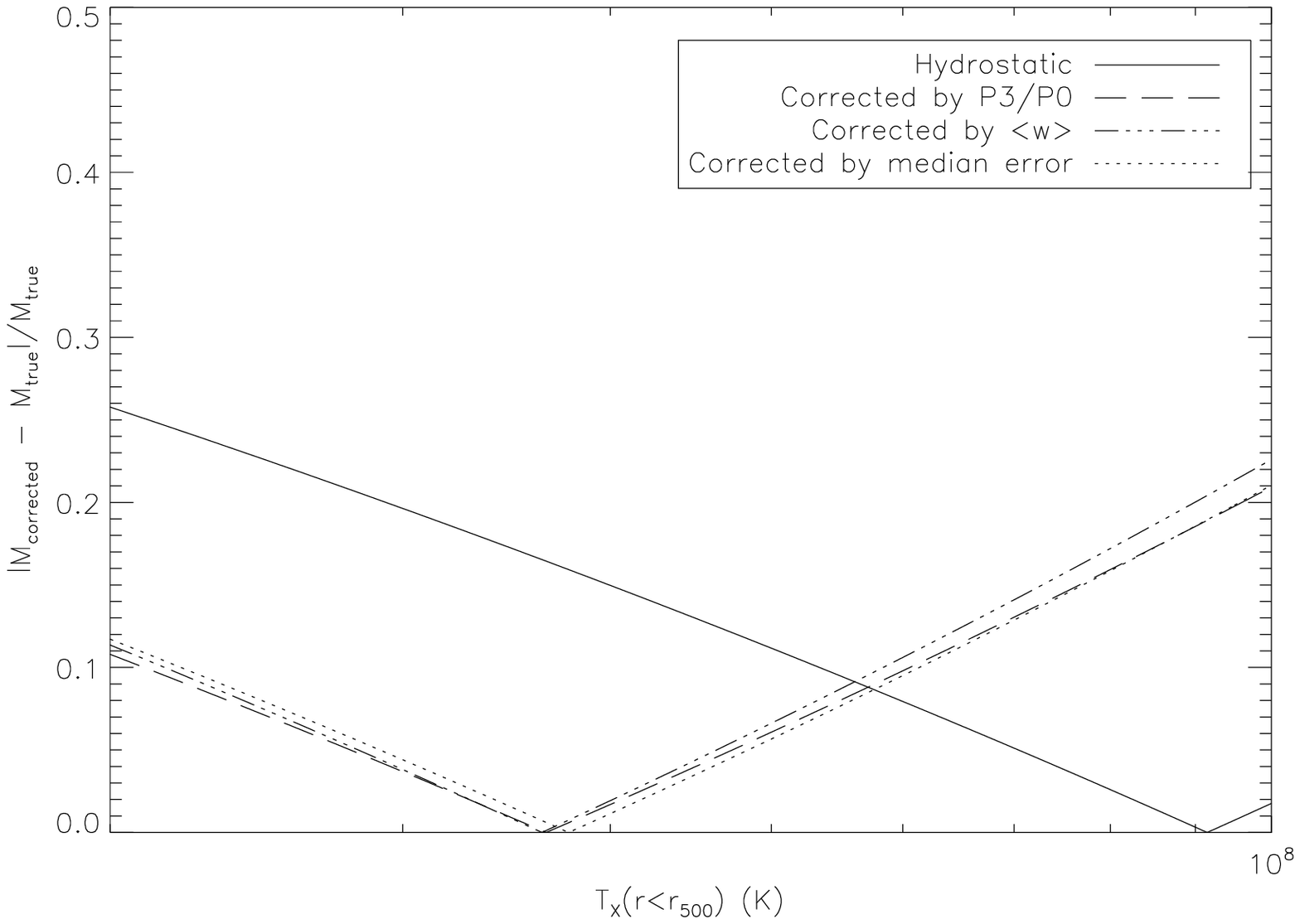}
\plottwo{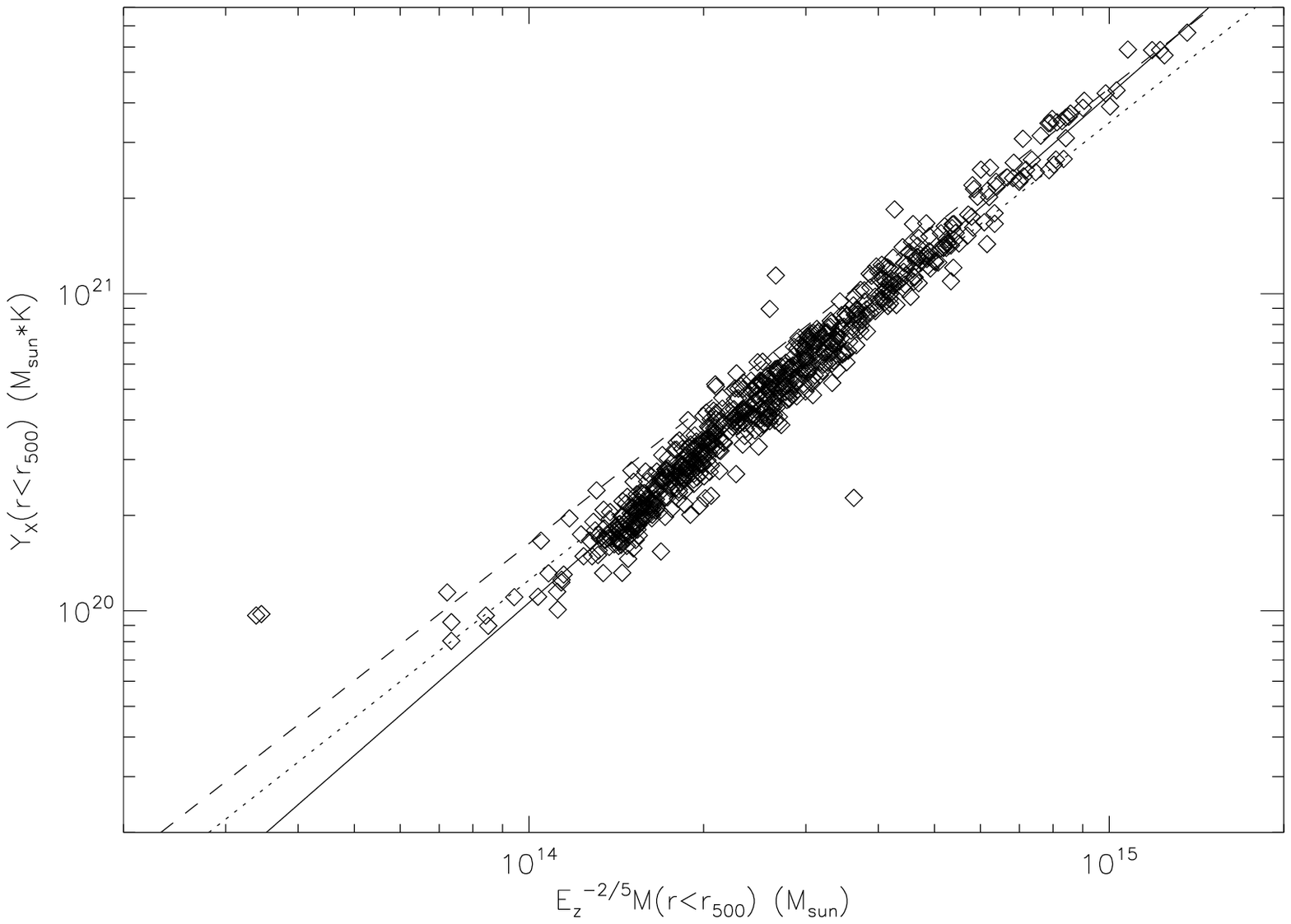}{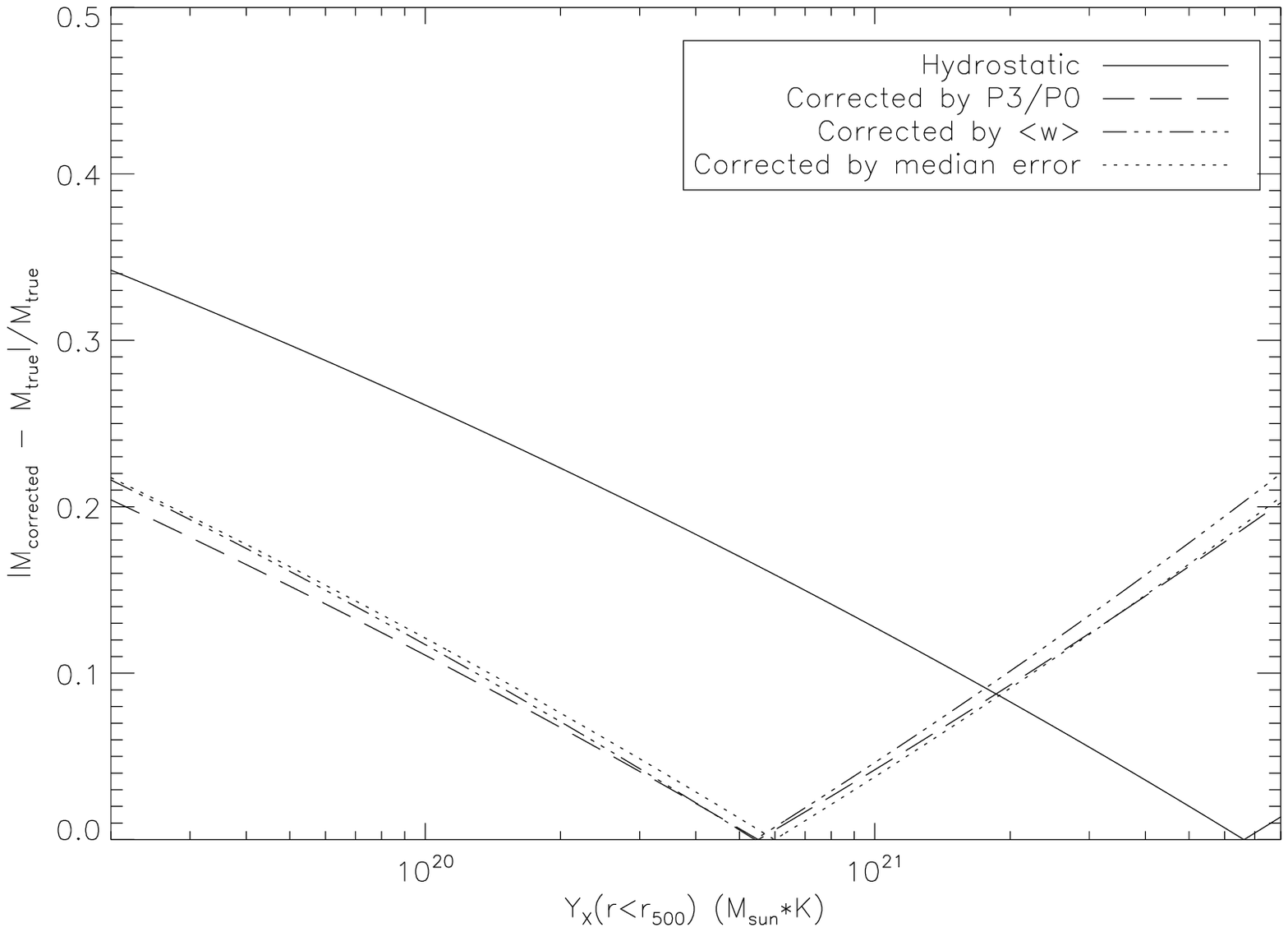}
\caption{ \textit{Left:} Comparison of the fits to the $T_X-M$ (\textit{top}) and $Y_X-M$ (\textit{bottom}) relations using the true masses (solid line), hydrostatic masses (dashed line), and corrected masses (dotted line).  For comparison, the points show the properties of the simulated clusters and their true masses.  \textit{Right:} The solid line shows the difference between the masses that would be estimated from the $T_X-M_{hydro}$ (\textit{top}) and $Y_X-M_{hydro}$ (\textit{bottom}) relations and the masses derived from $T_X-M_{true}$ and $Y_X-M_{true}$.  This difference is presented as an absolute fractional deviation.  Also shown are the absolute fractional deviations relative to the true relations if the hydrostatic masses are corrected by $P_3/P_0$ (long dashed), by $\langle w \rangle_{nocore}$ (dot dot dashed), and by the median error in the hydrostatic masses (dotted). }
\end{center}
\end{figure}
\clearpage

The advantage of correcting the hydrostatic masses using $P_3/P_0$ is shown clearly on the right side of Figure 12.  The solid line shows the absolute fractional difference between the masses that would be estimated from the $T_X-M_{hydro}$ and $Y_X-M_{hydro}$ relations and the masses derived from $T_X-M_{true}$ and $Y_X-M_{true}$, while the dashed line shows the same difference after the hydrostatic masses are corrected by $P_3/P_0$.
The corrected masses are closer to the true masses over most of the ranges in $T_X$ and $Y_X$, except at the highest masses.  At most the corrected masses are off by 20\% compared to the true masses, and unlike the hydrostatic masses, which for the most part show a systematic shift to lower masses, these can be either over or underestimates.  The same result can be seen if we compare the corrected masses for individual clusters: the corrected masses are closer to the true masses than the hydrostatic masses for about 75\% of the simulated clusters.  Also shown in Figure 12 are the resulting mass errors if we instead correct the hydrostatic masses based on centroid shift (dot dot dashed line) or based simply on the median error in the hydrostatic masses (dotted line).  The centroid shifts show a less significant correlation with mass error, and this correction performs worse than the power ratios over the full ranges of $T_X$ and $Y_X$.  It appears that while the centroid shifts are more robust against projection than the power ratios, they are less sensitive to deviations from hydrostatic equilibrium.  Correcting simply by the median error in the hydrostatic masses of course leads to an improvement in the mass estimates but performs worse than or similar to $P_3/P_0$ everywhere, showing the need for a structure dependent factor.

Here we have not explicitly shown the effect of cluster structure on gas mass fraction, another important cosmological tool.  A couple of recent studies have shown that gas mass can be estimated fairly accurately (within $\sim10$\%) with a relatively small bias for unrelaxed clusters \cite{H06,N07}.  In this work, we consider gas masses estimated using a universal temperature profile and deprojection under the assumption of spherical symmetry, and we find a significant correlation between cluster structure and deviation of the estimated gas masses from the true gas masses.  The errors in gas mass, though, are fairly small.  The gas masses of relaxed clusters are biased high by less than 5\% while the gas masses of disturbed clusters are high by $\sim$10\%.  However, the error in the hydrostatic masses and its structure dependence will translate directly in to error in the gas mass fractions.  Even considering only our relaxed subsample, the typical error in hydrostatic mass is 12\% and individual ``relaxed'' clusters have errors as large as 56\%.  For disturbed clusters, the $\sim$10\% overestimate in gas mass will add to the larger underestimate in hydrostatic mass.  A similar result was found by Nagai et al.~(2007a) for visually unrelaxed systems.

\subsection{ Comparison to Observations: Low Redshift }

An important question is how well the simulations reproduce the observed distribution of cluster morphologies.  Cluster structure and its evolution depend on both cosmology and gas physics and can act as a sensitive test of how well current simulations are reproducing the observed universe.  A detailed comparison to observations including the effects of noise, instrumental response, and selection is beyond the scope of the present paper but will be explored in future publications.  Here we limit our scope to a broad comparison of the structure of simulated clusters to low-redshift observations where the signal-to-noise is high and observational effects can be expected to be minimal.

We compare the power ratios of our simulated clusters to the \textit{ROSAT} observed sample from Buote \& Tsai (1996).  We limit both samples to include only clusters with redshifts less than or equal to 0.1 and luminosities greater than $10^{44}$ ergs s$^{-1}$.  For the simulations this includes three redshift bins at $z = $ 0.04, 0.08, and 0.1.  At these redshifts, the \textit{ROSAT} resolution ($\sim 20''$) is reasonable compared to the resolution of the simulations ($28''$ at z=0.04 and $12''$ at z=0.1), and the \textit{ROSAT} field-of-view, unlike \textit{Chandra}'s, is large enough to study cluster structure to reasonably large radii.  For direct comparison to the observations we consider the power ratios within the physical aperture radii of 0.5 Mpc and 1 Mpc.  Table 6 lists the median power ratios for these two radii.  We also list the probabilities from a Wilcoxon rank-sum test (e.g., Walpole \& Myers 1993) that the mean power ratios of the observed and simulated samples are the same and the probabilities from a Kolmogorov-Smirnov test (e.g., Press et al. 1992, \S 14.3) that the distributions of power ratios are the same.  Figure 13 compares the distributions of power ratios in the simulations (solid line) and observations (dashed line) for $R=0.5$ Mpc.

\clearpage
\begin{figure}
\epsscale{0.5}
\plotone{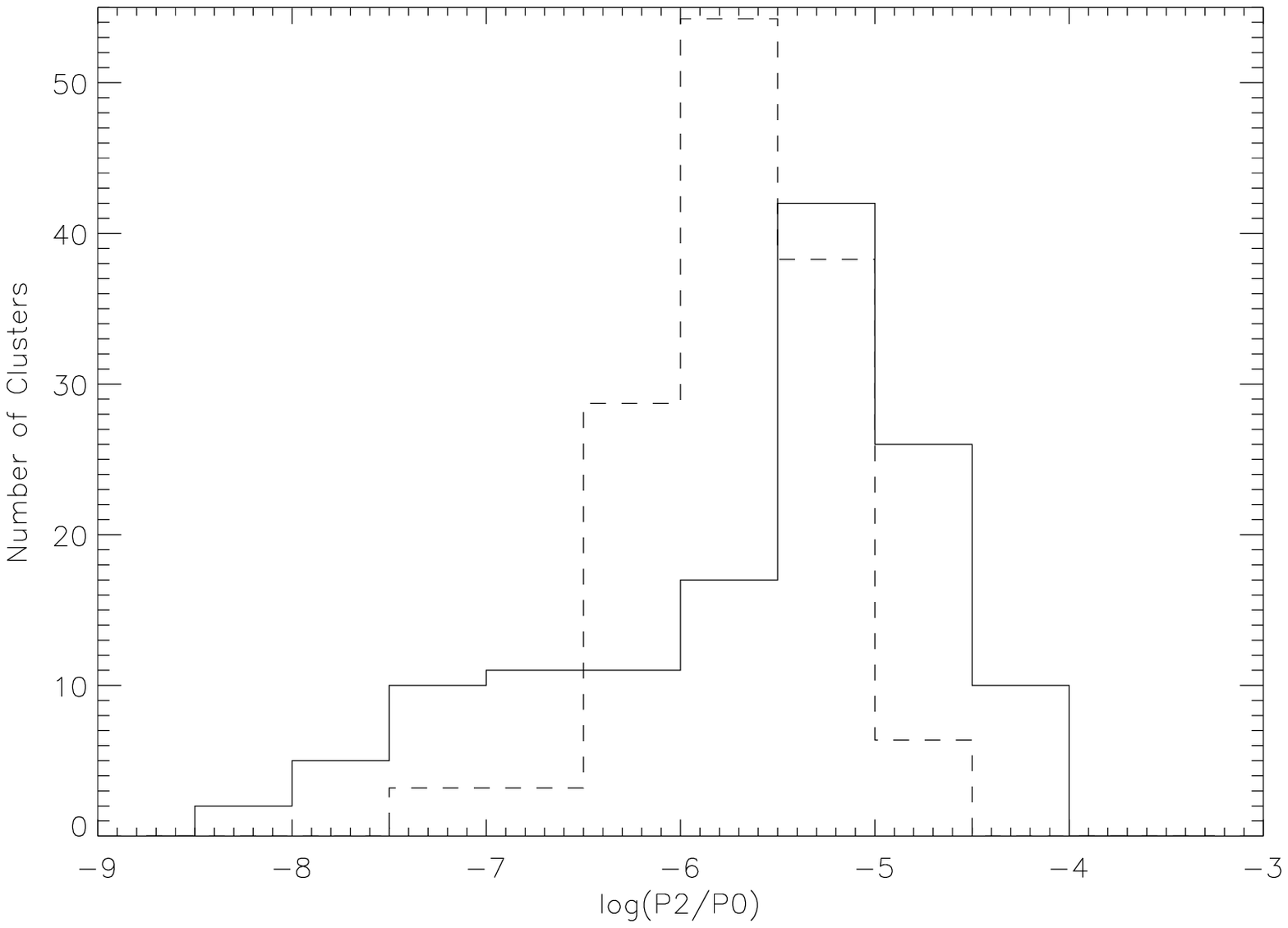}
\plotone{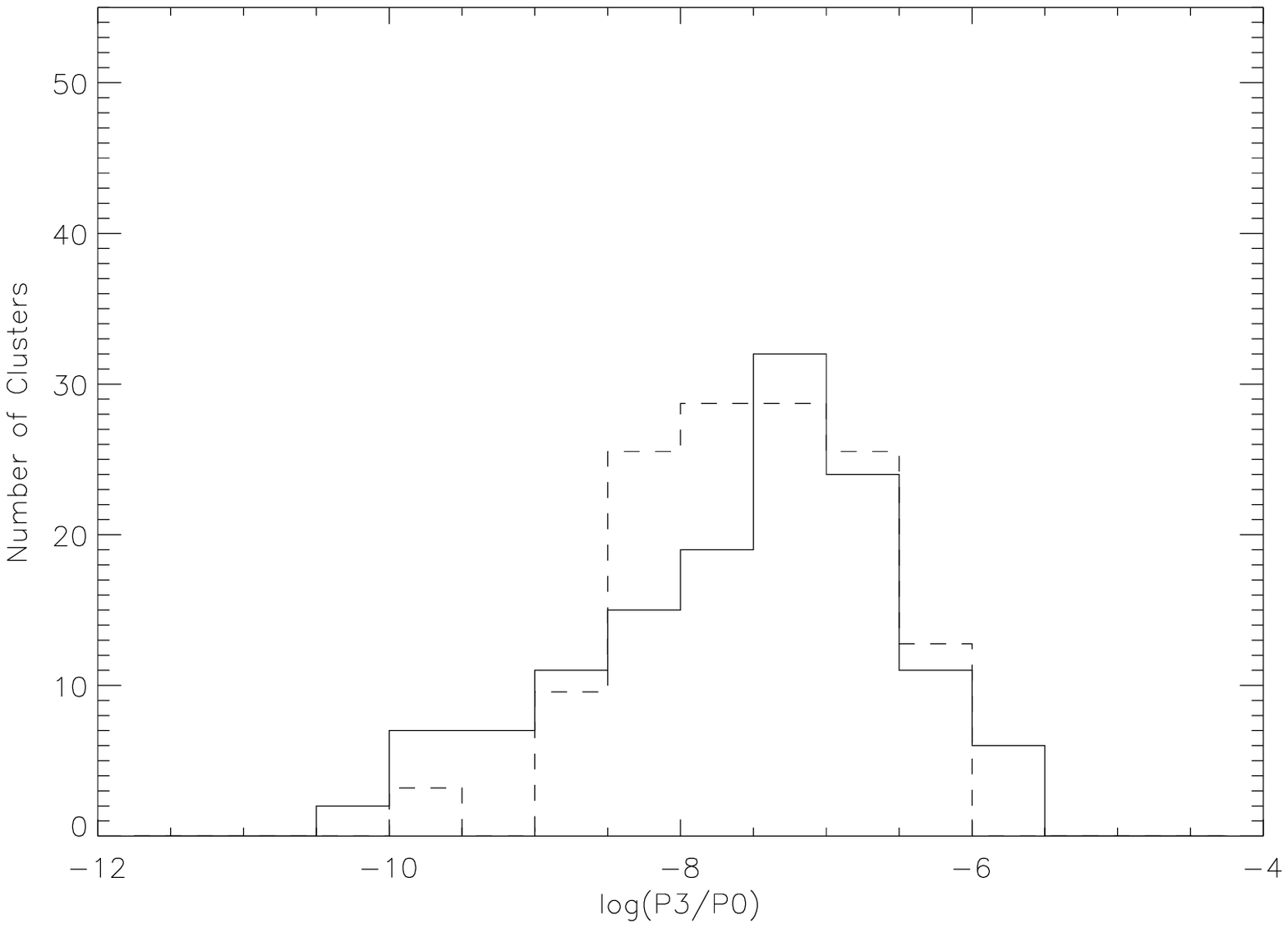}
\plotone{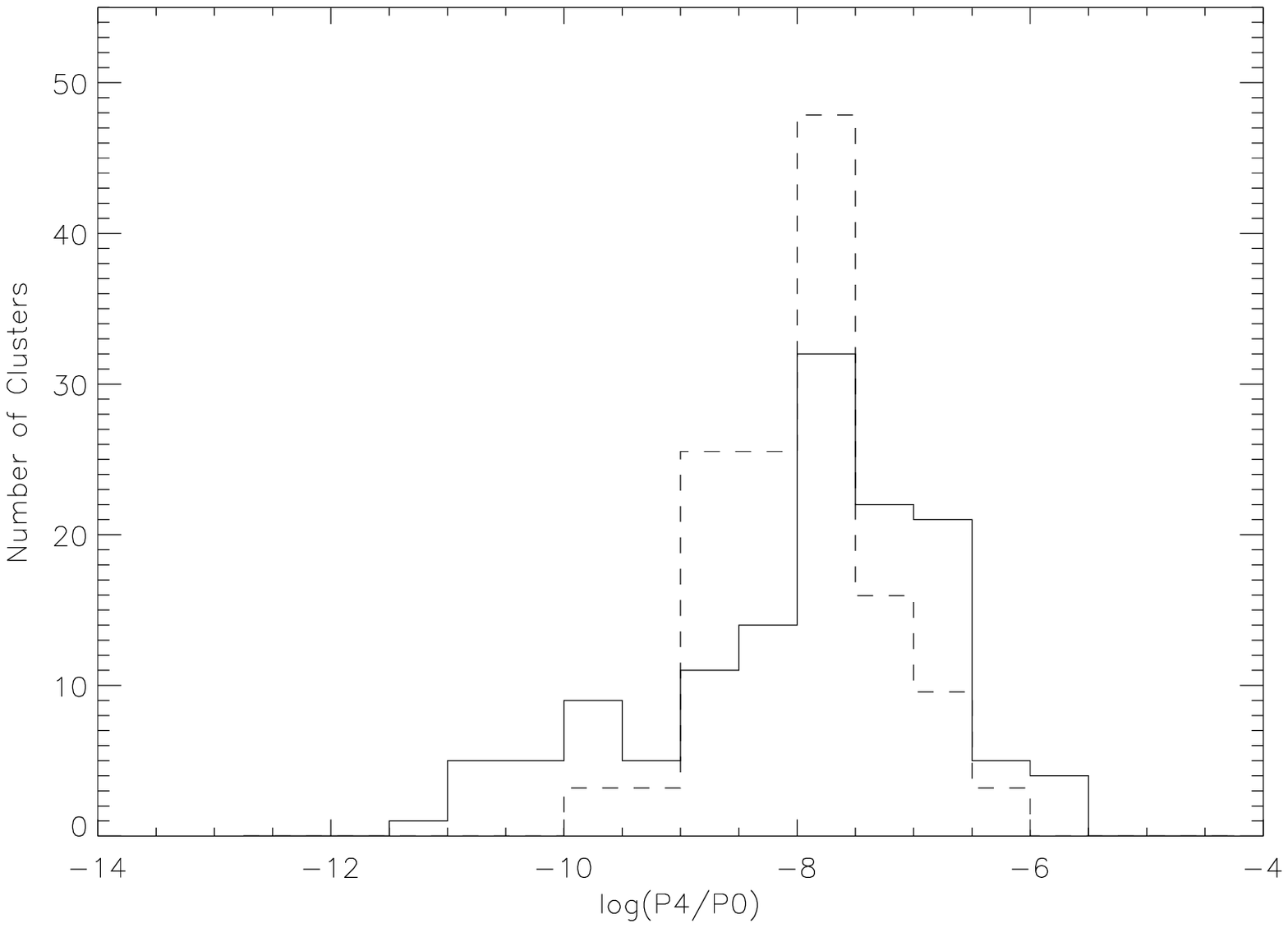}
\caption{ Comparison of the distributions of power ratios within $R=0.5$ Mpc in simulations (solid line) and observations (dashed line) for low-redshift clusters ($z \le 0.1$) with luminosities greater than $10^{44}$ ergs s$^{-1}$.  The number of observed clusters (42) is normalized to give the same total number of clusters as in the simulations.  The observational sample is taken from the \textit{ROSAT} sample of Buote \& Tsai (1996). }
\end{figure}
\clearpage

These comparisons show that clusters in our simulations have remarkably similar structures to observed clusters compared to what might have been expected.  The power ratio distributions have similar peaks, and we find no statistically significant differences in either the mean power ratios or the distributions of the two samples (with the possible exception of $P_2/P_0$ at $R=0.5$ Mpc).  However, investigation of Figure 13 reveals that the simulations show a broader distribution of power ratios than observed, particularly for the even power ratios.  The simulations appear to include both more clusters with very high power ratios and a tail of clusters with low power ratios extending to lower power ratios than are observed.  Investigation of those clusters in the simulations which have power ratios lower than what is observed reveal that these are typically low mass systems with luminosities that are high for their masses.  Clusters with power ratios higher than observed, on the other hand, tend to have larger masses.  For example, Figure 14 shows the $L_X-M$ relation for the simulated clusters in Figure 13 with very low power ratio (circles) and very high power ratio (squares) clusters highlighted.  

\clearpage
\begin{figure}
\epsscale{0.7}
\plotone{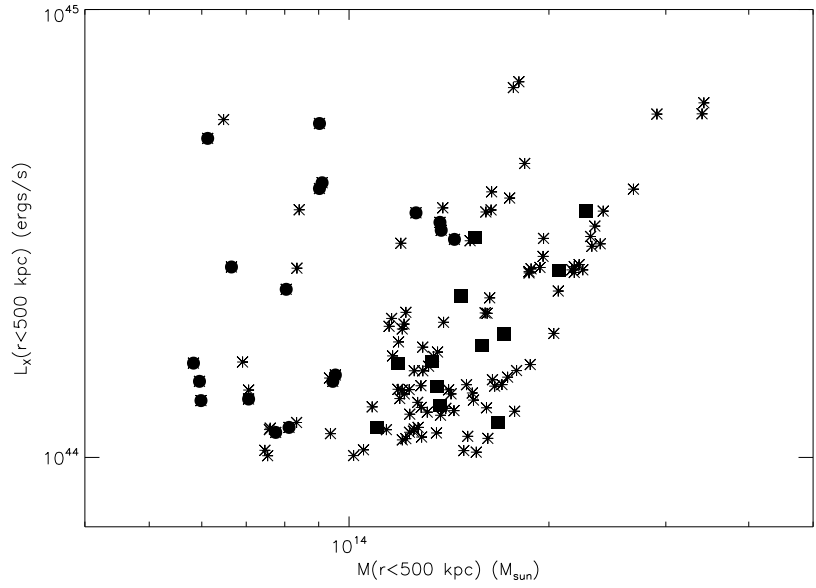}
\caption{ $L_X-M$ relation ($R=0.5$ Mpc) for simulated clusters with $z \le 0.1$ and luminosities greater than $10^{44}$ ergs s$^{-1}$.  Solid circles mark clusters with power ratios which are low compared to observed clusters, and solid squares mark clusters with power ratios which are high compared to observed clusters. }
\end{figure}
\clearpage

Figure 14 shows a similar trend for low power ratio clusters to be offset to higher luminosities as was seen in section 3.2.2, but here we use a stricter cut-off on morphology.  These observations strongly suggests the role of cool cores in low-mass clusters in producing at least the low power ratio tail of clusters.  In fact, all but one of the eighteen low power ratio clusters marked in Figure 14 are classified as a cool cores by the criteria defined in Burns et al.~(2007).  The radial dependence of the power ratios means that if clusters are more strongly peaked in the simulations compared to observations, the power ratios of single component clusters will be significantly lower.  At the high end, small but bright, strongly peaked substructures, bright shocks, or cold fronts could contribute to increasing the power ratios.  On the other hand, the median $P_2/P_0$ and $P_4/P_0$ of clusters in the simulations are slightly higher than observed.  Together with the fact that $P_3/P_0$ shows the best match to observations, this suggests that the simulated clusters are on average a bit more elliptical than observed.  As cooling acts to reduce cluster ellipticity (Kazantzidis et al. 2004; Rahman et al. 2006), our results could indicate the need for additional cooling in more massive clusters.  A couple of other studies have also found an offset between the observed and simulated ellipticites of clusters, but not in a consistent direction (Rahman et al. 2006; Flores et al. 2007).  This comparison likely depends on the specifics of the cooling and feedback prescriptions used.

Examples of clusters in the simulations with very low and very high power ratios are shown in Figures 15 and 16, respectively.  The three clusters with very small power ratios in Figure 15 are all low mass, cool core clusters with very strong surface brightness peaks and large dips in temperature in the core.  
The clusters in Figure 16 with high power ratios are unsurprisingly strongly merging with one or more bright substructures, shocks in both the surface brightness and temperature maps, and no cool core.  These clusters are viewed near the peak of the merger.  The third cluster, in particular, has a main component which is very disrupted showing no clear core.

\clearpage
\begin{figure}[h]
\epsscale{0.32}
\plotone{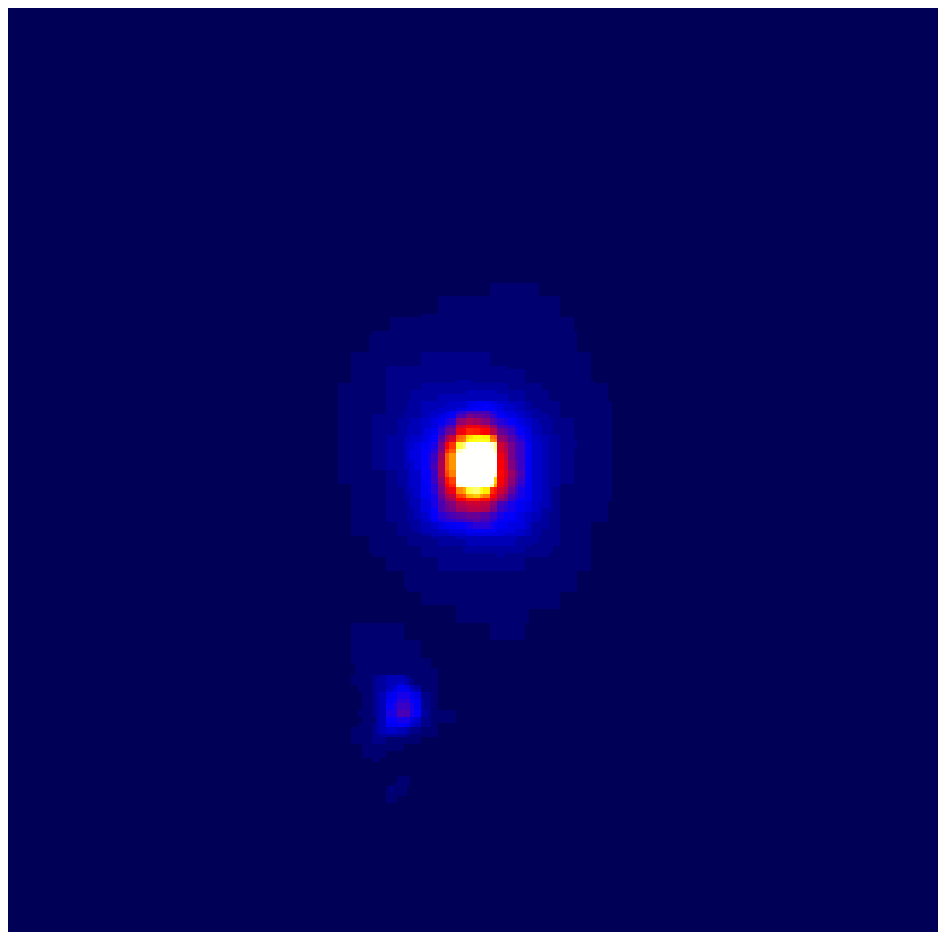}
\plotone{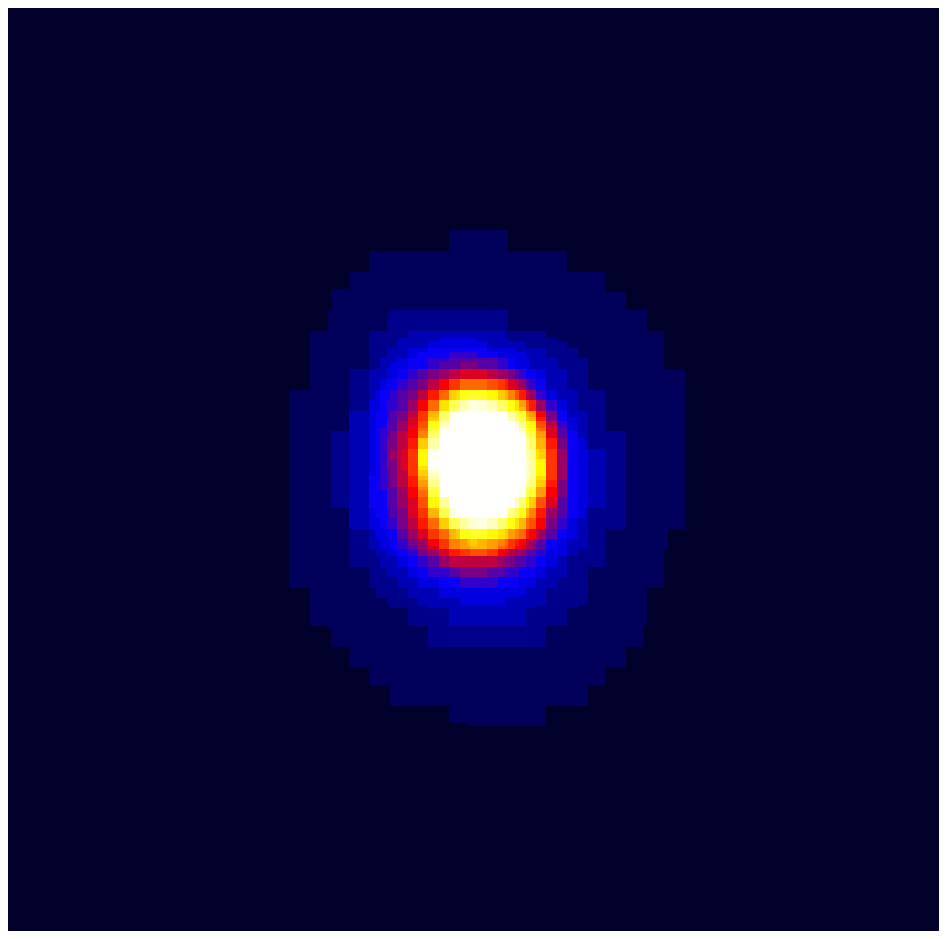}
\plotone{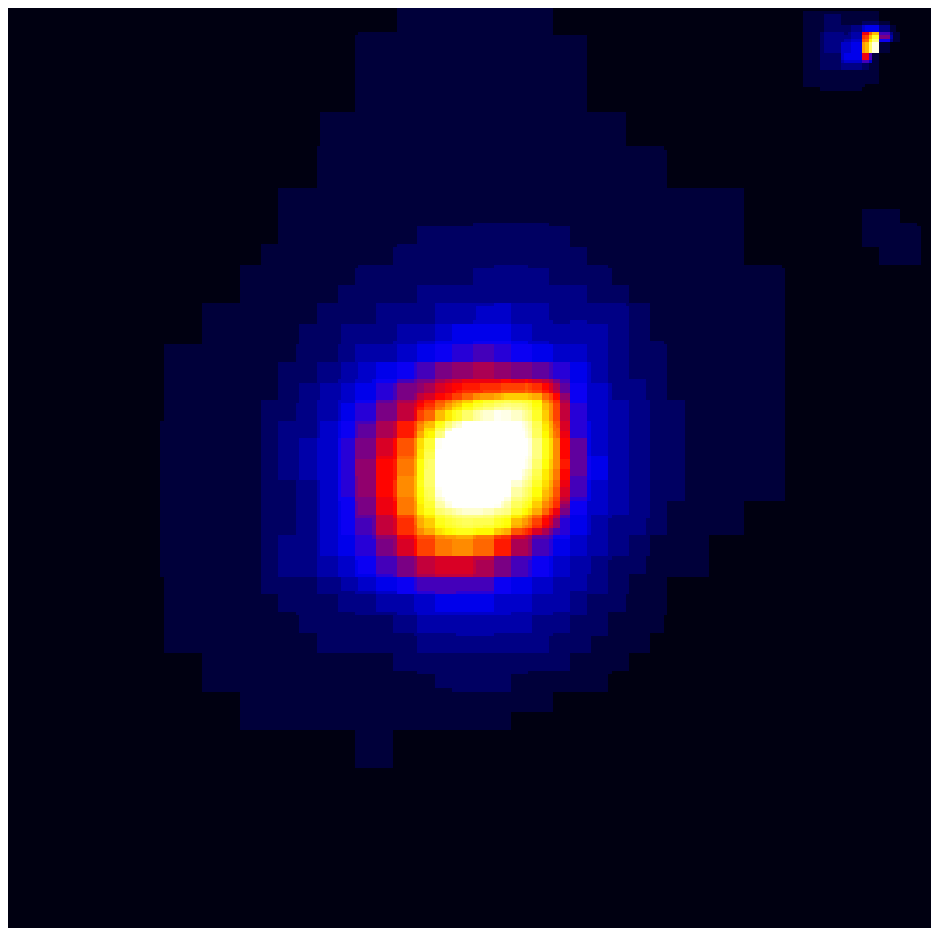}
\caption{ Examples of clusters in the simulations with very low power ratios, lower than typically seen in observations. Images are 2 Mpc on a side. }
\end{figure}

\begin{figure}[h]
\epsscale{0.32}
\plotone{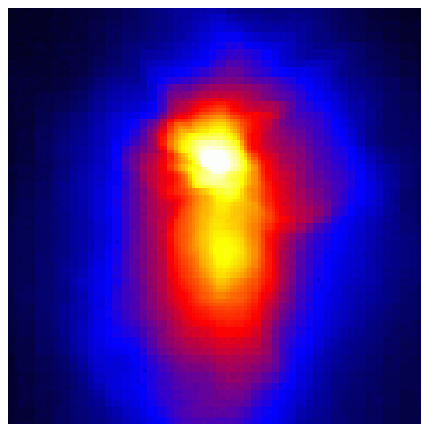}
\plotone{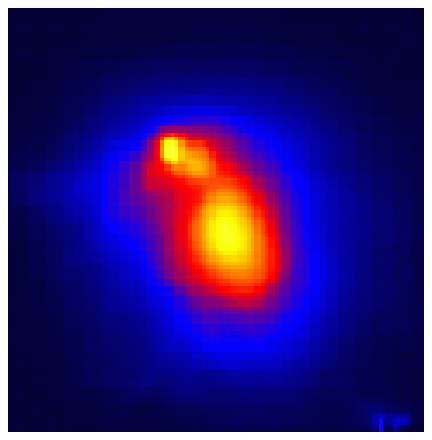}
\plotone{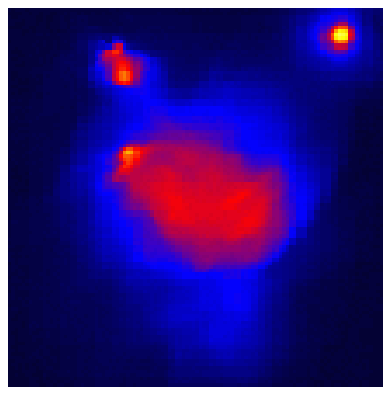}
\caption{ Examples of clusters in the simulations with very high power ratios, higher than typically seen in observations. Images are 2 Mpc on a side. }
\end{figure}
\clearpage

Observational effects which might contribute to the observed cluster morphology include resolution and noise, both of which could smooth out bright structures.  We tested both of these effects using the simulated clusters shown in Figures 15 and 16.  First the images were smoothed with a Gaussian filter with width equivalent to the \textit{ROSAT} resolution.  However, even at z=0.1, the \textit{ROSAT} resolution is within a factor of two of the resolution of the simulations and sufficiently small to resolve the typical scale of structure probed by the power ratios, and this smoothing has little effect.  We then added noise appropriate for a 10 ksec observation with \textit{ROSAT}, which is the typical exposure time of the clusters in the Buote \& Tsai (1996) sample.  The addition of noise has almost no effect on the high power ratio clusters in Figure 16, which is not surprising.  Jeltema et al. (2005) showed that for clusters with significant structure, the power due to noise is a very small fraction of the total power.  Noise also has little effect on $P_2/P_0$, because it is the lowest order power ratio; however, this is the ratio that shows the largest discrepancy between observations and simulations.  $P_3/P_0$ and $P_4/P_0$ for the relaxed clusters in Figure 15 do, for the most part, increase enough with the addition of noise to be consistent with the values of these ratios for relaxed clusters in the observed sample.

The above tests reveal that observational effects like noise can account for some difference in the power ratio distributions between observations and simulations, but that ``gastrophysics'' also plays a role.  It should be emphasized that the agreement in cluster structure between simulations and observations, at least at low redshift, is remarkably good with no statistically significant differences in either the means or distributions of the power ratios (except perhaps a mild offset in $P_2/P_0$).  The agreement indicates the general success of both the underlying cosmological model and the gas physics employed here.  The disagreements (like the broader distribution of power ratios in the simulations seen in Figure 13), on the other hand, suggest that with a more detailed treatment of observational effects cluster structure will provide a sensitive test of ``gastrophysics'' in simulations.  For example, our results suggest the need for a reduction in the peakiness of low mass cool cores but an increase in cooling in high mass clusters.  We may need an additional feedback mechanism like AGN or a slightly different balance between the cooling and heating prescriptions.  Another general success of these simulations is that they are the first set of simulations to produce both cool core and non-cool core clusters in the same volume (Burns et al.~2007).  However, the fraction of cool cores is low compared to observations of X-ray selected clusters, and there are very few high-mass cool cores in our sample (Burns et al.~2007), supporting our conclusions based on cluster structure.

In this section, we do not consider centroid shifts, because no low redshift observed sample of clusters using the same formulation of centroid shifts exists.  We note that our centroid shifts are very similar in range and median to those found by O'Hara et al. (2006), who define centroids within surface brightness contours rather than at at fixed radii.  For 45, $z < 0.2$ clusters, they find $\langle w \rangle$ between roughly 0.005 and 0.1 $r_{500}$ with a median value of $\sim0.02$.

\subsection{ Evolution of Structure }

Jeltema et al. (2005) showed using \textit{Chandra} observations that cluster structure evolves with redshift: high-redshift clusters are on average more disturbed than low-redshift clusters.  Specifically, the average $P_3/P_0$ power ratio increases with redshift.  The evolution of cluster structure has now been confirmed by Maughan et al. (2007) using centroid shifts and by Hashimoto et al. (2007) using asymmetry in the X-ray surface brightness distribution.  A direct comparison to these observations, particularly at high redshift, should include a more detailed treatment of selection and observational uncertainties, and we will defer an in depth discussion of this to a future paper.  In this section, we focus on whether or not cluster structure is observed to evolve in our simulations.

To quantify the significance of possible evolution in structure, we divided the clusters into three redshift bins of relatively equal time intervals: low redshifts ($z \le 0.25$), intermediate redshifts ($0.25 < z \le 0.65$), and high redshifts ($0.65 < z \le 1.5$).  We then used a rank-sum test to find the probability that the average power ratios and centroid shifts in these redshift bins are the same.  The results are given in Table 7.  For the power ratios, we list the results for both $R=r_{500}$ and, for better comparison to observations, $R=0.5$ Mpc.  We find a significant increase in the average centroid shifts with redshift, but no significant evolution in the power ratios.  For example, Figure 17 shows the median $P_3/P_0$ (top) and $\langle w \rangle_{no core}$ (bottom) within $R=r_{500}$ for each redshift bin in the simulations.  For comparison, the dashed lines indicate the lowest 20\% and highest 20\% of each structure measure versus redshift.  The median centroid shift does appear to increase slightly with redshift, while $P_3/P_0$ shows little or no evolution.  The right side of Figure 17 shows histograms of the distributions of $P_3/P_0$ and $\langle w \rangle_{no core}$ in the three redshift bins considered above; here the evolution in $\langle w \rangle_{no core}$ can also be seen.  These plots show, however, that any evolution in structure is mild compared to the range of cluster morphologies present at any redshift (i.e. there are relaxed and disturbed clusters at all redshifts).

\clearpage
\begin{figure}
\begin{center}
\epsscale{1.1}
\plottwo{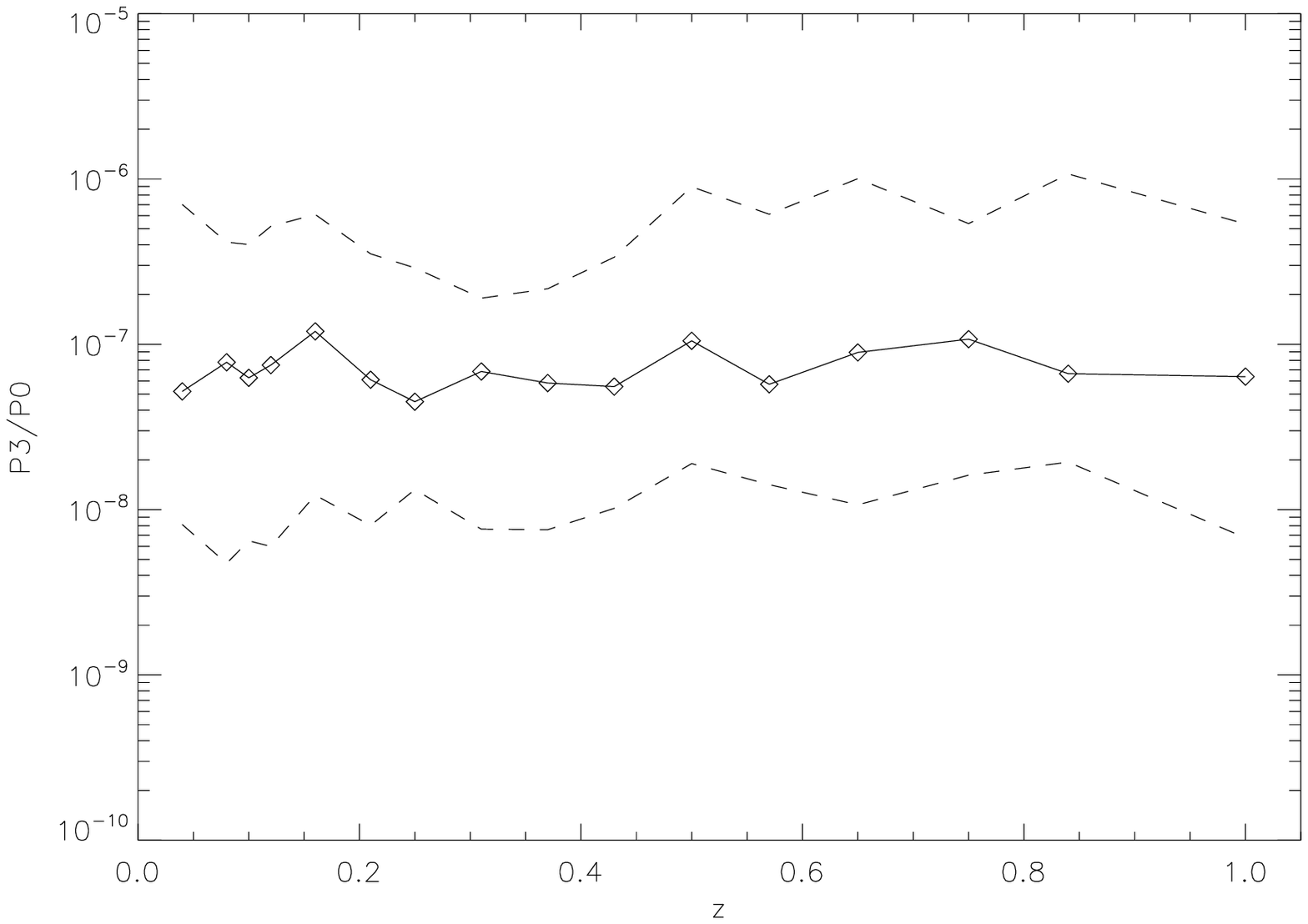}{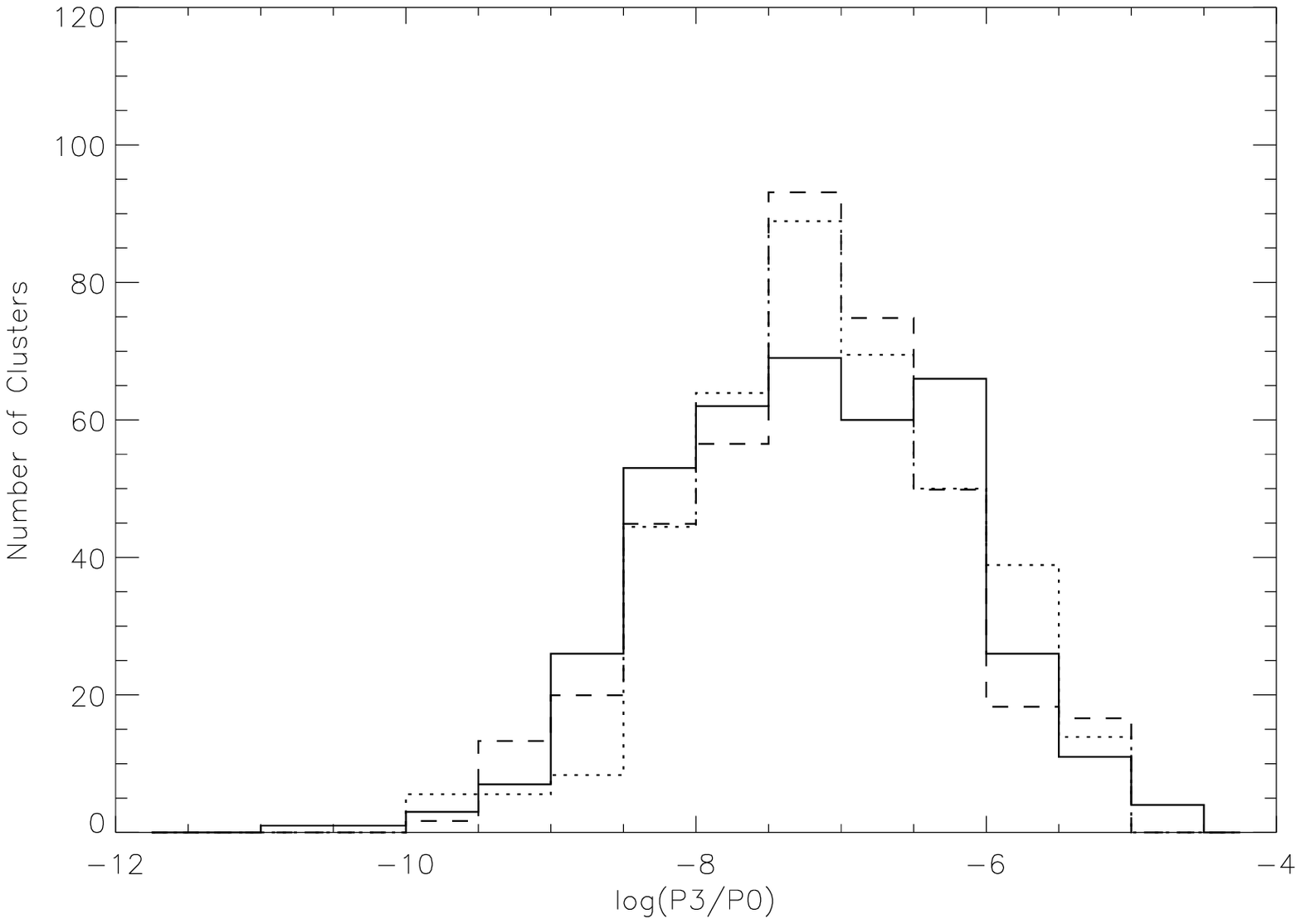}
\plottwo{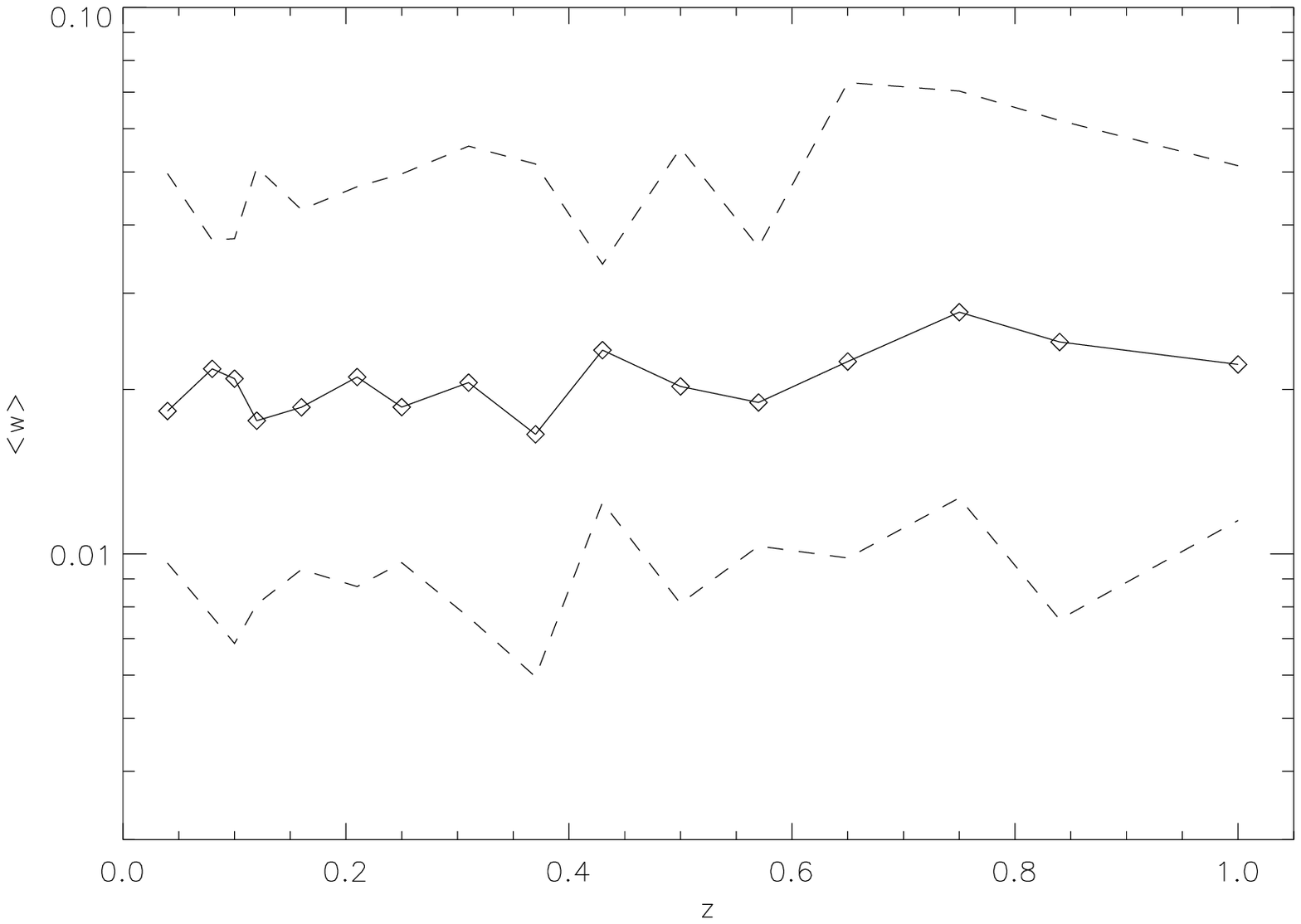}{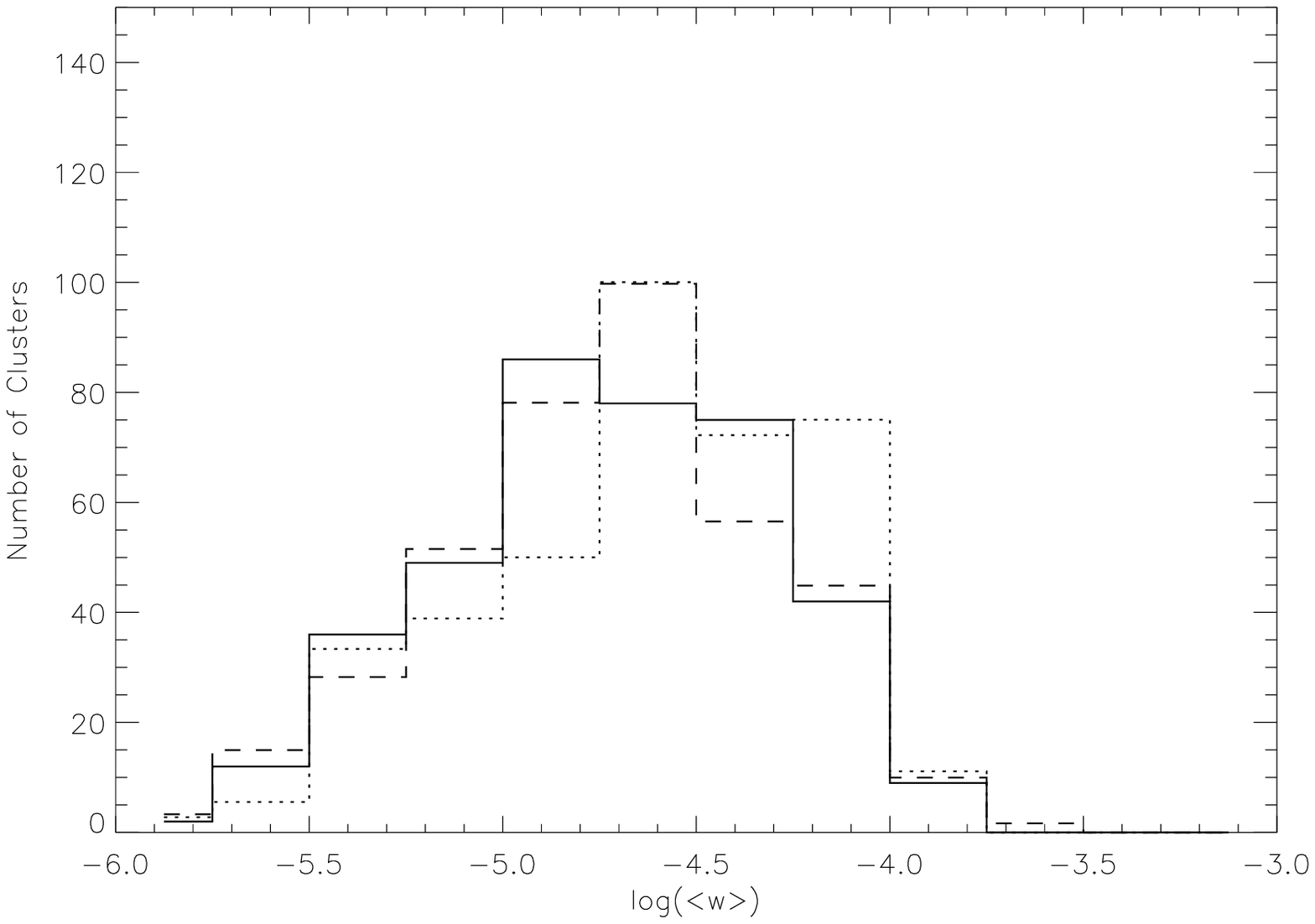}
\caption{ \textit{Left:} Median $P_3/P_0$ (top) and $\langle w \rangle_{no core}$ (bottom) within $R=r_{500}$ for each redshift bin in the simulations.  Also plotted are dashed lines indicating the lowest 20\% and highest 20\% of each structure measure versus redshift. \textit{Right:} Histograms showing the distribution of $P_3/P_0$ (top) and $\langle w \rangle_{no core}$ (bottom) in three redshift bins: $z \le 0.25$ (solid line), $0.25 < z \le 0.65$ (dashed), and $0.65 < z \le 1.5$ (dotted line). }
\end{center}
\end{figure}
\clearpage

Our simulated sample, however, is mass selected, while the observational samples detecting structure evolution are primarily composed of clusters from flux-limited X-ray surveys.  Jeltema et al. (2005) used a luminosity limit of $L_X > 2 \times 10^{44}$ ergs s$^{-1}$ in their sample selection, and for comparison we also list in Table 7 the probability of evolution in the simulations for this sample selection.  
With a luminosity rather than mass cut, we find significant evolution in all three power ratios and the centroid shifts for $R=r_{500}$.  This evolution appears to be mostly due to an increase in the average cluster structure from low to intermediate redshifts.  The stronger evolution of the power ratios in a luminosity-selected sample may be due to a preferential selection of relaxed, cool core systems (\S3.2) which show more significant cool cores at low redshifts.  For an aperture radius of $R=0.5$ Mpc, the same radius used by Jeltema et al. (2005), the power ratios do not show significant evolution (with the exception of $P_3/P_0$ between low and intermediate redshifts), but the use of a constant physical radius aperture could act to hide a mild evolution in cluster structure due to the fact that cluster virial radius decreases with increasing redshift.

Figure 18 compares the evolutions in $P_3/P_0$ (top) and $\langle w \rangle_{no core}$ (bottom) in the simulations to the current observational samples.  Here we use a luminosity cut of $L_X > 2 \times 10^{44}$ ergs s$^{-1}$ for both the observations and the simulations and appropriate choices of aperture radii to match the observations ($R=0.5$ Mpc for $P_3/P_0$ and $R=r_{500}$ for $\langle w \rangle_{no core}$).  The observed power ratios are taken from Buote \& Tsai (1996) for $z \le 0.1$ (lowest-redshift bin) and Jeltema et al. (2005) for higher redshifts (higher-redshift bins).  The observed centroid shifts are taken from Maughan et al. (2007).  The observational data points show the median value in redshift bins containing at least 20 clusters, with the exception of the highest-redshift $P_3/P_0$ bin which has 15 clusters.  As stated above, in this paper we will not seek a detailed quantitative comparison between structure evolution in simulations and observations, but since significant evolution in cluster structure is detected in observations, here we seek a qualitative comparison.

\clearpage
\begin{figure}
\epsscale{0.7}
\plotone{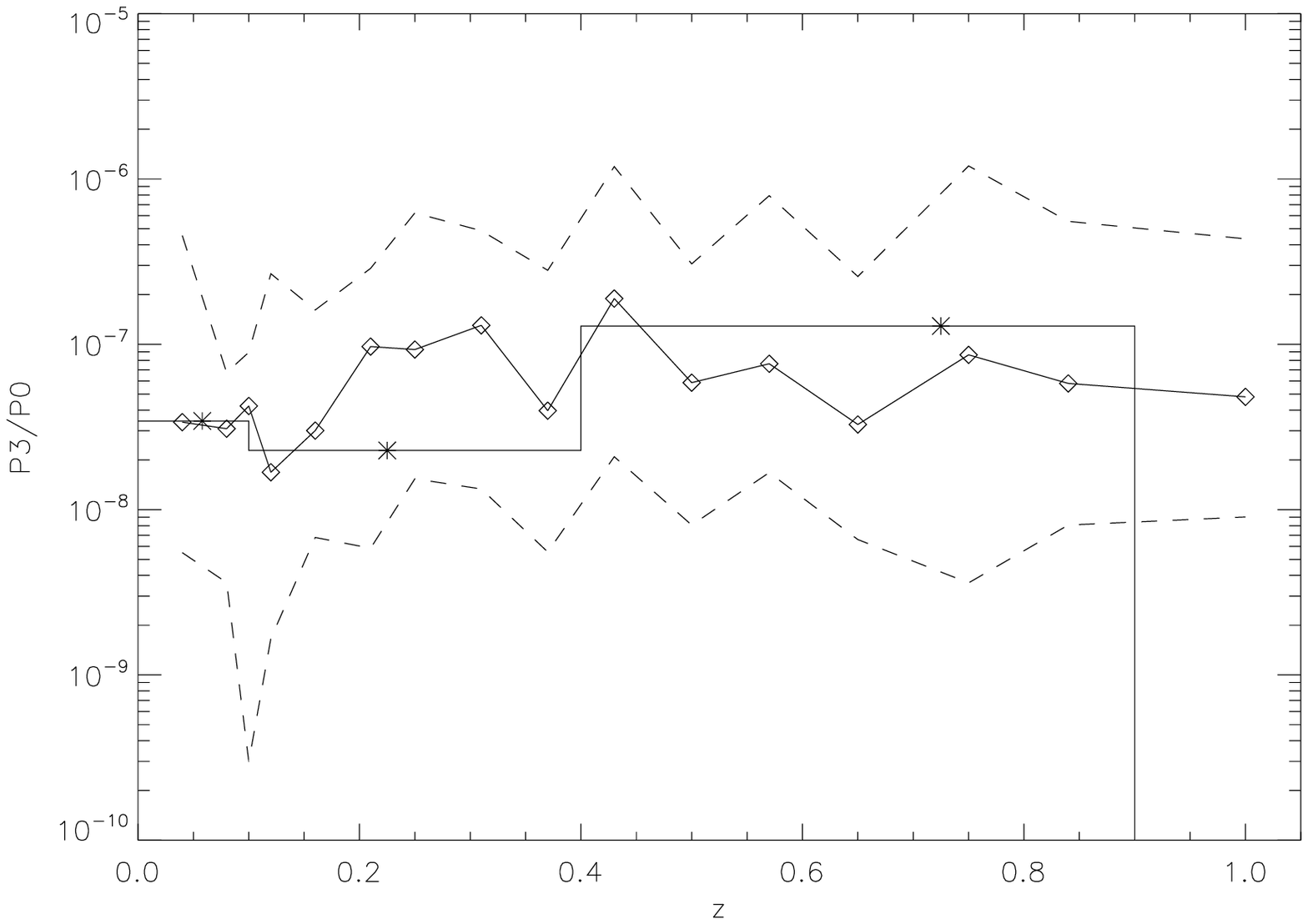}
\plotone{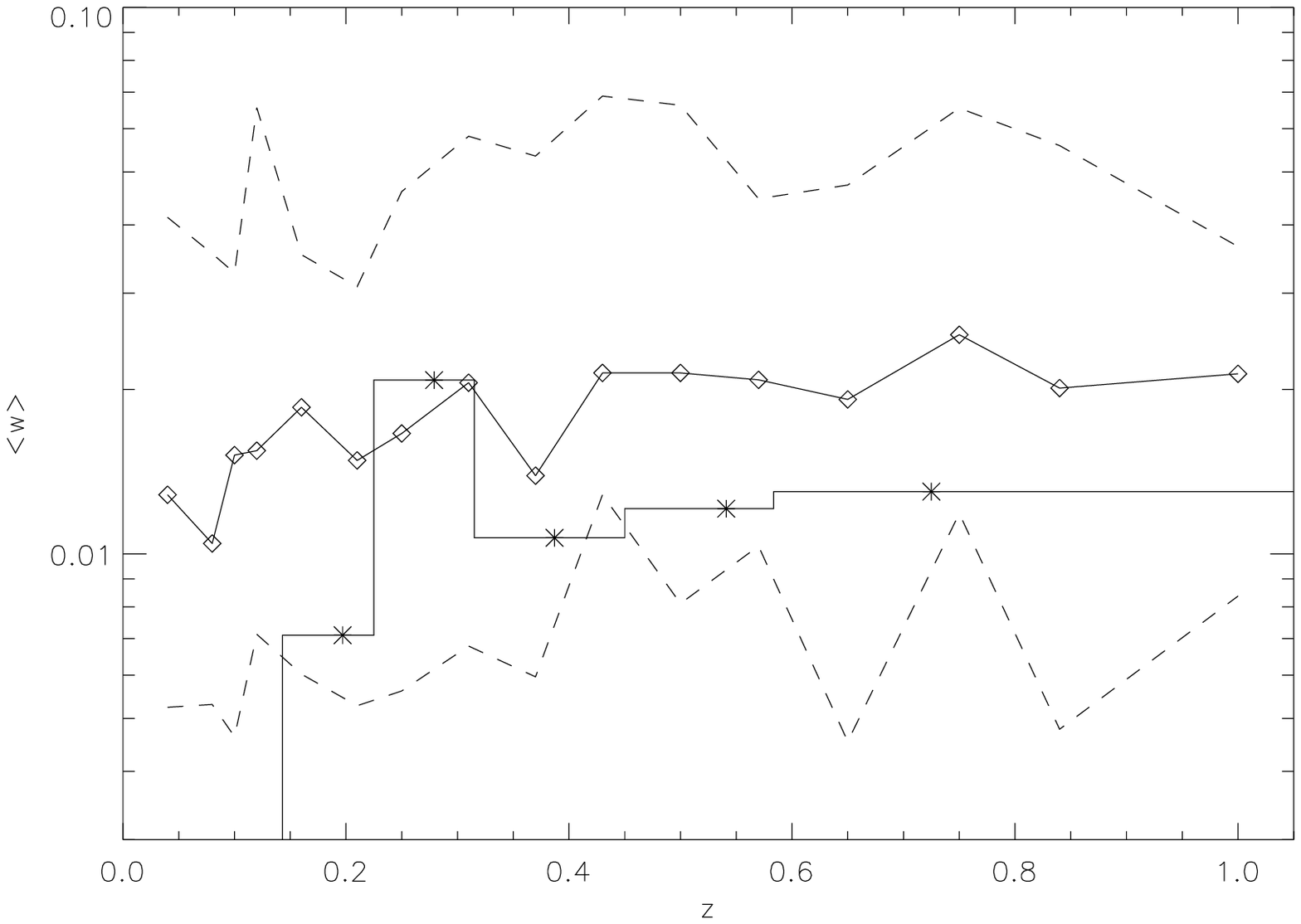}
\caption{ Evolutions in $P_3/P_0$ (top) and $\langle w \rangle_{no core}$ (bottom) in the simulations compared to the current observational samples.  We use a luminosity cut of $L_X > 2 \times 10^{44}$ ergs s$^{-1}$ for both the observations and the simulations and appropriate choices of aperture radii to match the observations ($R=0.5$ Mpc for $P_3/P_0$ and $R=r_{500}$ for $\langle w \rangle_{no core}$).  The observed power ratios are taken from Buote \& Tsai (1996) for $z \le 0.1$ (lowest-redshift bin) and Jeltema et al. (2005) for higher redshifts (higher-redshift bins).  The observed centroid shifts are taken from Maughan et al. (2007).  The observational data points (asterisks) show the median value in redshift bins containing at least 20 clusters, with the exception of the highest-redshift $P_3/P_0$ bin which has 15 clusters. These points are plotted at the median redshift of each bin with the histogram indicating the extent of the bin.  Simulations are plotted as in Figure 17. }
\end{figure}
\clearpage

The significant evolution in both $P_3/P_0$ and centroid shifts in the observations can be seen in Figure 18, but again this evolution is mild compared to the range of cluster morphologies present at any one redshift.  The observed $P_3/P_0$'s fall well within the range seen in our simulations at all redshifts; the main difference appears to be that in the simulations $P_3/P_0$ peaks at intermediate redshifts (at least for this choice of radius and luminosity cut) where the observed clusters have the highest $P_3/P_0$ at high redshift.  The centroid shifts observed by Maughan et al. (2007) for clusters in the \textit{Chandra} archive appear to be on average a bit lower than what is found in the simulations.  This offset is roughly a factor of two, similar to the (at best) marginally significant offset seen for $P_2/P_0$ in section 3.3, and within the scatter in $\langle w \rangle_{no core}$ seen in the simulations.  Since we have not used precisely the same derivation of centroid shifts as Maughan et al. (2007) (our definition and exclusion of the peak are different) nor have we reproduced all of the observational effects, we will not comment on the significance of this offset.  However, the observations do show a similar upward trend of the centroid shifts with redshift with a similar slope to what is found in our simulations.

In summary, on the observational side the results of Jeltema et al (2005) that clusters in X-ray selected samples are more disturbed at higher redshifts has been confirmed by two other studies using different samples and different structure measures (Maughan et al.~2007; Hashimoto et al.~2007), making this a fairly robust result.  
For a luminosity cut similar to that used by Jeltema et al.~(2005), we do find a significant evolution in the power ratios in our simulations; however, we find that this evolution depends both on the cluster selection and radius chosen.  In section 3.3, we found that at low redshift the distribution of power ratios in the simulations is a bit broader than observed likely due in part to our incomplete knowledge of gastrophysics.  It is possible that this greater spread in power ratios could hinder detection of a mild evolution in cluster structure.  We do find a significant increase in the average centroid shifts of clusters in our simulations with redshift independent of selection ($R=r_{500}$).  The centroid shifts appear to be more robust against projection (\S3.1) and less sensitive to deviations from hydrostatic equilibrium (\S3.2.3), possibly leading to a more robust evolution with redshift.  Evolution in a similar structure measure, the offset of the centroid from the surface brightness peak, was found by Kay et al. (2007) in the CLEF simulations, while Rahman et al. (2006) find indications of evolution in cluster ellipticity and multiplicity in low-redshift ($z \le 0.25$) simulations based on the same code used here.

We showed in section 3.2.3 that there is a strong correlation between cluster structure and bias in the hydrostatic mass relative to the true mass.  Combined with the evolution in structure with redshift, though mild, this will lead to systematically larger errors in mass estimates at high redshift.

\section{ SUMMARY AND CONCLUSIONS }

We use \textit{Enzo}, a hybrid Eulerian adaptive mesh refinement (AMR)/N-body code \\
\cite{NB99,O05}, to explore cluster morphology and its evolution in current generation cosmological simulations.  Rather than studying the impact of individual, idealized mergers on clusters, the large volume of these simulations allows us to study the development of cluster structure and its impact on observable cluster properties and scaling relations in a cosmological setting.  We also explore the effects of projection on apparent cluster morphology and how well the simulations reproduce observed cluster structure at low redshift.  We utilize and compare two commonly used measures of cluster structure, the power ratios and centroid shifts.  Our main conclusions are summarized below.

1) Projection:

Based on simulations of three orthogonal projections, we find that projection along the line of sight leads to significant scatter in the observed structure of individual clusters.  For example, less than 50\% of clusters which appear relaxed in one projection are truely relaxed, and 4-10\% of these ``relaxed'' clusters have very disturbed morphologies (for example, are undergoing major mergers) when viewed from another angle.

2) Correlations with cluster properties and scaling relations:

We do not find strong trends between cluster structure and mass, X-ray temperature, or $Y_X$, although they do correlate with centroid shift.  We do, however, find a trend for clusters with low power ratios to have higher luminosities and a significant offset of relaxed clusters toward higher luminosities in the $L_X-T_X$ and $L_X-M$ relations. This offset is partially corrected by removing the cluster cores from the luminosity, but a significant offset still remains.  When using the true cluster masses, we do not find a significant dependence in the $T_X-M$ and $Y_X-M$ relations on cluster structure.

3) Hydrostatic masses and the prospects for observable mass scaling relations:

We find that the hydrostatic masses typically underestimate true cluster mass and that the error in determining mass is strongly correlated with cluster structure.  For relaxed clusters, the hydrostatic masses are lower than the true masses by typically 12\%, while for disturbed clusters the typical error is doubled to 24\%.  This correlation shows that our structure measures do probe deviation from hydrostatic equilibrium.  We also find a weaker, but significant correlation of the deviation of hydrostatic mass from true mass with cluster mass.

The correlation of mass underestimate with cluster structure means that when the hydrostatic masses are used to form the mass-observable scaling relations ($T_X-M_{hydro}$ and $Y_X-M_{hydro}$), as would be done observationally, these relations depend significantly on cluster morphology.  The scatter in these relations also increases.  Even when using the most relaxed clusters to normalize the $T_X-M_{hydro}$ and $Y_X-M_{hydro}$ relations, they differ significantly from the true relations.  We can, however, use cluster structure to correct the hydrostatic masses.  Particularly when using the power ratios, correcting the hydrostatic masses improves the mass estimates for 75\% of the clusters, eliminates the offset between relaxed and disturbed clusters in the $T_X-M$ and $Y_X-M$ relations, and leads to good agreement with the true relations.  Some difference in slope versus the true relations remains due to the dependence of the hydrostatic mass error on cluster mass, and the scatter in the mass-scaling relations is only slightly reduced.  Here we do not remove substructures from our determination of average cluster properties, but this might allow the scatter in the mass-scaling relations to be reduced in high quality observations.

The underestimate in total mass under the assumption of hydrostatic equilibrium also leads to an overestimate of cluster gas mass fraction.  This error combines with a significant, but milder trend in cluster gas mass with structure.  Gas mass is systematically overestimated for more disturbed clusters, but only by on average $\sim$10\%.

4) Comparison to observations: 

In this paper, we do not attempt to reproduce all of the possible instrumental and observational effects, so we limit our comparison to observations of low-redshift cluster samples which have high S/N.  We find that there is no statistically significant difference in either the average power ratios or distribution in power ratios between simulations and observations (with the exception of a moderately significant difference in $P_2/P_0$ at small radii).  Our simulated centroid shifts are also similar to those found by O'Hara et al. (2006), but their method of calculating centroid shifts is slightly different.

Visual examination reveals that the distribution of power ratios in the simulations is broader than in observed clusters, including both clusters with higher power ratios and an extended tail of clusters with lower power ratios.  Some of this difference can be accounted for through the addition of noise to the simulations, but it is at least partially due to incomplete gas physics in the simulations.  In particular, our results suggest the need for a reduction in the peakiness of low mass cool cores but an increase in cooling in high mass clusters.  
The overall agreement between our simulations and observations indicates the general success of both the underlying cosmological model and the gas physics employed here, while the disagreements suggest that cluster structure will provide a sensitive test of ``gastrophysics'' in simulations.

5) Evolution in cluster structure:

Clusters at high redshift (above $z\sim0.5$), closer to the epoch of cluster formation, are observed to be on average more disturbed than low redshift clusters \cite{J05, M07, H07}.  In our simulations, we find a significant increase in the average centroid shifts with redshift, confirming that the evolution in cluster structure is also present in simulations.  If we select clusters based on luminosity, similar to the selection of observational samples, we also find a significant evolution in the power ratios of our simulated clusters; however, we find that this evolution depends both on the cluster selection and radius chosen.

In both observations and simulations, the evolution of cluster structure is mild compared to the range of cluster morphologies present at all redshifts, making its detection somewhat difficult and the comparison of evolution in observations and simulations sensitive to any differences in structure stemming from selection, gastrophysics, etc.  Though mild, the evolution in cluster structure combined with the strong correlation of structure with bias in the hydrostatic masses will lead to an increasing error in mass estimates with redshift.

6) Power ratios versus centroid shifts:

Overall, we find that the power ratios and centroid shifts perform relatively similarly as measures of cluster dynamical state.  Here we highlight a few of the areas where one measure performs better than another.  Centroid shift appears to be in some ways a more robust measure.  Clusters selected as relaxed or disturbed based on their centroid shifts suffer slightly less confusion due to projection, and the evolution in centroid shifts with redshift is more robust against sample selection.  The power ratios, on the other hand, appear to be a more sensitive measure.  They show a stronger correlation with deviations from hydrostatic equilibrium, shown in their stronger correlation to deviations in the hydrostatic masses from true mass.  They may also be more sensitive to differences in gas physics as they show a slightly stronger correlation to the location of clusters on scaling relations.  However, here we have not attempted a full comparison of these structure measures, and this comparison would benefit from simulations in different cosmologies and with different gas physics as well as further comparison to observations.

\section{ DISCUSSION }

In terms of cosmological studies, our results are encouraging in that cluster scaling relations like $T_X-M$ and $Y_X-M$ appear remarkably insensitive to cluster dynamical state.  The biggest source of error, as noted by previous authors \cite{Ka04,Rs06,N07}, is that estimates of the total mass based on the assumption of hydrostatic equilibrium are systematically low.  Here we also showed that the bias in hydrostatic mass is strongly correlated with structure.  We also note that the evolution in cluster structure with redshift combined with the correlation of mass error with structure leads to increasing errors in estimates of mass and gas mass fraction with redshift.  A bias exists even if only the most relaxed clusters are chosen to normalize the scaling relations or to measure gas mass fractions.  In addition, our results strongly caution against the accuracy that can be derived from individual clusters or small samples due to the large uncertainty in cluster structure from projection and the very large mass errors possible for individual ``relaxed'' clusters.

The goal of precision cosmology may then lead us to other means of normalizing the mass scaling relations, either though lensing studies (which have their own systematic biases) or, if deemed accurate enough, directly from simulations.  For example, the fitted relationship in Figure 9 between the power ratios and error in the hydrostatic masses can be used to eliminate systematic biases with cluster dynamical state.  This relationship is potentially testable with a study of a large sample of observed clusters of all morphologies to see if the normalizations of the mass scaling relations change with cluster structure as predicted in this paper.  The offset in the simulated mass scaling relations when using the hydrostatic masses rather than the true masses is enough to bring them in to agreement with observations (see also Nagai et al. 2007b).  It remains then to determine whether or not simulations are producing realistic estimates of the level of non-thermal pressure support in clusters.  In the future, it may be possible to directly observe the level of bulk motions in at least small samples of clusters through high resolution X-ray spectroscopy or possibly the combination of X-ray, SZ, and lensing data (e.g.~Puchwein \& Bartelmann 2007).  In the mean time, the models of cooling and feedback in current simulations are being tested in a number of ways.  Here we showed that cluster structure can provide a sensitive diagnostic of gas physics when compared to observations.

In this work, we have focused for the most part on cluster structure and its correlations in the simulations themselves, but in future work we will simulate mock observations of our clusters appropriate for the direct comparison to observed samples.  Here we have used relatively simple indicators of cluster dynamical state based on the X-ray surface brightness distribution.  The power ratios and centroid shift are more sophisticated than gross measures like ellipticity while they can still be measured fairly accurately in relatively shallow data (about the same depth needed for an accurate global temperature).  These measures are by no means the only possibilities (for example, in deeper data temperature structure could be a sensitive measure), but we have shown here that they are effective and correlated to dynamical state.

Another potentially important systematic effect of cluster structure on observational studies is its effect on cluster selection in surveys.  While properties like X-ray temperature and $Y_{X,SZ}$ when averaged over large radii like $r_{500}$ are relatively unbiased, mergers lead to spikes in both luminosity and central $Y_{SZ}$ (e.g. Ricker \& Sarazin 2001; Rowley et al. 2004; Motl et al. 2005; Poole et al. 2008).  For example, the most massive cluster in our simulations undergoes two major mergers between z=1.5 and z=0, both accompanied by dramatic increases in the X-ray luminosity near core passage when the cluster appears relaxed in all projections \cite{J07}.   In addition, cool core systems have luminosities which are high for their masses.  The bias introduced by cluster structure on cluster selection in X-ray and SZ surveys is not yet well understood (nor in optical and lensing surveys which may be effected by projection and central concentration).

\acknowledgments
We would like to thank Ben Maughan for kindly providing us with the details of his analysis and Daisuke Nagai, Brian O'Shea, and August Evrard for useful discussions relating to this work.  We would also like to sincerely thank the referee for their insightful comments on this work.  Support for this work was provided by the National Aeronautics and Space Administration through Chandra Award Numbers AR6-7014X and AR5-6016X.  J.O.B. and E.J.H acknowledge the support of NSF grant AST-0407368 and NASA grant NNX07AH53G.  T.E.J. is grateful for support from the Alexander F. Morrison Fellowship, administered through the University of California Observatories and the Regents of the University of California.  E.J.H. also acknowledges support from an NSF Astronomy and Astrophysics Postdoctoral Fellowship, NSF AST-0702923.

\clearpage

\begin{deluxetable}{lcccccccc}
\tabletypesize{\footnotesize}
\tablecaption{ Structure Measures }
\tablewidth{0pt}
\tablecolumns{9}
\tablehead{
\colhead{} & \colhead{$R=0.5$ Mpc} & \colhead{} & \colhead{} & \colhead{$R=r_{500}$} & \colhead{} & \colhead{} & \colhead{$R=1.0$ Mpc} \\
\colhead{} & \colhead{Median} & \colhead{Range} & \colhead{} & \colhead{Median} & \colhead{Range} & \colhead{} & \colhead{Median} & \colhead{Range}
}
\startdata
$P_2/P_0$ &5.57e-6 &5.28e-7-3.20e-5 &\phm{11} &2.56e-6 &2.70e-7-2.82e-5 &\phm{1} &2.05e-6 &1.60e-7-2.16e-5\\
$P_3/P_0$ &1.00e-7 &7.51e-9-1.03e-6 &\phm{11} &6.65e-8 &3.58e-9-1.07e-6 &\phm{1} &5.20e-8 &1.87e-9-1.37e-6 \\
$P_4/P_0$ &3.44e-8 &2.17e-9-4.28e-7 &\phm{11} &1.92e-8 &1.01e-9-5.10e-7 &\phm{1} &1.44e-8 & 6.0e-10-4.83e-7\\
$\langle w \rangle$     &0.0359 &0.00767-0.132 &\phm{11} &0.0192 &0.00458-0.0653 &\phm{1} &0.00996 &0.00251-0.0304\\
$\langle w \rangle_{nocore}$ &0.0379 &0.00852-0.138 &\phm{11} &0.0201 &0.00505-0.0692 &\phm{1} &0.0108 &0.00280-0.0342\\
\enddata
\tablecomments{ The ranges indicated in columns 3, 5, and 7 show the range spanned by clusters with power ratios/centroid shifts greater than the lowest 10\% and less than the highest 10\%.  The ranges are listed in this way to eliminate the most extreme clusters. }
\end{deluxetable}

\begin{deluxetable}{lcccccc}
\tabletypesize{\footnotesize}
\tablecaption{ Effects of Projection }
\tablewidth{0pt}
\tablecolumns{7}
\tablehead{
\colhead{} & \colhead{500 kpc} &\colhead{$r_{500}$} & \colhead{1 Mpc} & \colhead{$r_{500}$ ($\langle w \rangle_{no core}$)}
}
\startdata
\% disturbed &37.7 &30.9 &31.5 &29.6 \\
\% disturbed appear &55.4 &55.2 &52.7 &64.6 \\
disturbed in 1 proj. & & & & \\
\% relaxed &16.8 &11.3 &16.8 &8.6 \\
\% appear relaxed in 1 &10.5 &8.2 &10.6 &3.7 \\
proj. actually disturbed & & & & \\
\% appear relaxed in 1 &28.6 &38.1 &48.3 &45.0 \\
proj. actually relaxed & & & & \\
\enddata
\tablecomments{ Row 1 lists the percentage of clusters which are disturbed (i.e. have power ratios above our limits in at least one projection, see section 3.1 for definition).  Row 2 lists the percentage of disturbed clusters which appear disturbed in any one projection. Row 3 lists the fraction of clusters that are relaxed (see section 3.1 for definition) in all projections. Row 4 gives the fraction of clusters which appear relaxed in one projection but are disturbed in one of the other two projections, and row 5 lists the percentage of clusters that appear relaxed in one projection which are relaxed in all three projections.  }
\end{deluxetable}

\begin{deluxetable}{lcccc}
\tablecaption{ Correlation with Cluster Properties }
\tablewidth{0pt}
\tablecolumns{5}
\tablehead{
\colhead{Structure Measure} & \colhead{$M_{tot}(r \le r_{500})$} & \colhead{$L_X(r \le r_{500})$} & \colhead{$T_X(r \le r_{500})$} & \colhead{$Y_X(r \le r_{500})$} \\
\colhead{$(r \le r_{500})$} & \colhead{Probability} & \colhead{Probability} & \colhead{Probability} & \colhead{Probability}
}
\startdata
$P_2/P_0$ &0.94 &4.2e-6 &0.02 &8.7e-3 \\
$P_3/P_0$ &0.69 &2.3e-5 &0.63 &0.82 \\
$P_4/P_0$ &0.18 &5.9e-8 &0.70 &0.88 \\
$\langle w \rangle_{no core}$ &3.7e-3 &0.56 &0.030 &3.1e-3 \\
\enddata
\tablecomments{ Probability of no correlation between each cluster property and the power ratios and centroid shifts based on a Spearman rank-order correlation test. }
\end{deluxetable}

\clearpage
\begin{deluxetable}{lccccccccc}
\tabletypesize{\footnotesize}
\rotate
\tablecaption{ Mass Scaling Relations - Fixed Slope }
\tablewidth{0pt}
\tablecolumns{10}
\tablehead{
\colhead{} & \colhead{True Mass} & \colhead{} & \colhead{} & \colhead{Hydrostatic Mass} & \colhead{} & \colhead{} & \colhead{} & \colhead{Corrected Mass} & \colhead{} \\
\colhead{} & \colhead{all} & \colhead{} & \colhead{all} & \colhead{relaxed} & \colhead{disturbed} & \colhead{} & \colhead{all} & \colhead{relaxed} & \colhead{disturbed}
}
\startdata
$E_z^{-1}L_X-E_zM$ &$24.959\pm0.006$ &\phm{11} &$25.072\pm0.007$ &$25.106\pm0.014$ &$25.103\pm0.017$ &\phm{11} &$24.962\pm0.007$ &$25.033\pm0.014$ &$24.951\pm0.017$ \\
$T_X-E_zM$ &$-2.098\pm0.002$ &\phm{11} &$-2.043\pm0.003$ &$-2.060\pm0.004$ &$-2.019\pm0.008$ &\phm{11} &$-2.095\pm0.003$ &$-2.095\pm0.004$ &$-2.092\pm0.008$ \\
$Y_X-E_z^{-2/5}M$ &$-3.279\pm0.003$ &\phm{11} &$-3.143\pm0.006$ &$-3.192\pm0.008$ &$-3.070\pm0.015$ &\phm{11} &$-3.273\pm0.005$ &$-3.281\pm0.008$ &$-3.253\pm0.015$ \\
\enddata
\tablecomments{ Comparison of the normalizations of power law fits ($log(y) = A log(x) + B$) to the mass scaling relations using the true masses, the hydrostatic masses, and the hydrostatic masses corrected based on $P_3/P_0$.  For the hydrostatic and corrected masses separate fits are also given for the relaxed and disturbed subsamples.  The slopes have been fixed at their predicted self-similar values: 4/3 for $L_X-M$, 2/3 for $T_X-M$, and 5/3 for $Y_X-M$. }
\end{deluxetable}

\clearpage
\begin{deluxetable}{lccccccccccc}
\tabletypesize{\footnotesize}
\rotate
\tablecaption{ Mass Scaling Relations - Fit Slope }
\tablewidth{0pt}
\tablecolumns{12}
\tablehead{
\colhead{} & \colhead{} & \colhead{True Mass} & \colhead{} & \colhead{} & \colhead{} & \colhead{Hydrostatic Mass} & \colhead{} & \colhead{} & \colhead{} & \colhead{Corrected Mass} & \colhead{} \\
\colhead{} & \colhead{A} & \colhead{B} & \colhead{scatter} & \colhead{} & \colhead{A} & \colhead{B} & \colhead{scatter} & \colhead{} & \colhead{A} & \colhead{B} & \colhead{scatter}
}
\startdata
$E_z^{-1}L_X-E_zM$ &$1.369\pm0.035$ &$24.45\pm0.50$ &$0.131$ &\phm{11} &$1.216\pm0.036$ &$26.76\pm0.52$ &$0.152$ &\phm{11} &$1.223\pm0.036$ &$26.56\pm0.53$ &$0.155$ \\
$T_X-E_zM$ &$0.588\pm0.012$ &$-0.96\pm0.18$ &$0.078$ &\phm{11} &$0.527\pm0.014$ &$-0.04\pm0.21$ &$0.112$ &\phm{11} &$0.529\pm0.014$ &$-0.11\pm0.21$ &$0.110$ \\
$Y_X-E_z^{-2/5}M$ &$1.601\pm0.022$ &$-2.39\pm0.32$ &$0.045$ &\phm{11} &$1.435\pm0.032$ &$0.12\pm0.46$ &$0.085$ &\phm{11} &$1.442\pm0.032$ &$-0.09\pm0.46$ &$0.082$ \\
\enddata
\tablecomments{ Comparison of the power law fits ($log(y) = A log(x) + B$) to the mass scaling relations using the true masses, the hydrostatic masses, and the hydrostatic masses corrected based on $P_3/P_0$.  These fits include all clusters, and the slope is free to vary.  Scatter is measured as $[ \sum_i (log(x_i) + B/A - log(y_i)/A)^2/N ]^{1/2}$.}
\end{deluxetable}

\begin{deluxetable}{lccccccccc}
\tabletypesize{\footnotesize}
\tablecaption{ Comparison to Low Redshift Observations }
\tablewidth{0pt}
\tablecolumns{10}
\tablehead{
\colhead{} & \colhead{} & \colhead{$R=500$ kpc} & \colhead{} & \colhead{} & \colhead{} & \colhead{} & \colhead{$R=1$ Mpc} & \colhead{} & \colhead{}\\
\colhead{} & \colhead{Median Sims.} & \colhead{Median Obs.} & \colhead{Rank-Sum} & \colhead{KS} & \colhead{} & \colhead{Median Sims.} 
& \colhead{Median Obs.} & \colhead{Rank-Sum} & \colhead{KS}
}
\startdata
$P_2/P_0$ &4.4e-6 &2.1e-6 &0.02 &0.002 &\phm{11} &2.6e-6 &1.0e-6 &0.10 &0.07 \\
$P_3/P_0$ &4.3e-8 &3.4e-8 &0.40 &0.63 &\phm{11} &3.0e-8 &5.3e-8 &0.34 &0.14 \\
$P_4/P_0$ &2.1e-8 &1.1e-8 &0.17 &0.04 &\phm{11} &1.3e-8 &9.2e-9 &0.33 &0.12 \\
\enddata
\tablecomments{ Comparison of the power ratios of our simulated clusters to the \textit{ROSAT} observed sample from Buote \& Tsai (1996) at aperture radii of 0.5 Mpc and 1 Mpc. Columns 2 and 6 gives the median power ratios of the simulations compared to the medians of the observations in columns 3 and 7.  Columns 4 and 8 give the probability from a Wilcoxon rank-sum test  that the mean power ratios of the observed and simulated samples are the same.  Columns 5 and 9 give the probability from a Kolmogorov-Smirnov test that the distributions of power ratios are the same. }
\end{deluxetable}

\begin{deluxetable}{lcccccc}
\tabletypesize{\footnotesize}
\tablecaption{ Evolution }
\tablewidth{0pt}
\tablecolumns{7}
\tablehead{
\colhead{} & \colhead{} & \colhead{$R=r_{500}$} & \colhead{} & \colhead{} & \colhead{$R=0.5$ Mpc} & \colhead{} \\
\colhead{} & \colhead{Low vs. Mid} & \colhead{Mid vs. High} & \colhead{Low vs. High} & \colhead{Low vs. Mid} & \colhead{Mid vs. High} & \colhead{Low vs. High}
}
\startdata
Mass cut & & & & & & \\
$P_2/P_0$ &0.20$^*$ &0.061 &0.18 &0.31$^*$ &0.36$^*$ &0.19$^*$ \\
$P_3/P_0$ &0.40$^*$ &0.14 &0.17 &0.18 &0.28 &0.10 \\
$P_4/P_0$ &0.15$^*$ &0.077 &0.24 &0.38$^*$ &0.26 &0.30 \\
$\langle w \rangle_{no core}$ &0.44 &0.011 &0.0053 &- &- &- \\
\tableline
$L_X$ cut & & & & & & \\
$P_2/P_0$ &0.09 &0.16 &0.0092 &0.10 &0.015$^*$ &0.18$^*$ \\
$P_3/P_0$ &0.0043 &0.44 &0.0023 &0.0093 &0.13$^*$ &0.13 \\
$P_4/P_0$ &0.048 &0.23 &0.0058 &0.058 &0.12$^*$ &0.28 \\
$\langle w \rangle_{no core}$ &0.0036 &0.23$^*$ &0.023 &- &- &- \\
\enddata
\tablecomments{ Probability based on a rank-sum test that the means of each structure measure are the same between the specified redshift bins.  Three redshift bins are considered: low redshifts ($z \le 0.25$), intermediate redshifts ($0.25 < z \le 0.65$), and high redshifts ($0.65 < z \le 1.5$).  Probabilities are tabulated for clusters selected based both on a mass cut of $M > 2 \times 10^{14} M_{\odot}$ and a luminosity cut of $L_X > 2 \times 10^{44}$ ergs s$^{-1}$\\ (*) Denotes that the mean of the structure measure decreases between redshift bins rather than increasing. }
\end{deluxetable}


\begin{thebibliography}{9999}

\bibitem[Akritas \& Bershady(1996)]{AB96} Akritas, M.~G., \& 
Bershady, M.~A.\ 1996, \apj, 470, 706

\bibitem[Brickhouse et al.1995]{Br95} Brickhouse, N.~S., 
Raymond, J.~C., \& Smith, B.~W.\ 1995, \apjs, 97, 551 

\bibitem[Buote \& Tsai 1996]{B96} Buote, D.~A., \& Tsai, 
J.~C.\ 1996, \apj, 458, 27 

\bibitem[Buote \& Tsai 1995]{B95} Buote, D.~A., \& Tsai, 
J.~C.\ 1995, \apj, 452, 522 

\bibitem[Burns et al. 2007]{B07} Burns, J.~O., Hallman, E.~J., Gantner, B., Motl, P.~M., \& Norman, M.~L 2007, ApJ, submitted

\bibitem[Cen \& Ostriker 1992]{CO92} Cen, R., \& Ostriker, 
J.~P.\ 1992, \apjl, 399, L113 

\bibitem[Cohn \& White(2005)]{2005APh....24..316C} Cohn, J.~D., \& White, 
M.\ 2005, Astroparticle Physics, 24, 316

\bibitem[Dolag et al. 2005]{D05} Dolag, K., Vazza, F., 
Brunetti, G., \& Tormen, G.\ 2005, \mnras, 364, 753 

\bibitem[Eisenstein \& Hu 1999]{EH99} Eisenstein, D. J., \& Hu, W. 1999, ApJ, 511, 5

\bibitem[Flores et al. 2007]{F07} Flores, R.~A., Allgood, 
B., Kravtsov, A.~V., Primack, J.~R., Buote, D.~A., \& Bullock, J.~S.\ 2007, 
\mnras, 377, 883

\bibitem[Hallman et al. 2006]{H06} Hallman, E.~J., Motl, 
P.~M., Burns, J.~O., \& Norman, M.~L.\ 2006, \apj, 648, 852

\bibitem[Hallman et al. 2007]{Ha07} Hallman, E.~J., Motl, 
P.~M., Burns, J.~O., Norman, M.~L. \& Wagner, R.~P. 2007, ApJ, in prep.

\bibitem[Hashimoto et al. 2007]{H07} Hashimoto, Y., 
B{\"o}hringer, H., Henry, J.~P., Hasinger, G., \& Szokoly, G.\ 2007, \aap, 
467, 485

\bibitem[Jeltema et al. 2005]{J05} Jeltema, T.~E., 
Canizares, C.~R., Bautz, M.~W., \& Buote, D.~A.\ 2005, \apj, 624, 606 

\bibitem[Jeltema et al. 2008]{J07} Jeltema, T.~E.,
Hallman, E.~J., Burns, J.~0., \& Motl, P.~M. 2008, in prep.

\bibitem[Jones \& Forman 1999]{JF99} Jones, C., \& Forman, 
W.\ 1999, \apj, 511, 65 

\bibitem[Kay et al. 2004]{Ka04} Kay, S.~T., Thomas, P.~A., 
Jenkins, A., \& Pearce, F.~R.\ 2004, \mnras, 355, 1091 

\bibitem[Kay et al. 2007]{K07} Kay, S.~T., da Silva, 
A.~C., Aghanim, N., Blanchard, A., Liddle, A.~R., Puget, J.-L., Sadat, R., 
\& Thomas, P.~A.\ 2007, \mnras, 377, 317

\bibitem[Kazantzidis et al. 2004]{K04} Kazantzidis, S., 
Kravtsov, A.~V., Zentner, A.~R., Allgood, B., Nagai, D., \& Moore, B.\ 
2004, \apjl, 611, L73

\bibitem[Kolokotronis et al. 2001]{K01} Kolokotronis, V., 
Basilakos, S., Plionis, M., \& Georgantopoulos, I.\ 2001, \mnras, 320, 49 

\bibitem[Kravtsov et al. 2006]{K06} Kravtsov, A.~V., 
Vikhlinin, A., \& Nagai, D.\ 2006, \apj, 650, 128 

\bibitem[Mathiesen \& Evrard 2001]{ME01} Mathiesen, B.~F., 
\& Evrard, A.~E.\ 2001, \apj, 546, 100

\bibitem[Maughan et al. 2007]{M07} Maughan, B.~J., Jones, 
C., Forman, W., \& Van Speybroeck, L.\ 2007, ApJS, accepted

\bibitem[Maughan 2007]{M07b} Maughan, B.~J.\ 2007, ApJ, 668, 772 

\bibitem[Mazzotta et al. 2004]{Ma04} Mazzotta, P., Rasia, 
E., Moscardini, L., \& Tormen, G.\ 2004, \mnras, 354, 10

\bibitem[Mohr et al. 1993]{M93} Mohr, J.~J., Fabricant, 
D.~G., \& Geller, M.~J.\ 1993, \apj, 413, 492 

\bibitem[Mohr et al. 1995]{M95} Mohr, J.~J., Evrard,
A.~E., Fabricant, D.~G., \& Geller, M.~J.\ 1995, \apj, 447, 8

\bibitem[Motl et al. 2004]{M04} Motl, P.~M., Burns, J.~O., 
Loken, C., Norman, M.~L., \& Bryan, G.\ 2004, \apj, 606, 635 

\bibitem[Motl et al. 2005]{M05} Motl, P.~M., Hallman, 
E.~J., Burns, J.~O., \& Norman, M.~L.\ 2005, \apjl, 623, L63 

\bibitem[Nagai et al. 2007a]{N07} Nagai, D., Vikhlinin, A., 
\& Kravtsov, A.~V.\ 2007a, \apj, 655, 98

\bibitem[Nagai et al. 2007b]{N07b} Nagai, D., Kravtsov, 
A.~V., \& Vikhlinin, A.\ 2007b, ApJ, 668, 1

\bibitem[Norman \& Bryan 1999]{NB99} Norman, M.~L., \& 
Bryan, G.~L.\ 1999, ASSL Vol.~240: Numerical Astrophysics, 19 

\bibitem[O'Hara et al. 2006]{O06} O'Hara, T.~B., Mohr, 
J.~J., Bialek, J.~J., \& Evrard, A.~E.\ 2006, \apj, 639, 64 

\bibitem[O'Shea et al. 2005]{O05} O'Shea, B. W., Bryan, G., Bordner, J., Norman, M. L., Abel, T., Harkness, R., \& Kritsuk, A. 2005, in Adaptive Mesh Refinement: Theory and Applications, ed. T. Plewa, T. Linde \& V. G. Weirs (Berlin: Springer), 341

\bibitem[Poole et al. 2006]{P06} Poole, G.~B., Fardal, 
M.~A., Babul, A., McCarthy, I.~G., Quinn, T., \& Wadsley, J.\ 2006, \mnras, 
373, 881

\bibitem[Poole et al. 2007]{P07} Poole, G.~B., Babul, A., 
McCarthy, I.~G., Fardal, M.~A., Bildfell, C.~J., Quinn, T., \& Mahdavi, A.\ 
2007, MNRAS, 380, 437

\bibitem[]{593}
Press, W. H., Teukolsky, S. A., Vetterling, W. T., \& Flannery, B. P. 1992, Numerical Recipes in C (2d ed.; New York: Cambridge University Press)

\bibitem[Puchwein \& Bartelmann(2007)]{PB07} Puchwein, E., 
\& Bartelmann, M.\ 2007, A\&A, 474, 745

\bibitem[Rahman et al. 2006]{R06} Rahman, N., Krywult, J., 
Motl, P.~M., Flin, P., \& Shandarin, S.~F.\ 2006, \mnras, 367, 838 

\bibitem[Rasia et al. 2004]{Rs04} Rasia, E., Tormen, G., \& 
Moscardini, L.\ 2004, \mnras, 351, 237

\bibitem[Rasia et al. 2006]{Rs06} Rasia, E., et al.\ 2006, 
\mnras, 369, 2013

\bibitem[]{608}
Richstone, D., Loeb, A., \& Turner, E. L. 1992, ApJ, 393, 477

\bibitem[Ricker \& Sarazin 2001]{RS01} Ricker, P.~M., \& 
Sarazin, C.~L.\ 2001, \apj, 561, 621

\bibitem[Ritchie \& Thomas 2002]{RT02} Ritchie, B.~W., \& 
Thomas, P.~A.\ 2002, \mnras, 329, 675

\bibitem[Roettiger et al. 1996]{R96} Roettiger, K., Burns, 
J.~O., \& Loken, C.\ 1996, \apj, 473, 651 

\bibitem[Rowley et al. 2004]{R04} Rowley, D.~R., Thomas, 
P.~A., \& Kay, S.~T.\ 2004, \mnras, 352, 508 

\bibitem[Schindler \& Mueller 1993]{SM93} Schindler, S., \& 
Mueller, E.\ 1993, \aap, 272, 137 

\bibitem[Schuecker et al. 2001]{S01} Schuecker, P., Bohringer, H., Reiprich, T.H., \& Feretti, L. 2001, A\&A, 378, 408

\bibitem[Vikhlinin et al. 2005]{V05} Vikhlinin, A., 
Markevitch, M., Murray, S.~S., Jones, C., Forman, W., \& Van Speybroeck, 
L.\ 2005, \apj, 628, 655 

\bibitem[Voit 2005]{Vo05} Voit, G.~M.\ 2005, Reviews of 
Modern Physics, 77, 207

\bibitem[]{635}
Walpole, R. E., \& Myers, R. H. 1993, Probability and Statistics for Engineers and Scientists (5th ed.; New York: Macmillan Publishing Company)

\end{thebibliography}
\end{document}